\documentclass[onecolumn,11pt]{article}
\usepackage{jheppub}
\usepackage[font={small}]{caption}
\usepackage{subfigure}
\usepackage{graphicx}% Include figure files
\usepackage{dcolumn}% Align table columns on decimal point
\usepackage{float}
\usepackage[normalem]{ulem}

\usepackage{mathrsfs} 

\usepackage{multicol}
\usepackage{cancel}
\usepackage{mathtools}
\usepackage{hyperref}

\usepackage{makecell}

\usepackage{subfigure}
\usepackage{xcolor}
\def\({\left(} \def\){\right)}
\def\[{\left[} \def\]{\right]}

\newcommand{\eg}{{\it e.g.,}\ }
\newcommand{\ie}{{\it i.e.,}\ }

\newcommand{\bea}{\begin{eqnarray}}
\newcommand{\eea}{\end{eqnarray}}
\def\p{\partial}
\usepackage{changepage}

\newcommand{\beq}{\begin{equation}}
\newcommand{\eeq}{\end{equation}}

\def\le{\left(}
\def\ri{\right)}

\renewcommand{\eqref}[1]{(\ref{#1})}

\begin{document}

\title{The Cosmological Switchback Effect II}

\author{Stefano Baiguera$^{1}$ and Rotem Berman$^{1}$}

\affiliation{$^1$Department of Physics, Ben-Gurion University of the Negev, \\ David Ben Gurion Boulevard 1, Beer Sheva 84105, Israel}

\emailAdd{baiguera@post.bgu.ac.il}
\emailAdd{bermar@post.bgu.ac.il}

\abstract{\sloppy 
Recent developments in static patch holography proposed that quantum gravity in de Sitter space admits a dual description in terms of a quantum mechanical theory living on a timelike surface near the cosmological horizon.
In parallel, geometric observables associated with the Einstein-Rosen bridge of a black hole background were suggested to compute the computational complexity of the state dual to a gravitational theory.
In this work, we pursue the study of the complexity=volume and complexity=action conjectures in a Schwarzschild-de Sitter geometry perturbed by the insertion of a shockwave at finite boundary times.
This analysis extends previous studies that focused either on the complexity=volume 2.0 conjecture, or on the case of a shockwave inserted along the cosmological horizon.
We show that the switchback effect, describing the delay in the evolution of complexity in reaction to a perturbation, is a universal feature of the complexity proposals in asymptotically de Sitter space.
The geometric origin of this phenomenon is related to the causal connection between the static patches of de Sitter space when a positive pulse of null energy is inserted in the geometry.}

\maketitle

\setcounter{tocdepth}{1}

\section{Introduction}

Holography has provided a remarkable tool relating gravitational physics inside a spacetime region with a dual quantum theory living on its boundary \cite{tHooft:1993dmi,Susskind:1994vu}.
While this framework has been successfully applied to anti-de Sitter (AdS) space \cite{Maldacena:1997re}, there are several hints towards the application of the
holographic principle to geometries with positive cosmological constant, such as de Sitter (dS) space (\eg see \cite{Galante:2023uyf} for a recent review). 
In this paper, we pursue the static patch holography approach (described below) by investigating the reaction of dS space to matter perturbations, as measured by holographic complexity (\eg see \cite{Chapman:2021jbh} for a review).

\paragraph{Holography in de Sitter space.}
Gibbons and Hawking observed that the cosmological horizon of dS space has a thermodynamical interpretation in terms of entropy, similar to the black hole setting \cite{PhysRevD.15.2738}.
Inspired by the AdS case, the idea that dS space might have a dual interpretation in terms of a (Euclidean) conformal field theory (CFT) living on its (spacelike) boundaries $\mathcal{I}^{\pm}$ led to the development of the dS/CFT correspondence \cite{Fischler:1998st,Balasubramanian:2001nb,Witten:2001kn,Strominger:2001pn}.
The major drawbacks of this program are that the CFT presents unconventional features, it is hard to probe the event horizon and its features, and there is not a true identification between unitary quantum systems.\footnote{This is in contrast with the central dogma of dS space, which requires that the cosmological horizon should be described from its inside as a unitary quantum system with a finite number of degrees of freedom \cite{Shaghoulian:2021cef}. }
Inspired by recent developments on the topic, in this work we will focus instead on the so-called \textit{static patch holography} approach, which assumes that the dual quantum theory lives on  the \textit{stretched horizon}, a timelike surface located just inside the cosmological horizon \cite{Bousso:1999dw,Banks:2000fe,Bousso:2000nf,Banks:2001yp,Banks:2002wr,Dyson:2002nt,Dyson:2002pf,Banks:2005bm,Banks:2006rx,Anninos:2011af,Banks:2018ypk,Banks:2020zcr,Susskind:2021omt,Susskind:2021dfc}.\footnote{An alternative approach, that we will not consider in this work, is based on the embedding of dS space inside AdS background, thus providing a standard asymptotic boundary where a dual theory can be defined \cite{Freivogel:2005qh,Lowe:2010np,Fischetti:2014uxa,Anninos:2017hhn}.
Other approaches have been studied in \cite{Alishahiha:2004md,Nomura:2017fyh,Nomura:2019qps,Murdia:2022giv}. }

This framework presents several advantages. 
First, it naturally associates the bulk time running along the stretched horizon with the time coordinate of the dual quantum theory.
Second, it nicely fits with the discovery that a dual quantum system to dS space could be constructed in terms of an operator algebra on the worldline of an observer in the static patch \cite{Chandrasekaran:2022cip,Witten:2023qsv,Witten:2023xze,Mirbabayi:2023vgl}. 
The timelike boundary is also necessary to define a sensible themodynamics \cite{Banihashemi:2022htw}, and to build a concrete dual theory to three-dimensional dS space in terms of $T\bar{T}$ deformations of two-dimensional CFTs, followed by the addition of a cosmological constant term \cite{Lewkowycz:2019xse,Shyam:2021ciy,Coleman:2021nor,Batra:2024kjl}.
This approach also captures logarithmic corrections to the entropy \cite{Anninos:2020hfj}, which were interpreted holographically in terms of an extension of the Ryu-Takayanagi formula \cite{Maldacena:2012xp,Susskind:2021esx,Shaghoulian:2022fop}.
Finally, the static patch holography framework was used to match the two-point functions and the spectrum of a one-dimensional double scaled SYK model with a gravity model in three-dimensional dS space \cite{Narovlansky:2023lfz,Rahman:2023pgt,Verlinde:2024znh,Verlinde:2024zrh,Rahman:2024vyg}, and to reproduce features related to energy conservation and scrambling \cite{Milekhin:2023bjv}.\footnote{The von Neumann algebra of double scaled SYK model was studied in \cite{Xu:2024hoc}.}

\paragraph{Holographic complexity proposals.}
In quantum information, computational complexity heuristically counts the number of unitary operators required to perform a certain task, or to engineer a certain state.
This quantity started to play a prominent role in high-energy physics with the observation that the dynamics of the Einstein-Rosen bridge (ERB) in a black hole setting could not be captured by entanglement entropy.
Instead, complexity was proposed to be the right quantity describing the interior growth of the wormhole \cite{Susskind:2014moa}.
Since then, several conjectures have been introduced to find an appropriate geometric quantity in the bulk associated with the complexity
of the thermofield double state.
In this work, we will focus on two holographic proposals: complexity=volume (CV) and complexity=action (CA).
The CV conjecture relates complexity to the induced maximal volume $\mathcal{B}$ on a codimension-one slice anchored at the boundary $\Sigma$ \cite{Susskind:2014rva}
\beq
\mathcal{C}_V (\Sigma) = \max_{\Sigma= \partial \mathcal{B}} \frac{\mathcal{V}(\mathcal{B})}{G_N \ell} \, ,
\label{eq:intro_CV}
\eeq
where $\ell$ is an appropriate length scale in the bulk geometry (typically, the (A)dS radius).
The CA proposal computes the gravitational on-shell action $I_{\rm WDW}$ inside the Wheeler-De Witt (WDW) patch, \ie the causal domain of dependence of the ERB \cite{Brown:2015bva,Brown:2015lvg}
\beq
\mathcal{C}_A = \frac{I_{\rm WDW}}{\pi \hbar} \, .
\eeq
Together with complexity=volume 2.0 (CV2.0) \cite{Couch:2016exn},
CV and CA belong to a large class of proposals known as complexity=anything (CAny) \cite{Belin:2021bga,Belin:2022xmt,Jorstad:2023kmq}.
The guiding principle to build these holographic conjectures is that they all reproduce the following two characteristic features of computational complexity: \textbf{(1)} a linear increase for late times, and \textbf{(2)} the switchback effect, \ie a delay in their growth as a consequence of inserting a perturbation in the system.
These universal properties were all shown to be valid for black holes in asymptotically AdS space, with the perturbation being modelled by a shockwave of null matter inserted from the boundary \cite{Stanford:2014jda,Brown:2015bva,Brown:2015lvg,Couch:2016exn,Lehner:2016vdi,Carmi:2016wjl,Chapman:2016hwi,Carmi:2017jqz,Chapman:2018dem,Chapman:2018lsv,Belin:2021bga,Belin:2022xmt,Jorstad:2023kmq}.

\paragraph{Holographic complexity in de Sitter space.}
One can get valuable insights on the properties and differences among the complexity proposals by moving away from AdS space towards backgrounds with different features and asymptotics (for some examples, see \cite{Alishahiha:2018tep,Auzzi:2018pbc,Auzzi:2018zdu,Chapman:2018bqj,Braccia:2019xxi,Sato:2019kik,Goto:2018iay,Auzzi:2021nrj,Baiguera:2021cba,Auzzi:2021ozb,Sato:2021ftf}).
Inspired by recent trends, the notion of holographic complexity was extended to dS space by requiring that the geometric observables defined above are anchored at the stretched horizon, instead of the asymptotic boundary of AdS space \cite{Susskind:2021esx}.\footnote{For an alternative approach to holographic complexity in dS space using dS/dS correspondence, see \cite{Geng:2019ruz}.}

The first distinguishing feature of holographic complexity in dS space is the so-called \textit{hyperfast growth}, \ie it admits a divergent rate at finite boundary time \cite{Jorstad:2022mls}.
The geometric reason for this behaviour is that both the codimension-one surfaces and the WDW patch include divergent contributions coming from timelike infinities $\mathcal{I}^{\pm}$.
From the perspective of the quantum theory, this phenomenon was interpreted to arise from circuits which involve a large number of qubits in each step of the time evolution \cite{Lin:2022nss}. 
The hyperfast growth also occurs in two dimensions \cite{Chapman:2022mqd,Anegawa:2023wrk}, in models of inflation where a bubble of dS is contained inside AdS \cite{Auzzi:2023qbm}, and in the presence of shockwaves \cite{Baiguera:2023tpt,Anegawa:2023dad}.
Notable exceptions to this trend are provided by a certain class of codimension-one CAny observables, which exhibit a persistent linear growth \cite{Aguilar-Gutierrez:2023zqm}; and by gravitational observables accessing both the cosmological and the black hole regions of a Schwarzschild-de Sitter background \cite{Aguilar-Gutierrez:2024rka}.

The second distinguishing feature of holographic complexity in dS space is the switchback effect.
The insertion of a gravitational shockwave at finite boundary time in an asymptotically dS geometry induces a transition between two black holes with different masses \cite{Hotta_1993,PhysRevD.47.3323,Sfetsos:1994xa,Aalsma:2021kle}.
In this setting, it was shown in \cite{Baiguera:2023tpt} that the CV2.0 conjecture admits a pleateau around $t=0$ when complexity is approximately constant, similar to the AdS case. Furthermore, the duration of the plateau regime increases when the shockwave is inserted at earlier times, and shows signatures of scrambling characteristic to chaotic systems.
In the case of CV, CV2.0 and CA conjectures, it was shown in \cite{Anegawa:2023dad} that the hyperfast growth is always delayed by the insertion of a shockwave with small energy (\ie inserted along the cosmological horizon).
In the same setting, reference \cite{Aguilar-Gutierrez:2023pnn} revealed that the switchback effect is also displayed by the above-mentioned codimension-one CAny observables which do not admit hyperfast growth.
The same result applies to the case of multiple shockwaves.

\paragraph{Novelties of this work and main results.}
This paper is a direct continuation of the analysis performed in \cite{Baiguera:2023tpt}, with the aim to show that the switchback effect is a universal feature that also happens in the case of CV and CA conjectures.
The novelties, compared to reference \cite{Anegawa:2023dad}, are that we will consider a shockwave inserted at finite boundary time, and we will provide analytic expressions valid in generic dimensions $d \geq 2$.\footnote{We work in $(d+1)$--dimensional asymptotically dS space.}

We anticipate the main results of this work.
Following the same trend as the CV2.0 case, both the CV and CA proposals admit a time interval during the evolution when holographic complexity is approximately constant, before admitting a hyperfast growth at finite time.
When the shockwave is inserted at early times in the past, the duration $t_{\rm pl}$ of this plateau regime asymptotically approaches a linear increase parametrized by
\beq
t_{\rm pl} = 4 (t_w - t_*) \, , 
\label{eq:linear_plateau_intro}
\eeq
where $-t_w$ is the insertion time of the shockwave, and $t_*$ is the time it takes the system to scramble a perturbation.
A general analytic description of the scrambling time can be achieved by considering the following double-scaling limit
\beq
\varepsilon \rightarrow 0 \, , \qquad
\rho \rightarrow 1 \, , \qquad
\frac{1-\rho}{\varepsilon} \quad \mathrm{fixed} \, ,
\label{eq:double_scaling_intro}
\eeq
where $\varepsilon$ is a parameter describing the energy carried by the shockwave, and $\rho$ determines the location of the stretched horizon (when $\rho \rightarrow 1$, it approaches the cosmological horizon).
The assumptions \eqref{eq:double_scaling_intro} are physically relevant because the holographic boundary is located very close to the cosmological horizon (as required by static patch holography), the perturbation induced by the shockwave is small, but these two regimes are fine-tuned in such a way that both are relevant.
In this setting, the scrambling time in dimensions $d\geq 2$ reads
\begin{subequations}
\beq
t^{\rm SdS_{d+1}}_* =  \frac{1}{2 \pi T_{c1}} 
\log \left[   \frac{1-\rho}{\alpha \varepsilon} \le r_{c1}- r_{h1} \ri  \right]
+ \mathcal{O} (1-\rho, \varepsilon) \, ,
\label{eq:scrambling_time_SdS_intro}
\eeq
where $\alpha$ is defined by
\beq
r_{c2} = r_{c1} + \alpha \varepsilon + \mathcal{O}(\varepsilon^2) \, ,
\label{eq:rc2_rc1_intro}
\eeq
\end{subequations}
and where $r_{c1} \leq r_{c2}$ are the cosmological horizons of the black hole before and after the shockwave (respectively), $r_{h1} \geq r_{h2}$ are the black hole horizons, and $T_{c1}$ is the temperature of the cosmological horizon.
Analytic expressions away from the double-scaling limit \eqref{eq:double_scaling_intro} can be achieved in three dimensions, as we will find in eq.~\eqref{eq:duration_plateau_SdS3}.

For comparison, the three-dimensional AdS-Vaidya geometry admits a plateau regime whose scrambling time, in the limit of light shocks, reads \cite{Chapman:2018lsv}
\beq
t_*^{\rm Vaidya} \underset{\varepsilon \ll 1}{\approx} \frac{1}{2 \pi T_1} \log \le \frac{2}{\varepsilon} \ri \, ,
\label{eq:scrambling_Vaidya}
\eeq
where $\varepsilon$ is related to the jump in the mass of the black hole, and it has still the interpretation of energy carried by the shockwave.
We observe the following facts about eq.~\eqref{eq:scrambling_time_SdS_intro}:
\begin{itemize}
    \item Similar to the Vaidya case, there is a logarithmic dependence on $\varepsilon$, and the result is inversely proportional to the Hawking temperature.
    These features are usually typical of chaotic systems.
    \item There is a novel logarithmic dependence $\log(1-\rho)$ on the location of the stretched horizon. When the latter is taken closer to the cosmological horizon, the scrambling time increases.
    \item The scrambling time depends on quantities associated with both the event horizons, since Schwarzschild-de Sitter black holes are not in thermal equilibrium.\footnote{Indeed, stationary observers in Schwarzschild-de Sitter background experience thermal radiation from both the horizons, unless they are very close to one of them. In such case, the corresponding horizons provides a dominant flux of thermal radiation.} 
    \item The expression contains universal information about dS geometries, since it applies to the CV, CV2.0 and CA cases.
    Despite the different geometrical objects involved in these conjectures,
    we might ultimately interpret this universality as a consequence of the causal properties of dS space. 
    Indeed, the Penrose diagram grows taller after the insertion of matter perturbations \cite{Gao:2000ga}, allowing for causal contact between the static patches.
\end{itemize}

\paragraph{Outline.}
The paper is organized as follows.
In section \ref{sec:geometric_preliminaries} we briefly review black hole solutions in asymptotically dS space perturbed by a shockwave, including an analysis of their causal structure. Of great importance is the introduction of the stretched horizon, which defines a notion of time for the putative dual quantum theory.
Section \ref{sec:CA} contains the evaluation of the CA conjecture, including the time dependence of the WDW patch, a generic analytic computation and a numerical analysis in certain examples.
We show evidence for the existence of the switchback effect.
Next, we investigate the CV conjecture in section \ref{sec:CV_conjecture}, by studying the time evolution of the maximal surface and of its induced volume.
We show that the switchback effect is realized in this setting, too.
We summarize our results and discuss possible future developments in section \ref{sec:discussion}.
Appendices \ref{app:details_CA} and \ref{app:details_CV} are devoted to additional technical details on the evaluation of CA and CV proposals, respectively.

\section{Geometric preliminaries} 
\label{sec:geometric_preliminaries}

Black hole solutions in asymptotically dS space present an interesting causal structure, composed in the general case by an inflating region with a cosmological horizon and by a black hole patch with a corresponding event horizon. 
We review the main features of these geometries in subsection \ref{ssec:SdS_BH}.
We then proceed to perturb them with the insertion of a spherically-symmetric shockwave in subsection \ref{ssec:perturbation_shocks}, which describes a transition between black holes with different masses.
In view of the computation of geometric observables in the context of static patch holography, we define in subsection \ref{ssec:stretched_horizon} 
the stretched horizon, \ie the location where a  putative dual quantum theory should be defined.

\subsection{Schwarzschild-de Sitter black hole}
\label{ssec:SdS_BH}

\subsubsection{General dimensions}

Schwarzschild-de Sitter (SdS) black hole in $d+1$ dimensions provides a maximally symmetric solution of vacuum Einstein's equation in the presence of a positive cosmological constant, coming from the action \cite{Kottler,PhysRevD.15.2738,Spradlin:2001pw}
\beq
I = \frac{1}{16 \pi G_N} \int d^{d+1} x \, \sqrt{-g} \, \le R - 2 \Lambda \ri \, , \qquad
\Lambda = \frac{d(d-1)}{2 L^2} \, ,
\label{eq:action_EH_gend}
\eeq
In terms of static coordinates, the metric reads
\beq
ds^2 = - f(r) dt^2 + \frac{dr^2}{f(r)} + r^2 d\Omega_{d-1}^2 \, ,  \qquad
f (r) = 1 - \frac{2 m}{r^{d-2}} - \frac{r^2}{L^2} \, ,
\label{eq:asympt_dS}
\eeq
where $L$ is the dS curvature radius, $m$ a parameter related to the asymptotic mass of the black hole (\eg see \cite{Balasubramanian:2001nb,Ghezelbash:2001vs} for more details), $d\Omega_{d-1}^2$ is the line element of the spherical sections $S^{d-1}$, and $f(r)$ is referred to as the blackening factor.
In this coordinate system, $r=\infty$ represents the location of timelike infinities $\mathcal{I}^{\pm}$, while $r=0$ corresponds to the singularity of the black hole.

In any number of dimensions, the choice $m=0$ leads to empty dS space with cosmological horizon of radius $L$.
In this work we will generically assume $d\geq2$ and $m \in (0,m_{\rm cr})$, where the critical mass and radius are defined as follows:
\beq
m_{\rm cr} \equiv \frac{r_{\rm cr}^{d-2}}{d}  \, , \qquad
r_{\rm cr} \equiv L \sqrt{\frac{d-2}{d}} \, .
\label{eq:critical_mass_SdS}
\eeq
In this regime, the blackening factor admits two real roots $r_h <r_c$ corresponding to a black hole horizon (the smaller one) and a cosmological horizon (the larger one).

The case $d=2$ is special because the critical mass \eqref{eq:critical_mass_SdS} vanishes, there is only a cosmological horizon, and the black hole singularity disappears.
We will treat this case separately below.
Another peculiar configuration is the Nariai geometry, obtained as the limiting case $m=m_{\rm cr}$ where the two roots of the blackening factor approach each other, \ie $r_h \rightarrow r_c$.
Since the blackening factor $f(r)$ is infinitesimal in this regime, the proper distance between the two event horizons does not vanish.
An appropriate analysis of this near-horizon limit requires a rescaling of the coordinates, and one can ultimately map the Nariai geometry to $\mathrm{dS}_2 \times S^{d-1}$ (\eg see \cite{Anninos:2012qw,Svesko:2022txo,Maldacena:2019cbz} for more details).
We leave the study of this configuration for future investigations.

The SdS background presents a non-vanishing temperature and entropy associated with the thermal radiation from both the event horizons, given by \cite{PhysRevD.15.2738} 
\beq
     T_{h(c)} = \frac{1}{4 \pi} \left|\frac{\partial f(r)}{\partial r}\right|_{r=r_{h(c)}}  
    \qquad
    S_{h(c)} =  \frac{\Omega_{d-1} r_{h(c)}^{d-1}}{4 G_N} \, .
\label{eq:temp_entropy_SdS_general}  
\eeq
In terms of the horizon radii, the Hawking temperatures read \cite{Morvan:2022ybp}
\beq 
T_h  = d \, \frac{r_{\rm cr}^2 - r_h^2 }{4 \pi r_h L^2}  \, , \qquad
    T_c  = d \, \frac{r_c^2 -r_{\rm cr}^2 }{4 \pi r_c L^2}  \, .
   \label{eq:hor_cosm_temp_SdS}
\eeq
For any mass parameter in the range $m \in (0, m_{\rm cr})$, we find $T_h > T_c$, implying that the background is out of equilibrium.\footnote{There are few cases when an asymptotically dS geometry is in thermal equilibrium and presents a unique global temperature: in empty dS space ($m=0$), in three dimensions ($d=2$), and in the Nariai limit ($m \rightarrow m_{\rm cr}$) \cite{Anninos:2012qw,Bousso:1997wi}.}

\begin{figure}[ht]
    \centering
    \includegraphics[scale=0.45]{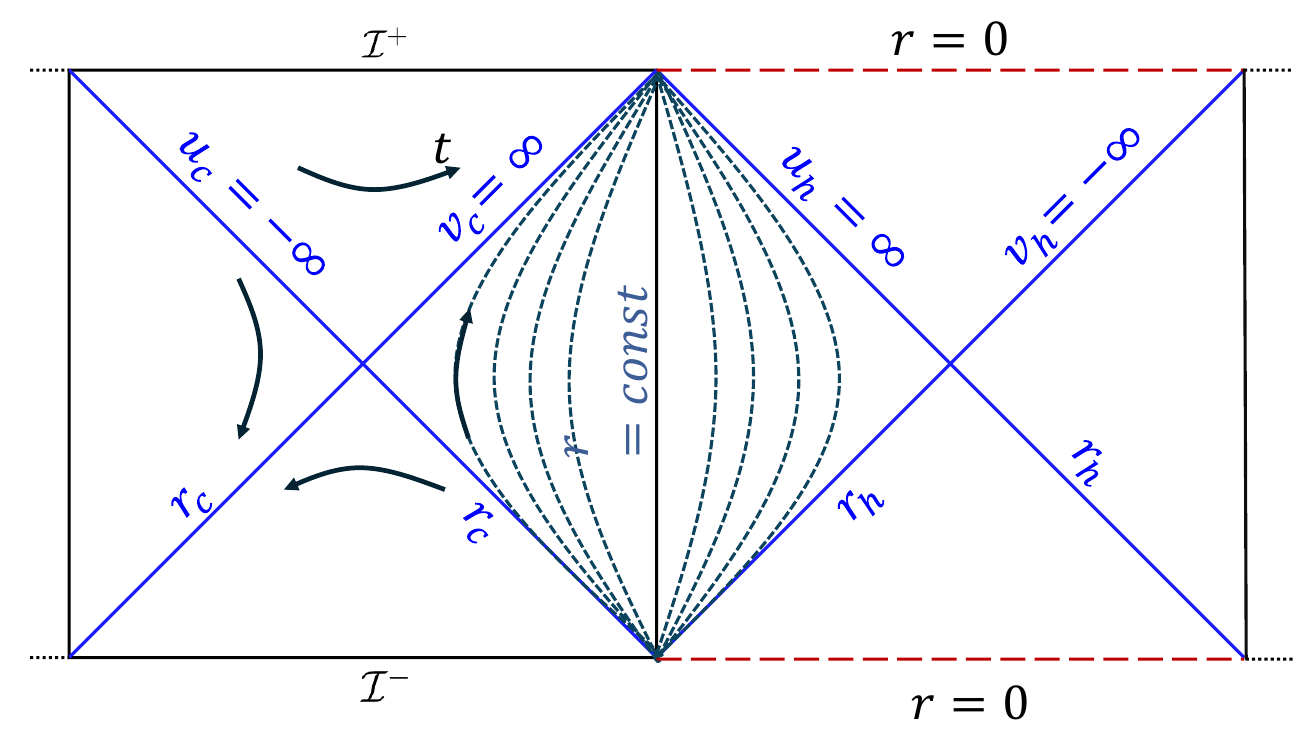}
    \caption{Penrose diagram of SdS$_{d+1}$ space in dimensions $d\geq 3,$ in the regime $m \in (0,m_{\rm cr}).$ 
    $r_h$ denotes the black hole horizon and $r_c$ the cosmological horizon. Black arrows denote the orientation of the Killing vector $\partial_t$.  }
    \label{fig:Penrose_SdS}
\end{figure}

Next, we discuss the causal structure of the spacetime.
A priori, the coordinate system in eq.~\eqref{eq:asympt_dS} only covers the static patch, \ie the region outside the black hole and inside the cosmological horizon.
In order to analytically extend the geometry beyond the horizons, we introduce the Eddington-Finkelstein (EF) (also called null) coordinates
\beq
u = t - r^* (r) \, , \qquad
v = t + r^* (r) \, ,
\label{eq:general_null_coordinates}
\eeq
defined in terms of the tortoise coordinate
\beq
r^*(r) = \int_{r_{0}}^r \frac{d r'}{f(r')} \, .
\label{eq:general_tortoise_coordinate}
\eeq
In this expression, $r_0$ is an arbitrary constant that we can always select such that $r^*(r\rightarrow \infty)=0$.\footnote{When $d=2$ or if the mass vanishes, the same choice of the integration constant also implies $r^*(r\rightarrow 0)=0$.}
After performing the change of variables \eqref{eq:general_null_coordinates}, the metric in EF form reads
\beq
ds^2 = - f(r) du^2 - 2 du dr + r^2 d\Omega_{d-1}^2 \, .
\label{eq:generic_null_metric_noshock}
\eeq
The maximal analytic extension of the geometry is achieved by using Kruskal coordinates $(U,V)$, which are defined in the static patch by the following transformations
\beq
 U_c =  e^{\frac{u}{\ell}}  \, , \quad
V_c = - e^{-\frac{v}{\ell}}  \, ,\qquad U_h =  -e^{-\frac{u}{\ell}}  \, , \quad
V_h = e^{\frac{v}{\ell}}  \, .
\label{eq:Kruskal_coord_sec2} 
\eeq
where $\ell$ is an arbitrary length scale (for instance, one can choose $\ell=L$).
To cover all the geometry, here we introduced two sets of Kruskal variables: $(U_h,V_h)$ cover the patch with $r\in (0,r_c)$, while $(U_c,V_c)$ cover the region with $r\in (r_h,\infty)$.
Both the coordinate systems are well-defined in the static patch $r \in (r_h, r_c)$, where one is allowed to move from one chart to the other. 
The Penrose diagram of the SdS background is depicted in fig.~\ref{fig:Penrose_SdS}.
In the following, we will refer to the left side of the causal diagram (containing the cosmological horizon) as the \textit{cosmological patch}, and to the right side (containing the black hole horizon) as the \textit{black hole patch}.

Next, we focus on the case $d=2$, where most of the concrete examples in this work will be given.

\subsubsection{Three dimensions}

\begin{figure}[ht]
    \centering
    \includegraphics[scale=0.5]{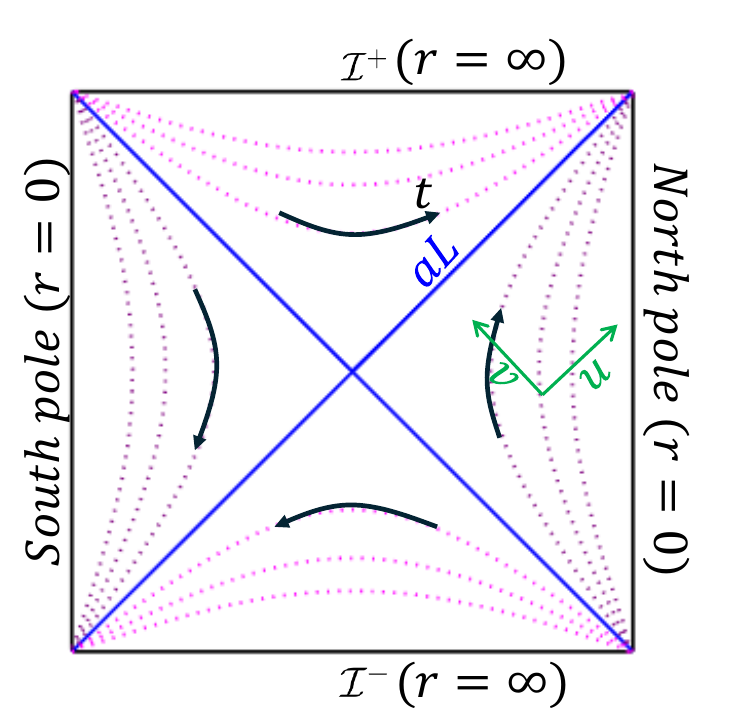}
    \caption{Penrose diagram of three-dimensional SdS background. The blue lines represent the cosmological horizons ($r=L$), the horizontal black lines are the future and past timelike infinity $\mathcal{I}^{\pm}$ (at $r=\infty$) 
    and the vertical black lines represent the north pole (right) and the south pole (left), located at $r=0$ along the worldline of an observer.  }
    \label{fig:Penrose_dS}
\end{figure}

In dimension $d=2$, the blackening factor $f(r)$ in eq.~\eqref{eq:asympt_dS} simplifies to 
\beq
f (r) = 1 - 8 G_N \mathcal{E} - \frac{r^2}{L^2}  \, ,
\label{eq:metric_SdS3}
\eeq
where we conventionally rescaled the mass parameter in terms of the energy $\mathcal{E}$ of the solution as $m\equiv4 G_N \mathcal{E}$ \cite{DESER1984405,Spradlin:2001pw}.
As anticipated below eq.~\eqref{eq:critical_mass_SdS}, this geometry admits a single (cosmological) event horizon located at 
\beq
r_c = a L  \, , \qquad
a \equiv \sqrt{1 - 8 G_N \mathcal{E}} \, .
\label{eq:cosmological_horizon_SdS3}
\eeq
The geometry is in thermal equilibrium with Hawking temperature and entropy determined by
\beq
T_{\rm SdS_3} = \frac{a}{2 \pi L} \, , \qquad
S_{\rm SdS_3} = \frac{\pi a L}{2 G_N} \, .
\label{eq:Hawking_temperature_SdS3}
\eeq
The case $\mathcal{E}=0$ corresponds to empty dS space and it is the solution with maximal entropy.
Notice that whenever a mass $m \ne 0$ is introduced in empty dS space, it is possible to extract entropy, but only up to a maximal value (such that the square root in eq.~\eqref{eq:cosmological_horizon_SdS3} remains real).
In three dimensions, the worldline associated with matter located at the origin $r=0$ of dS space creates a defect. 
Indeed, the SdS$_3$ black hole can be obtained as a discrete quotient of empty dS space with a conical deficit.
This identification can be made explicit by performing the following change of coordinates
\beq
\tilde{t} = a \, t \, , \qquad 
\tilde{r} = \frac{r}{a} \, , 
\qquad  \tilde{\theta} = a \, \theta \, ,
\label{eq:map_SdS3_dS3}
\eeq
which maps the SdS$_3$ geometry to dS$_3$ with cosmological horizon of length $L$.
For this reason, the Penrose diagram of the three-dimensional black hole solution is the same as empty dS space, see fig.~\ref{fig:Penrose_dS}.
The only difference is the existence of a conical singularity arising due to the change in the periodicity of the angular coordinate, which now presents a deficit angle $\alpha_{\rm def} = 2 \pi \le 1 - a \ri $ at the origin.

The topology of the background is $\mathbb{R} \times S^d$.
The vertical black lines on the far right (left) side of the picture represent the worldlines of two observers located on the north (south) pole, located at $r=0$.
A horizontal cross section in the causal diagram corresponds to the spatial sphere $S^d$.
The region delimited by $r\in [0,r_c)$ is the static patch, \ie the portion in causal contact with an observer at the corresponding pole.
The curves at constant radial coordinate depicted in the static patch in fig.~\ref{fig:Penrose_dS} represent the timelike trajectories of inertial observers, and will be used in subsection \ref{ssec:stretched_horizon} to define the stretched horizon.

In this geometry, null directions are identified by the EF coordinates \eqref{eq:general_null_coordinates} with tortoise coordinate
\beq
r^*(r) = \frac{L}{2 a} 
\log \left| \frac{a L+ r}{a L  - r} \right| \, .
\label{eq:rstar_SdS}
\eeq

\subsection{Perturbation with shockwaves}
\label{ssec:perturbation_shocks}

The main goal of this work is to understand how SdS space reacts to perturbations. 
This operation can be performed at the level of the bulk geometry by inserting a shockwave sourced by null matter propagating along a spherically symmetric null surface \cite{Hotta_1993,PhysRevD.47.3323,Sfetsos:1994xa,Aalsma:2021kle}.
The result of this procedure is to induce a transition between a black hole with mass $m_1$ and another black hole with mass $m_2$, where the label 1 refers to the region before the shockwave insertion, while 2 denotes the region after the shockwave.
From a physical perspective, these backgrounds can be envisioned as toy models for the outside of a spherically symmetric star, where the shockwave carries a certain amount of mass away from the star.

By requiring that the pulse of null matter satisfies the null energy condition (NEC), we obtain that the mass always decreases in SdS space \cite{Baiguera:2023tpt}.
Equivalently, there is a transition between event horizons such that $r_{c1} \leq r_{c2}$ and $r_{h1} \geq r_{h2}$.
In particular, the increase of the radius of the cosmological horizon will be crucial for the existence of special configurations of the geometric objects characterizing holographic complexity in this work.

The metric with the shockwave perturbation reads
\bea
& ds^2 = - F(r,u) du^2 - 2 dr du + r^2 d\Omega_{d-1}^2 \, , &
\label{eq:dS_metric_shock_wave}
\\
& F(r,u) = f_1 (r) \le 1-\theta (u-u_s) \ri + 
f_2 (r) \theta(u-u_s) \, , &
\eea
where $u_s$ is the constant null coordinate along which the shockwave propagates, and $\theta$ is the Heaviside distribution.
The regions before and after the shockwave are separately described by the metric \eqref{eq:asympt_dS}, each with its mass parameter.
In other words, the blackening factor is given by
\beq
\begin{aligned}
 &   u < u_s \, : \quad
    F(r,u) = f_1(r) = 1 - \frac{2m_1}{r^{d-2}} - \frac{r^2}{L^2} \, , & \\
    &   u > u_s \, : \quad
    F(r,u) = f_2(r) = 1 - \frac{2m_2}{r^{d-2}} - \frac{r^2}{L^2} \, . &
\end{aligned}
\label{eq:blackening_factor_shock_SdS}
\eeq
Accordingly, the tortoise coordinate is defined by
\beq
r^*(r)  = \int_{r_{0,1}}^r \frac{dr'}{f_1(r')} \le 1-\theta (u-u_s) \ri + \int_{r_{0,2}}^r \frac{dr'}{f_2(r')}  \theta (u-u_s)  \, ,
\label{eq:general_def_tortoise_coordinate_shock}
\eeq
where eq.~\eqref{eq:general_tortoise_coordinate} has been separately used in each side of the geometry.

\begin{figure}[ht]
    \centering
   \includegraphics[scale=0.41]{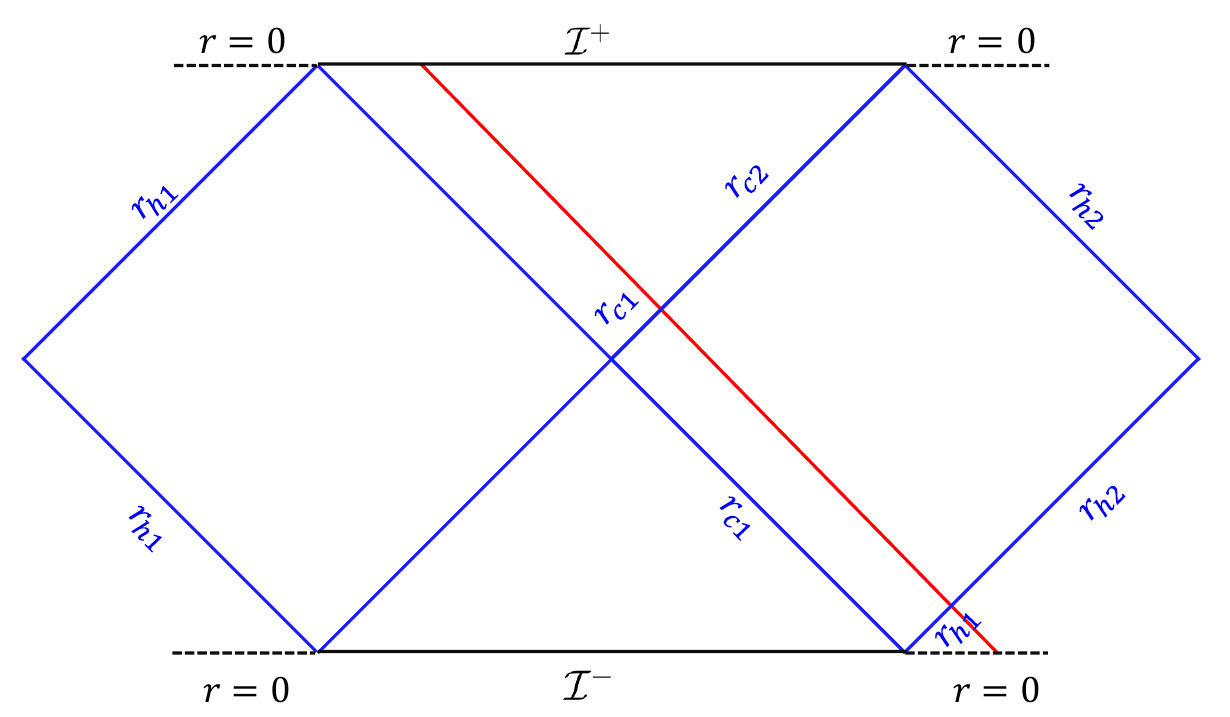}
    \caption{Penrose diagram of the SdS black hole in the presence of a shockwave in $d+1$ dimensions for $d>2$. $r_{c1}$, $r_{c2}$ and $r_{h1}$, $r_{h2}$ corresponds to the cosmological and black hole horizons before and after the shockwave, respectively. }
    \label{fig:Penrose_shock} 
\end{figure}

The Penrose diagram of this geometric setting is depicted in fig.~\ref{fig:Penrose_shock}. 
Let us stress that the coordinates $r$ and $u$ are continuous across the shockwave, while the time $t$ and the outgoing null coordinate $v$ are discontinuous, as a consequence of the jump in the blackening factor.

Finally, it is convenient to introduce the following dimensionless parameter $\varepsilon$ associated with the energy carried by the shockwave:
\beq
 \varepsilon \equiv  \begin{cases}
1- \mathcal{E}_2/\mathcal{E}_1  & \mathrm{SdS}_3  \\
1- m_2/m_1  & \mathrm{SdS}_{d+1} \quad (d \geq 3) . \\
\end{cases}
\label{eq:generic_epsilon_geometries}
\eeq
The NEC implies that $\varepsilon \in [0,1]$.
In particular, a light shockwave corresponds to $\varepsilon \ll 1$.

\subsection{Stretched horizons}
\label{ssec:stretched_horizon}

\subsubsection{Definition in asymptotic dS geometries}
\label{ssec:stretched_hor_dS}

The stretched horizon $r_{\rm st}$ is defined as a timelike surface at constant $r$ in the coordinate system with metric \eqref{eq:asympt_dS}.
According to static patch holography in empty dS space, the stretched horizon should be located just inside the cosmological horizon, and it plays an important role as the location where a putative dual quantum system is defined \cite{Dyson:2002pf,Susskind:2011ap,Susskind:2021esx,Susskind:2021dfc,Shaghoulian:2021cef,Susskind:2021omt,Shaghoulian:2022fop}.
In the case of the SdS black hole, timelike surfaces at constant $r$ are parametrized by 
\beq
r_{\rm st} = (1-\rho) r_h + \rho r_c \, , \qquad
\rho \in [0,1] \, ,
\label{eq:stretched_horizon_SdS4}
\eeq
such that $r_{\rm st}$ approaches the black hole horizon when $\rho \rightarrow 0$, and the cosmological horizon when $\rho \rightarrow 1$.
Of course the latter limit is of greater interest in the context of static patch holography, but it is useful to keep $\rho$ arbitrary in order to interpolate the region between the two event horizons.
To justify the choice \eqref{eq:stretched_horizon_SdS4}, we remark that the central dogma for black holes and for inflationary geometries state that the unitary evolution of the dual quantum system should be encoded by the region inside the cosmological horizon and outside the black hole one.
Furthermore, freely falling observers in the static patch (satisfying the requirement $f'(r)=0$) evolve along worldlines which coincide with the stretched horizon.

\begin{figure}[ht]
    \centering
    \includegraphics[width=0.32 \textwidth]{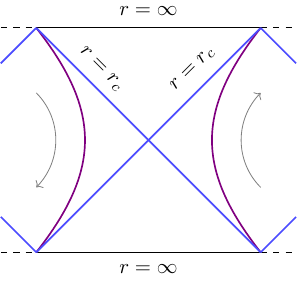}
    \caption{Symmetric configuration of the stretched horizons (in purple) located in the cosmological patch of the SdS geometry. The grey arrows represent the orientation of the Killing vector $\p_t$.  }
    \label{fig:stretched_horizon}
\end{figure}

In principle, stretched horizons can be defined in multiple ways inside the SdS background, since the geometry can be periodically extended, resulting in the existence of several static patches \cite{Aguilar-Gutierrez:2024rka}.
In view of the applications to static patch holography, in the remainder of this work we will only consider the case where the stretched horizon is located in the cosmological patch of the SdS background.
Moreover, we will adopt for simplicity the symmetric configuration where $r= r_{\rm st}$ on both sides of the cosmological patch. 
Since all the geometric observables (such as codimension-one maximal surfaces and the WDW patch) are anchored to the stretched horizons, it is not restrictive to only study the spacetime region in the cosmological patch contained between the two stretched horizons, as depicted in fig.~\ref{fig:stretched_horizon}.
For this reason, from now on we will \textit{cut away} from any Penrose diagram the spacetime region lying beyond the stretched horizons, and ignore the black hole patch.

We identify the time coordinates running upwards along the left and right stretched horizons as the \textit{boundary times} $(t_L, t_R)$ of the dual quantum theory.
The orientation of the Killing vector $\p_t$, associated with the invariance of the metric \eqref{eq:asympt_dS} under translations of the bulk time $t_{\rm bulk}$, determines the following relations
\beq
t_{\rm bulk} = \begin{cases}
    -t_L & \text{on the left stretched horizon} \\
    t_R & \text{on the right stretched horizon}
\end{cases}
\eeq
Furthermore, the Killing vector $\partial_t$ generates a boost symmetry such that the conjectured dual state is invariant under the shift
\beq
t_L \rightarrow t_L + \Delta t \, , \qquad
t_R \rightarrow t_R - \Delta t \, .
\label{eq:time_shift_symmetry}
\eeq
While the two boundary times $(t_L, t_R)$ are independent, we can synchronize them by means of a spacelike codimension-one surface connecting the stretched horizons.
Using the boost invariance \eqref{eq:time_shift_symmetry}, it is then always possible to choose 
\beq
\frac{t}{2} \equiv t_R = t_L \, ,
\label{eq:symmetric_times}
\eeq
that will be referred to as \textit{symmetric time configuration} in the remainder of the paper. 

\begin{figure}[ht]
    \centering
    \includegraphics[scale=0.41]{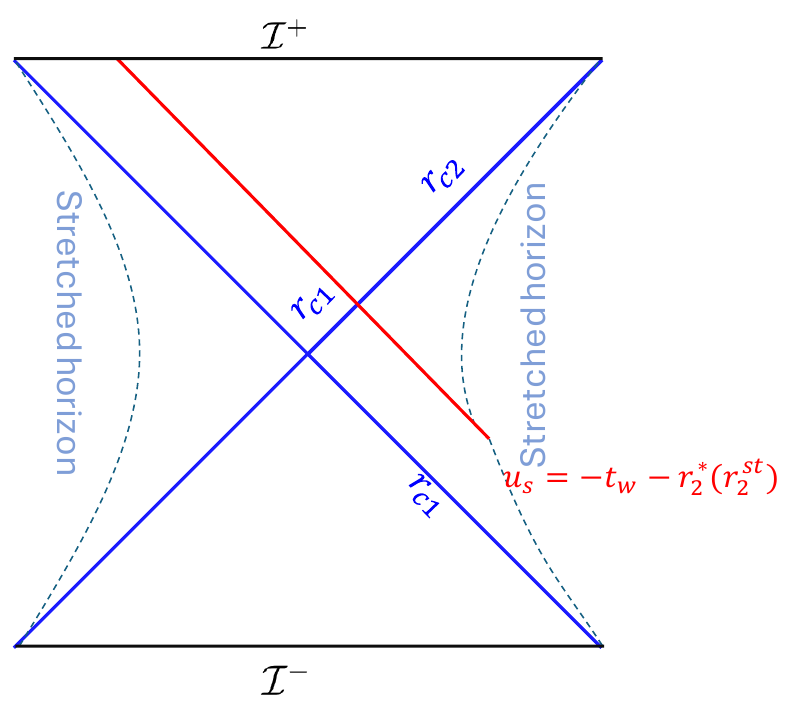}
    \caption{Penrose diagrams of SdS black hole in the presence of a shockwave. The location of the right stretched horizon before and after the shockwave insertion is determined according to any of the prescriptions outlined in subsection \ref{ssec:stretched_shocks}. }
    \label{fig:Penrose_shock_stretch} 
\end{figure}

In this setting, it is important to observe that the shockwave intersects just the right stretched horizon.
This defines a boundary time coordinate $t_R= -t_w$ associated with the insertion of the pulse of null energy, that we choose to satisfy the identity
\beq
u_s = - t_w - r^*_2 (r^{\rm st}_2)  \, ,
\label{eq:constant_us_shock}
\eeq
where $u_s$ is the constant value of the null coordinate along which the shock propagates.
Since we will describe in subsection \ref{ssec:stretched_shocks} different prescriptions to define the stretched horizon, including cases where the time coordinate is not continuous along it, we specify that the time $-t_w$ is always measured after the shockwave insertion.
Following the idea that we only depict the region in the geometry contained within the stretched horizons, the Penrose diagram corresponding to this setting is reported in fig.~\ref{fig:Penrose_shock_stretch}. 

\subsubsection{Definition in the presence of shockwaves}
\label{ssec:stretched_shocks}

When the asymptotic dS background \eqref{eq:asympt_dS} is perturbed by a shockwave, there exist various guiding principles (motivated by certain physical reasons) that make the definition of the stretched horizon more involved.
In the following, we are going to present the main possibilities.
First of all, one may simply require that the stretched horizon is still defined by a surface at constant radial coordinate, unaffected by its intersection with the shockwave.
While this is a reasonable working assumption, it has the undesired property that a stretched horizon located within the static patch $r_{\rm st} < r_{c1}$ before the shockwave insertion, cannot approach the larger cosmological horizon $r_{c2}$ after crossing the shockwave.\footnote{Notice that this remark only applies when the shockwave is inserted at finite boundary time $-t_w$. When the shock propagates along the cosmological horizon ($t_w \rightarrow \infty$), $r_c$ remains the same in all the geometry, and therefore there is no jump of the stretched horizon. This is the case studied in \cite{Anegawa:2023dad}.}
Since the degrees of freedom of the dual quantum theory predicted by static patch holography live inside the cosmological horizon and very close to it, we will instead define the stretched horizons in such a way to avoid the previous issue. 

We propose the following prescriptions:
\begin{itemize}
    \item \textbf{Constant redshift.} The stretched horizon is a surface of constant cosmological redshift with respect to an observer located on a timelike surface at constant radius.
    In practice, one can devise a physical experiment where light rays are exchanged between a source and a detector, in such a way to keep the cosmological redshift fixed.
    This possibility was carefully analyzed in section~3.1 of \cite{Baiguera:2023tpt}. 
    \item \textbf{Continuous time across the shock.}
    The stretched horizon is located at (a different) constant value of the radial coordinate before and after the shockwave insertion, in such a way that the bulk time coordinate running along it is continuous when crossing the shock. This prescription was studied in section~3.2 of \cite{Baiguera:2023tpt}.
    \item \textbf{Constant proper acceleration.}
    The stretched horizon is a surface generated by a Killing flow where the proper acceleration is radially directed and has constant norm.
    Contrarily to the previous prescriptions, this condition is local.
\end{itemize}

In reality, the specific prescription of the stretched horizon is not important, since they all lead to qualitatively similar results.
Instead, we stress the universal features that they share, which will play an important role for our investigations:
\begin{itemize}
    \item The stretched horizon is located at a fixed radial coordinate in the early past and in the far future, \ie when $|t_L|=|t_R| \gg L$.
    \item There is a limit of certain parameters $\rho_i$ such that the stretched horizons approach the respective cosmological horizon, before and after the shockwave insertion.  
\end{itemize}

While we argued that the specific choice of the stretched horizon among the previous possibilities is not important, for concreteness we will work with the constant redshift prescription.
The shockwave does not enter the left side of the cosmological patch, therefore we will simply keep the radial coordinate of the left stretched horizon to be a constant defined below, \ie $r=r^{\rm st}_1$. 
On the right side, the analysis of section~3.1 in \cite{Baiguera:2023tpt} gives
\beq
r_{\rm st} = 
\begin{cases}
   r^{\rm st}_1 \equiv  (1-\rho_1) r_{h1} + \rho_1 r_{c1} & \text{if} \,\,\, t_R \ll L \\
    R_{\rm st} (t) & \text{otherwise} \\
     r^{\rm st}_2 \equiv  (1-\rho_2) r_{h2} + \rho_2 r_{c2} & \text{if} \,\,\, t_R \geq -t_w \\
\end{cases}
\label{eq:stretched_constant_redshift}
\eeq
where $R_{\rm st}(t)$ is a time-dependent function that can be computed numerically.
In particular, one can show that whenever one of the $\rho_i$ goes to 1, the same is also true for the other parameter.
Since the time-dependence at times $t_R \lesssim -t_w$ makes the shape of the stretched horizon complicated, in the remainder of the paper we will only focus on the regime $t_R \geq -t_w$.
This case is also more relevant to study the influence of the shockwave on geometric observables.

\section{Complexity=action}
\label{sec:CA}

We evaluate the CA conjecture using static patch holography in the SdS background perturbed by a shockwave, see eq.~\eqref{eq:dS_metric_shock_wave}.
After highlighting the geometric features of the WDW patch in subsection \ref{ssec:WDW_patch}, we compute the action in subsection \ref{ssec:general_CA}, providing several plots in concrete examples in subsection \ref{ssec:examples_CA}.
In subsection \ref{ssec:CA_switchback_plateau}, we show evidence of a \textit{cosmological} switchback effect for asymptotically dS geometries in two ways: by studying the time duration of a regime where CA conjecture is approximately constant, and by computing the complexity of formation.

\subsection{WDW patch}
\label{ssec:WDW_patch}

The WDW patch is defined as the bulk domain of dependence of a spacelike surface attached to the stretched horizons defined in subsection \ref{ssec:stretched_horizon}.
In the case of the asymptotically dS geometry with shockwave \eqref{eq:dS_metric_shock_wave}, a detailed analysis of the WDW patch was performed in section~4 of reference \cite{Baiguera:2023tpt}, to which we refer the reader for an exhaustive treatment.
In this work, we will introduce the definitions and review the main results necessary to study the time evolution of CA.

Holographic complexity conjectures involving the WDW patch are divergent in asymptotically dS space whenever the top (bottom) joint delimiting the null boundaries intersects future (past) timelike infinity $\mathcal{I}^+ (\mathcal{I}^-)$.
For this reason, we introduce cutoff surfaces located at constant radial coordinate $r=r_{\rm max}$ both in the future and past regions of the cosmological patch of SdS space.\footnote{A priori, one can pick two different regulators $r_{\rm max,1} \ne r_{\rm max,2}$ at past and future infinity. It turns out that such a choice does not lead to additional meaningful insights compared to the results obtained in this work.} 
For practical convenience, we parametrize
\beq
r_{\rm max} = \frac{r_{c1}}{\delta} \, .
\label{eq:definition_rmax_shocks}
\eeq
Divergences arise when the regulator is removed, \ie in the limit $\delta \rightarrow 0$.

\begin{figure}[ht]
    \centering
    \includegraphics[scale=0.48]{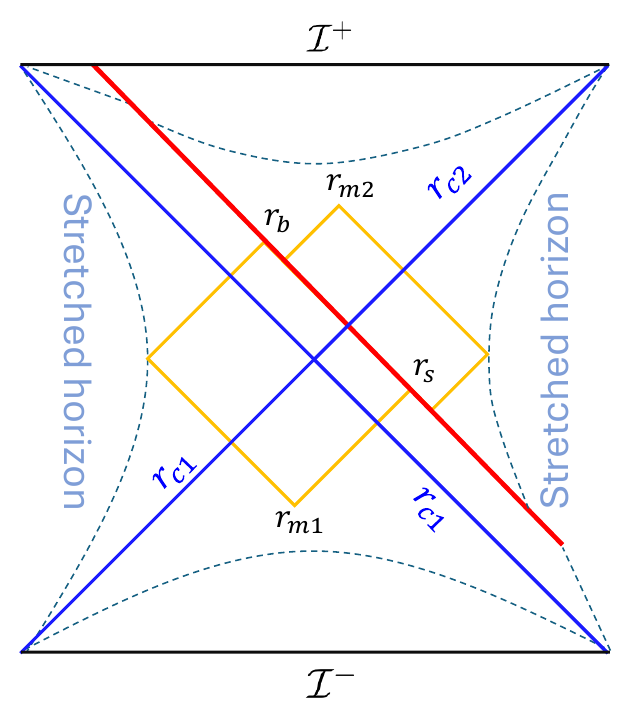} 
    \caption{Representation of the WDW patch in SdS black hole perturbed by a shockwave. The cosmological horizons before and after the shockwave insertion are $r_{c1}$ and $r_{c2}$, respectively.
    Special positions of the Penrose diagram are denoted as $r_{m1}, r_{m2}$ (bottom and top joints of the WDW patch) and as $r_s, r_b$ (intersections of the past and future null boundaries of the WDW patch with the shockwave).}
   \label{fig:WDW_patches}
  \end{figure}

\subsubsection{Special positions of the WDW patch}

The WDW patch is depicted in fig.~\ref{fig:WDW_patches}.
We highlight the following special positions in the Penrose diagram, which are relevant to the time evolution of CA conjecture:
\begin{itemize}
    \item $r_s$ is the intersection between the bottom-right boundary of the WDW patch and the shockwave $(r_s \leq r_{c2})$; 
    \item $r_b$ is the intersection between the top-left boundary of the WDW patch and the shockwave $(r_b \geq r_{c1})$;
    \item $r_{m1}$ is the past joint of the WDW patch;
    \item $r_{m2}$ is the future joint of the WDW patch.
\end{itemize}
These special positions are implicitly defined by the following equations, obtained by computing the null coordinates delimiting the boundaries of the WDW patch:
\begin{subequations}
\bea
&    t_R+t_w = 2r^*_2(r_s) - 2 r^*_2 (r^{\rm st}_2) \, , &
\label{eq:identity_WDW_times1}\\
&   t_L - t_w =  r^*_1(r^{\rm st}_1) + r^*_2(r^{\rm st}_2) - 2 r^*_1(r_b) \, , &
\label{eq:identity_WDW_times2}\\
&   t_L - t_w = 2 r^*_1 (r_{m1})   - 2 r^*_1 (r_s) - r^*_1(r^{\rm st}_1) + r^*_2(r^{\rm st}_2)  \, , & 
\label{eq:identity_WDW_times3} \\
&   t_R + t_w =  2 r^*_2 (r_{b})  - 2 r^*_2 (r_{m2}) 
\label{eq:identity_WDW_times4} \, .  &
    \eea    
\end{subequations}
The time derivatives of these identities determine the time evolution of the WDW patch. At constant $t_L$ we get
\beq
\frac{d r_s}{d t_R} =  \frac{f_2 (r_s)}{2} \, , \qquad
\frac{dr_b}{d t_R} = 0 \, , \qquad 
\frac{d r_{m1}}{d t_R} =  \frac{f_1 (r_{m1})}{2} \frac{f_2 (r_s)}{f_1 (r_s)} \, , \qquad
\frac{d r_{m2}}{d t_R} = - \frac{f_2 (r_{m2})}{2}  \, ,
\label{eq:derivatives_tR}
\eeq
while at constant $t_R$ we find
\beq
\frac{d r_s}{d t_L} = 0 \, , \qquad
\frac{dr_b}{d t_L} = - \frac{f_1 (r_b)}{2}  \, , \qquad 
\frac{d r_{m1}}{d t_L} =   \frac{f_1 (r_{m1})}{2}  \, , \qquad
\frac{d r_{m2}}{d t_L} =  - \frac{f_2 (r_{m2})}{2} \frac{f_1 (r_b)}{f_2 (r_b)} \, .
\label{eq:derivatives_tL}
\eeq

\subsubsection{Universal critical times of the WDW patch}
\label{ssec:critical_times_WDW}

\begin{figure}[ht]
    \centering
    \subfigure[$t=t_{c1}$.]{\label{subfig:WDW_patch_tc1} \includegraphics[scale=0.6]{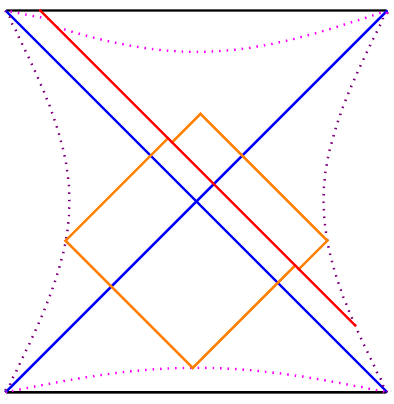}} \quad
      \subfigure[$t=t_{c2}$.]{\label{subfig:WDW_patch_tc2} \includegraphics[scale=0.6]{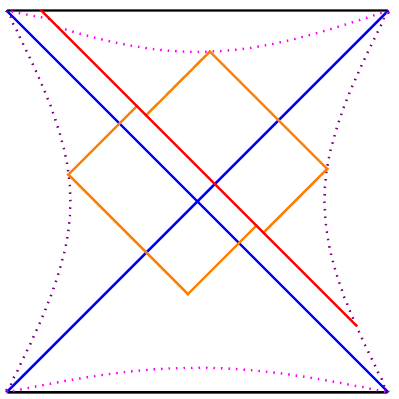}} 
    \quad
     \subfigure[ $t=t_{c3}$.]{\label{subfig:WDW_tc3} \includegraphics[scale=0.383]{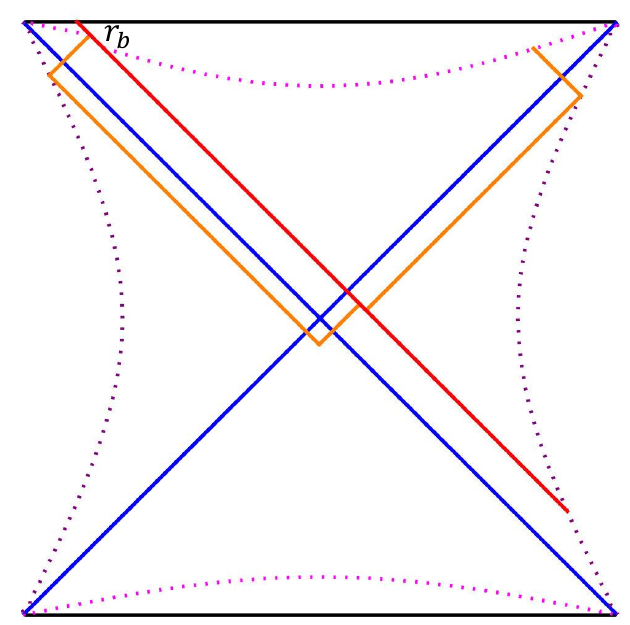}} 
    \caption{Critical times of the WDW patch. (a) Time $t_{c1}$ when the past joint $r_{m1}$ intersects the surface located at $r=r_{\rm max}$ close to $\mathcal{I}^-$.
    (b) Time $t_{c2}$ when the future joint $r_{m2}$ intersects the 
    surface located at $r=r_{\rm max}$ close to $\mathcal{I}^+$.
    (c) Time $t_{c3}$ when the special position $r_b$ of the WDW patch intersects the surface located at $r=r_{\rm max}$ close to $\mathcal{I}^+$. }
    \label{fig:critical_times}
\end{figure}

The above-mentioned special positions in the Penrose diagram define critical times occurring when the shape of the WDW patch changes.
Here we only focus on the \textit{universal} critical times reported in
fig.~\ref{fig:critical_times}, \ie that occur during the time evolution independently of the energy $\varepsilon$ of the shockwave and the time $-t_w$ when it is inserted on the right stretched horizon.
There exist certain special configurations of the WDW patch, always occurring when the shockwave is inserted very far in the past ($t_w \gg L$), where the future and past joints of the WDW patch move inside the cosmological horizon.
We will reserve their treatment for subsection \ref{ssec:special_configurations_WDW}.

For simplicity, let us assume from now on that the boundary times along the stretched horizons are symmetric, as in eq.~\eqref{eq:symmetric_times}.
First of all, there is a trivial critical time $t_{c0}$ corresponding to the instant when the shockwave is inserted from the right stretched horizon, that is
\beq
t_{c0} = -2 t_w \, . 
\label{eq:critical_time_tc0}
\eeq
According to the prescription where the cosmological redshift is constant (presented in subsection \ref{ssec:stretched_horizon}), the stretched horizon is time-dependent at times $t<t_{c0}$.
In this work, we will avoid the technical difficulties involved with such regime and focus only on the case $t \geq t_{c0}$, when the WDW patch crosses the shockwave.

Next, let us compute the critical times represented by the configurations in fig.~\ref{fig:critical_times}:
\begin{enumerate}
    \item The critical time $t_{c1}$ happens when the bottom joint of the WDW patch crosses the past cutoff surface, \ie when $r_{m1}=r_{\rm max}$.
    Summing and subtracting eqs.~\eqref{eq:identity_WDW_times1} and \eqref{eq:identity_WDW_times3} leads to the identities
    \begin{subequations}
        \bea
& t_w = r^*_1(r_s) +r^*_2(r_s)+ \frac{1}{2} r^*_1(r^{\rm st}_1) - \frac{3}{2}r^*_2(r^{\rm st}_2)  -r^*_1(r_{\rm max}) \, ,  &\label{eq:rstc1_identity1} \\
& t_{c1} = 2 t_w - 4 r^*_1(r_s) + 4 r^*_1(r_{\rm max}) 
- 2 r^*_1(r^{\rm st}_1) + 2 r^*_2(r^{\rm st}_2) \, . &\label{eq:tc1}
\eea
    \end{subequations}
    Concretely, one solves the first equation for $r_s$ at fixed $t_w$, and then plugs the result inside the second equation to find the critical time. 
    \item The critical time $t_{c2}$ corresponds to the instant when the top joint of the WDW patch crossing the future cutoff surface, \ie when $r_{m2}=r_{\rm max}$.
    Similar manipulations of eqs.~\eqref{eq:identity_WDW_times2} and \eqref{eq:identity_WDW_times4} give
    \begin{subequations}
        \bea
& t_w = r^*_1(r_b) + r^*_2(r_b) - r^*_2(r_{\rm max})  - \frac{1}{2} r^*_1(r^{\rm st}_1) - \frac{1}{2} r^*_2(r^{\rm st}_2) \, ,  &\label{eq:rbtc2_identity1} \\
& t_{c2} = - 2 t_w + 4 r^*_2(r_b) - 4 r^*_2(r_{\rm max})   \, . &
\label{eq:tc2}
\eea
    \end{subequations}
    \item The critical time $t_{c3}$ happens when the special position $r_b$ of the WDW patch reaches the cutoff surface close to $\mathcal{I}^+$, in other words $r_b(t_{c3})=r_{\rm max}$. Using eq.~\eqref{eq:identity_WDW_times2}, this gives
\beq
t_{c3} = 2 t_w  + 2 r^*_1(r^{\rm st}_1) + 2r^*_2(r^{\rm st}_2) - 4 r^*_1(r_{\rm max}) \, .
\label{eq:critical_time_tc3}
\eeq
\end{enumerate}

The critical times defined above satisfy the hierarchy summarized in table~\ref{tab:critical_times}.
Depending on the insertion time of the shockwave, $t_{c0}$ and $t_{c1}$ can change the ordering, while the other critical times always have a definite order.
Since we are only interested in the regime $t \geq t_{c0}$ after the shockwave insertion, we will focus on the case reported in the first row of the table. 
We checked that the configuration in the second row leads to similar qualitative features for the CA conjecture, therefore it does not add other meaningful physical insights on the problem. 
Finally, we anticipate that the critical times $t_{c1}, t_{c2}$ will play a crucial role in subsection~\ref{ssec:CA_switchback_plateau} to show the existence of the switchback effect for CA conjecture.

\begin{table}[ht]   
\begin{center}    
\begin{tabular}  {|c|c|c|c|c|} \hline  \textbf{Choice of parameters} & \multicolumn{4}{c|}{\textbf{Time ordering}} \\ \hline
\rule{0pt}{3ex} $t_w \gg L $  & $t_{c0}$ &  $t_{c1}$ &  $t_{c2}$ &  $t_{c3}$   \\[0.1cm]
\hline
\rule{0pt}{3ex}
$t_w \ll L $  & $t_{c1}$ &  $t_{c0}$ &  $t_{c2}$ &  $t_{c3}$  \\[0.2cm]
\hline
\end{tabular}   
\caption{Hierarchy between the critical times.} 
\label{tab:critical_times}
\end{center}
\end{table}

\subsubsection{Special configurations of the WDW patch}
\label{ssec:special_configurations_WDW}

\begin{figure}[ht]
    \centering
   \subfigure[]{\label{fig:alternative1_WDWpatch} \includegraphics[scale=0.64]{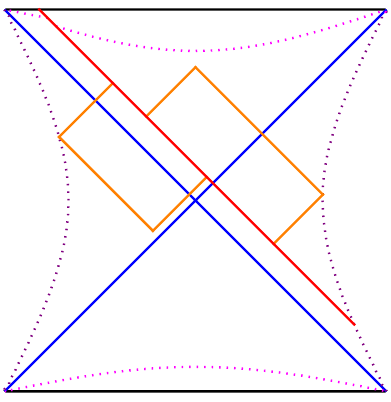}} \quad
    \subfigure[]{\label{fig:alternative2_WDWpatch} \includegraphics[scale=0.64]{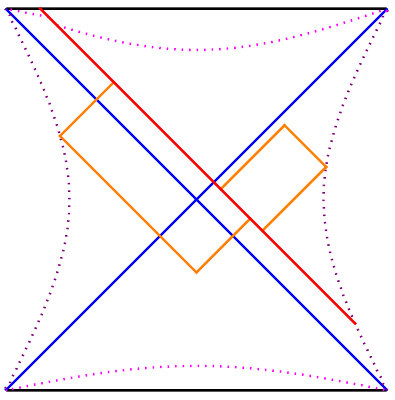}} \quad
    \subfigure[]{\label{fig:alternative3_WDWpatch} \includegraphics[scale=0.64]{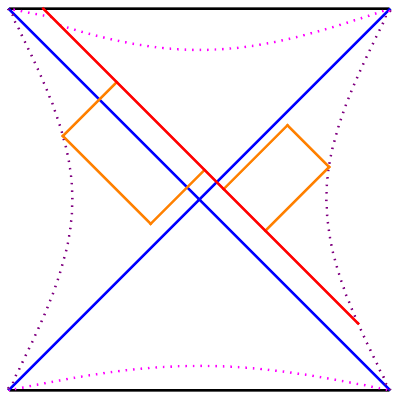}}
    \caption{Special configurations of the WDW patch. (a) The position $r_s$ of the WDW patch sits in the range $ [r_{c1}, r_{c2}].$ (b) The position $r_b$ sits in the range $ [r_{c1}, r_{c2}].$ 
    (c) Both $r_s, r_b$ are located in the region $[r_{c1}, r_{c2}].$}
    \label{fig:alternative_WDWpatch}
\end{figure}

The insertion of a shockwave in an asymptotically dS geometry brings the two static patches in causal contact \cite{Gao:2000ga}.
Technically, this feature is associated with the fact that the NEC implies $r_{c1} \leq r_{c2}$, contrarily to the AdS counterpart \cite{Chapman:2018dem,Chapman:2018lsv}.
This phenomenon ultimately leads to the existence of \textit{special} configurations of the WDW patch, shown in fig.~\ref{fig:alternative_WDWpatch}, corresponding to the top (or bottom) joints moving inside the cosmological horizon.
We will instead refer to a \textit{standard} configuration whenever both the top and bottom vertices of the WDW patch sit outside the cosmological horizon.

The transition from a standard to a special configuration of the WDW patch occurs in correspondence of the following critical times:\footnote{For the purposes of this work, the only relevant information carried by these critical times is the hierarchy described in table~\ref{tab:special_critical_times} below. Therefore, we refer the reader to sections 4.4 and 4.5 of \cite{Baiguera:2023tpt} for the explicit equations, while here we only report the definitions.}
\begin{itemize}
    \item The critical time $t_{c,s}$ happens when $r_s=r_{m1}=r_{c1}$.
When $t \geq t_{c,s}$, the bottom joint $r_{m1}$ moves inside the past cosmological horizon (see fig.~\ref{fig:alternative1_WDWpatch}). 
\item The critical time $t_{c,b}$ happens when $r_b=r_{m2}=r_{c2}$. 
When $t \leq t_{c,b}$, the top joint $r_{m2}$ moves inside the future cosmological horizon (see fig.~\ref{fig:alternative2_WDWpatch}).
\end{itemize}

The existence and the hierarchy between these novel critical times vary with the choice of the parameters $(\rho, \varepsilon, t_w)$.
If the inequality $t_{c,s} \leq t_{c,b}$ is true, then there is a time interval $t\in [t_{c,s}, t_{c,b}]$ when both the top and bottom joints of the WDW patch sit inside the cosmological horizon, as depicted in fig.~\ref{fig:alternative3_WDWpatch}.
Only during this regime (if it occurs), there may exist two more critical times $t_{c, \rm st1}$ and $t_{c, \rm st2}$ such that the following conditions are simultaneously met:
\begin{itemize}
    \item The special positions of the WDW patch satisfy $r_s=r_b$. 
    \item The bottom joint of the WDW patch crosses the stretched horizon, \ie  $r_{m1}=r^{\rm st}_1.$ 
    \item The top joint of the WDW patch crosses the stretched horizon, \ie $r_{m2}=r^{\rm st}_2.$
\end{itemize}

\begin{figure}[ht]
    \centering
    \includegraphics[scale=0.45]{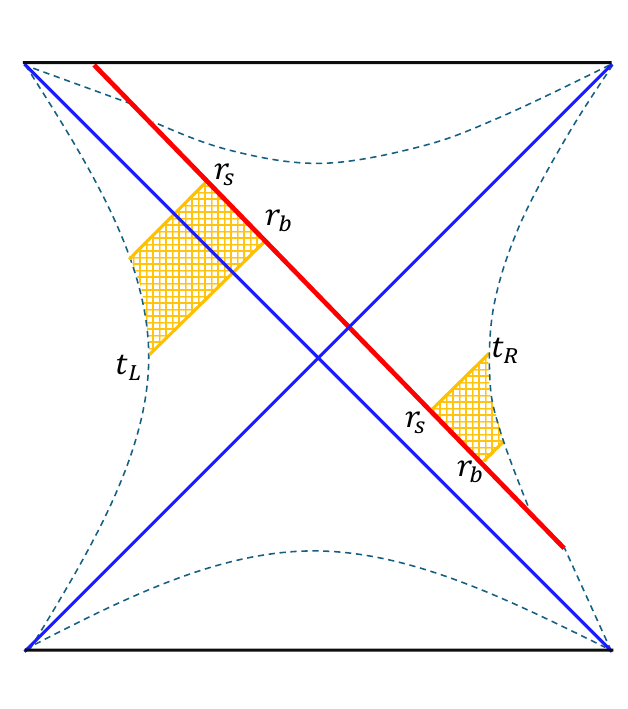}
    \caption{Shape of the WDW patch during the regime $t\in [t_{c, \rm st1}, t_{c, \rm st2}]$. }
    \label{fig:alternative4_WDWpatch}
\end{figure}

We stress that the critical times $t_{c, \rm st1}$ and $t_{c, \rm st2}$ always come in pair. In other words, if one of them exist, so does the other.
During the time interval $t \in [t_{c, \rm st1}, t_{c, \rm st2}]$, the WDW patch takes the shape depicted in fig.~\ref{fig:alternative4_WDWpatch}.

The critical times introduced in this subsection satisfy the hierarchies summarized in table \ref{tab:special_critical_times}.
The first row is true because the critical times $t_{c, \rm st}$ only exist when both the special positions $r_s, r_b$ of the WDW patch are located inside the cosmological horizon.
The second row holds because the critical time $t_{c1} (t_{c2})$ requires the bottom (top) joint of the WDW patch to reach the cutoff surface near $\mathcal{I}^- (\mathcal{I}^+)$, therefore cannot coexist with a configuration where both of them sit behind the cosmological horizon.
Finally, we point out that for large enough $t_w$, the configuration in fig.~\ref{fig:alternative4_WDWpatch} always exists.
This is crucial to realize the switchback effect in asymptotically dS geometries, as we will show in subsection \ref{ssec:CA_switchback_plateau}, and it is a consequence of the causal connection between stretched horizons generated by the shockwave insertion. 

\begin{table}[ht]   
\begin{center}    
\begin{tabular}  {|c|c|c|c|c|} \hline  \textbf{Comparison} & \multicolumn{4}{c|}{\textbf{Time ordering}} \\ \hline
\rule{0pt}{3ex} Hierarchy with fig.~\ref{fig:alternative1_WDWpatch} and \ref{fig:alternative2_WDWpatch} & $t_{c,s}$ &  $t_{c,\mathrm{st1}}$ &  $t_{c,\mathrm{st2}}$ &  $t_{c,b}$   \\[0.1cm]
\hline
\rule{0pt}{3ex} Hierarchy with fig.~\ref{subfig:WDW_patch_tc1} and \ref{subfig:WDW_patch_tc2}
 & $t_{c1}$ &  $t_{c,\mathrm{st1}}$ &  $t_{c,\mathrm{st2}}$ &  $t_{c2}$  \\[0.2cm]
\hline
\end{tabular}   
\caption{Hierarchies involving the critical times $t_{c,\mathrm{st1}},t_{c,\mathrm{st2}}$ defined by the shape shown fig.~\ref{fig:alternative4_WDWpatch} for the WDW patch. 
In the first row, the comparison is done with respect to the critical times $t_{c,s}$ and $t_{c,b}$ defining the beginning of the regimes in fig.~\ref{fig:alternative1_WDWpatch} and \ref{fig:alternative2_WDWpatch}, respectively. In the second row, the comparison is done with the standard configurations in fig.~\ref{subfig:WDW_patch_tc1} and \ref{subfig:WDW_patch_tc2}.} 
\label{tab:special_critical_times}
\end{center}
\end{table}

\subsection{General computation of the action}
\label{ssec:general_CA}

We apply the CA conjecture to compute holographic complexity in the shockwave geometry \eqref{eq:dS_metric_shock_wave} as the on-shell gravitational action $I_{\rm WDW}$ evaluated in the WDW patch
\beq
\mathcal{C}_A = \frac{I_{\rm WDW}}{\pi} \, , \qquad
I_{\rm WDW} = \sum_{\mathcal{X}}  I_{\mathcal{X}} \, , \qquad
\mathcal{X} \in \lbrace \mathcal{B}, \rm GHY, \mathcal{N}, \mathcal{J} ,\rm ct  \rbrace \, .
\label{eq:CA_conjecture}
\eeq
The terms $I_{\mathcal{X}}$ composing the gravitational action, listed in references \cite{Lehner:2016vdi,Carmi:2016wjl}, are the following:
\begin{itemize}
    \item The bulk term is given by the Einstein-Hilbert action 
    \beq
I_{\mathcal{B}} = \frac{1}{16 \pi G_N} \int_{\rm WDW} d^{d+1} x \, \sqrt{-g}	 \,  \le R - 2 \Lambda \ri \, ,
\label{eq:general_bulk_action}
\eeq
where $R$ is the Ricci scalar and $\Lambda$ the cosmological constant.
\item The Gibbons-Hawking-York (GHY) term, evaluated on codimension-one timelike (spacelike) boundaries, reads
\beq
I_{\rm GHY} = \frac{\varepsilon_{t,s}}{8 \pi G} \int_{\mathcal{B}_{t,s}} d^d x \, \sqrt{h} \, K \, ,
\label{eq:GHY_term}
\eeq
where $h$ is the induced metric determinant and $K$ the trace of the extrinsic curvature. The overall sign is $\varepsilon_t=1$ if the boundary $\mathcal{B}_t$ is timelike and $\varepsilon_s=-1$ if $\mathcal{B}_s$ is spacelike.
\item The term denoted with $I_{\mathcal{N}}$ in eq.~\eqref{eq:CA_conjecture} is evaluated on codimension-one null surfaces.
For the purposes of this work it is sufficient to state that it is proportional to the acceleration $\kappa$ along the congruence of null geodesics composing the null boundary, therefore it vanishes whenever we use an affine parametrization.
\item Codimension-two joint terms arise from the intersection of two codimension-one surfaces. They are given by \cite{Lehner:2016vdi}
 \beq
I_{\mathcal{J}} =  \frac{\varepsilon_{\mathfrak{a}}}{8 \pi G} \int_{\mathcal{J}} d^{d-1} x \, 
 \sqrt{\gamma} \, \mathfrak{a}  \, ,
 \label{eq:joints_action}
\eeq
where $\mathfrak{a}$ is defined below for the case of interest in this work, \ie when at least one codimension-one null boundary is involved. 
Here $\gamma$ is the induced metric along the codimension-two joint, while the pre-factor $\varepsilon_{\mathfrak{a}} = \pm 1$ depends on the orientation of the null normals to the intersecting surfaces, according to the prescription defined in references \cite{Lehner:2016vdi,Carmi:2016wjl}.
The explicit expression for the integrand is
\beq
\mathfrak{a} = \begin{cases}
\log \left| \mathbf{t} \cdot \mathbf{k}  \right|  & \mathrm{if} \,\, \mathbf{t} \,\, \mathrm{timelike}  \\
\log \left| \mathbf{n} \cdot \mathbf{k}  \right|  & \mathrm{if} \,\, \mathbf{n} \,\, \mathrm{spacelike}  \\
\log \left| \frac{1}{2} \mathbf{k}_L \cdot \mathbf{k}_R  \right|  & \mathrm{if} \,\, \mathbf{k}_L, \mathbf{k}_R \,\, \mathrm{null}  \\
\end{cases}
\label{eq:integrand_a_actionjoints}
\eeq
\item To ensure the invariance of the full action under reparametrizations, one needs to include a counterterm on codimension-one null boundaries defined as
\beq
I_{\rm ct} = \frac{1}{8 \pi G} \int_{\mathcal{B}_n} 
d\lambda d^{d-1} x \, \sqrt{\gamma} \,  \Theta \, \log |\ell_{\rm ct} \Theta| \, ,
\label{eq:counterterm_action}
\eeq 
where $\lambda$ is the parameter along the congruence of null geodesics composing the surface, and $\gamma$ the induced metric determinant along the orthogonal directions.\footnote{The parameter $\lambda$ chosen to compute eq.~\eqref{eq:counterterm_action} must be the same used to compute $I_{\mathcal{N}}$.} Finally, $\Theta$ is the expansion of the geodesics, and $\ell_{\rm ct}$ an arbitary length scale. Its value will be fine-tuned to impose certain properties for the gravitational observable under consideration. 
\end{itemize}

\begin{figure}[ht]
    \centering
     \subfigure[]{\includegraphics[scale=0.63]{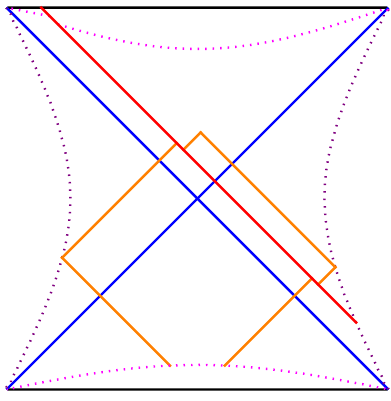}} \quad
      \subfigure[]{\label{subfig:intermediate_regime_dS}  \includegraphics[scale=0.63]{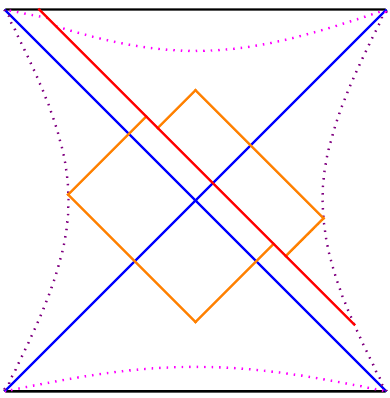}} \\
      \subfigure[]{\label{fig:WDW3reg}\includegraphics[scale=0.63]{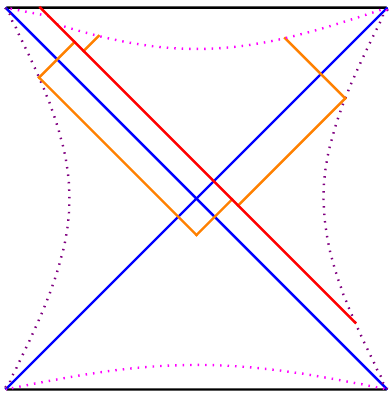}} \quad
        \subfigure[]{\label{fig:WDW4reg}\includegraphics[scale=0.63]{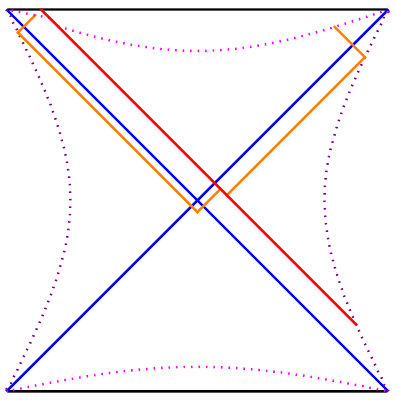}}
    \caption{Time evolution of the WDW patch in asymptotically dS space with a shockwave.
    (a) Times $t_{c0} \leq t < t_{c1}$ after the shockwave insertion, when the bottom joint of the WDW patch sits behind the past cutoff surface. 
    (b) Intermediate times  $t_{c1} \leq t < t_{c2}$ when both the vertices of the WDW patch are located outside the cosmological horizons.
    (c) Times  $t_{c2} \leq t < t_{c3}$ when the top joint sits  behind the future cutoff surface.
    (d) Late times $t \geq t_{c3}$ when $r_b$ sits behind the future cutoff surface. }
    \label{fig:time_evolution_WDW_reg}
\end{figure}

The next step is to compute the time evolution of CA conjecture, in particular by calculating its rate in the symmetric case \eqref{eq:symmetric_times}.
The discussion outlined in subsection \ref{ssec:WDW_patch} has shown that the WDW patch evolves according to the plot reported in fig.~\ref{fig:time_evolution_WDW_reg}, where we only consider boundary times after the shockwave insertion, \ie $t \geq t_{c0}$.
The strategy will be to compute holographic complexity during the \textit{intermediate} regime $t \in [t_{c1}, t_{c2}]$, when the top and bottom joints of the WDW patch do not reach the cutoff surfaces close to timelike infinity, see fig.~\ref{subfig:intermediate_regime_dS}.
This setting is important because it contains a regime where complexity shows a plateau, responsible for the geometric realization of the switchback effect. Furthermore, the special configurations of the WDW patch discussed in subsection \ref{ssec:special_configurations_WDW} can only occur during this intermediate step of the time evolution.
The computation of CA in the other regimes is performed in appendix \ref{app:details_CA}.

\subsubsection{Bulk term}

Since the Einstein-Hilbert Lagrangian is constant in the geometry \eqref{eq:dS_metric_shock_wave}, the bulk term in the action is directly proportional to the CV2.0 conjecture, \ie the spacetime volume of the WDW patch
\beq
I_{\mathcal{B}} =
\frac{d}{8 \pi G_N L^2} V_{\rm WDW} = \frac{d}{8 \pi} \, \mathcal{C}_{2.0V} \, .
\label{eq:bulk_CV20_relation}
\eeq
This identity immediately entails that we can use the computations performed in section~5 of  \cite{Baiguera:2023tpt}, where the time dependence of CV2.0 was studied in details, to find the corresponding evolution of the bulk action.
We will not need to use the explicit expression of the integrated bulk term $I_{\mathcal{B}}$; the interested reader can find it by taking eqs.~(5.2), (5.3), (5.7) and (5.8) in \cite{Baiguera:2023tpt}, and then applying the identity~\eqref{eq:bulk_CV20_relation}.
Here we directly report the rate of growth of the bulk term in the symmetric case \eqref{eq:symmetric_times}, that is
\beq
\begin{aligned}
\frac{d I_{\mathcal{B}}}{dt} = & \frac{\Omega_{d-1}}{16 \pi G_N L^2} 
\left[ r_{m2}^{d} \le 1 + \frac{f_1(r_b)}{f_2(r_b)}  \ri
- r_{m1}^d \le  1 + \frac{f_2(r_s)}{f_1(r_s)} \ri \right. \\
& \left. + r_b^d \le 1 - \frac{f_1(r_b)}{f_2(r_b)} \ri
- r_s^d \le 1 - \frac{f_2(r_s)}{f_1(r_s)}   \ri
\right] \, .
\label{eq:rate_Ibulk}
\end{aligned}
\eeq
This expression is valid during the intermediate time regime $t\in [t_{c1}, t_{c2}]$ for any configuration of the WDW patch, either standard or special (see discussion in subsections \ref{ssec:critical_times_WDW} and \ref{ssec:special_configurations_WDW}).
The only exception is provided by the setting where both the top and bottom joints sit inside the cosmological horizon, as depicted in fig.~\ref{fig:alternative4_WDWpatch}.
In the latter case, the rate reads
\beq
\begin{aligned}
\frac{d I_{\mathcal{B}}}{dt} = & - \frac{\Omega_{d-1}}{16 \pi G_N L^2} 
\left[ (r^{\rm st}_{2})^{d} \le 1 + \frac{f_1(r_b)}{f_2(r_b)}  \ri
- (r^{\rm st}_{1})^d \le  1 + \frac{f_2(r_s)}{f_1(r_s)} \ri \right. \\
& \left. + r_b^d \le 1 - \frac{f_1(r_b)}{f_2(r_b)} \ri
- r_s^d \le 1 - \frac{f_2(r_s)}{f_1(r_s)}   \ri
\right] \, .
\label{eq:rate_Ibulk_alt}
\end{aligned}
\eeq
As a rule of thumb, we get this expression by performing the limit $r_{m1} \rightarrow r^{\rm st}_1, r_{m2} \rightarrow r^{\rm st}_2$ of eq.~\eqref{eq:rate_Ibulk}, and then reversing the overall sign.
This is nothing but a way to account for the joints of the WDW patch moving behind the stretched horizon, and the position of $r_s, r_b$ being exchanged in this case.

\subsubsection{Boundary terms}

For notational convenience, we define the combination of boundary terms as
\beq
I_{\rm bdy} \equiv I_{\rm GHY} + I_{\mathcal{N}} + I_{\mathcal{J}} + I_{\rm ct} \, .
\label{eq:def_Ibdy}
\eeq
The details of its computation are presented in appendix~\ref{app:ssec:bdy_CA_intermediate}, with the total boundary term obtained in eq.~\eqref{eq:total_boundary_intermediate_regime}.
Using the derivatives~\eqref{eq:derivatives_tR}--\eqref{eq:derivatives_tL}, we then get the rate
\beq
\begin{aligned}
 \frac{d I_{\rm bdy}}{dt} & = \frac{\Omega_{d-1}}{32 \pi G_N} \left\lbrace  
(d-1) f_1(r_{m1}) \le  \frac{f_2(r_s)}{f_1(r_s)} +1 \ri  (r_{m1})^{d-2}  \log \left| \frac{(r_{m1})^2}{f_1(r_{m1}) \ell_{\rm ct}^2 (d-1)^2} \right|  \right. \\
& \left.  - (d-1) f_2(r_{m2}) \le  \frac{f_1(r_b)}{f_2(r_b)} +1 \ri  (r_{m2})^{d-2}  \log \left| \frac{(r_{m2})^2}{f_2(r_{m2}) \ell_{\rm ct}^2 (d-1)^2} \right| 
\right. \\
& \left. - f_1'(r_{m1}) \, (r_{m1})^{d-1} \le \frac{f_2(r_s)}{f_1(r_s)} +1 \ri
+ f_2'(r_{m2}) \, (r_{m2})^{d-1} \le \frac{f_1(r_b)}{f_2(r_b)} +1 \ri
\right. \\
& \left. - (d-1) f_2(r_s) \, (r_s)^{d-2} \log \left| \frac{f_2(r_s)}{f_1(r_s)} \right|
+ (r_s)^{d-1} \le \frac{f_1'(r_s) f_2(r_s)}{f_1(r_s)} - f_2'(r_s) \ri \right. \\
& \left. + (d-1) f_1(r_b) \, (r_b)^{d-2} \log \left| \frac{f_1(r_b)}{f_2(r_b)} \right|
- (r_b)^{d-1} \le \frac{f_2'(r_b) f_1(r_b)}{f_2(r_b)} - f_1'(r_b) \ri
\right\rbrace \, .
\end{aligned}
\label{eq:rate_bdy_intermediate}
\eeq
As required by the reparametrization invariance of the action, this expression is independent of the ambiguity in normalizing the normals to the null boundaries of the WDW patch, while it depends on the counterterm length scale $\ell_{\rm ct}$. 
If the critical times $t_{c, \rm st}$ defined in section \ref{ssec:special_configurations_WDW} exist, during the regime delimited by them we get the rate
\beq
\begin{aligned}
& \frac{d I_{\rm bdy}}{dt} = \frac{\Omega_{d-1}}{32 \pi G_N} \left\lbrace  
  (d-1) f_2(r_s) \, (r_s)^{d-2} \log \left| \frac{f_2(r_s)}{f_1(r_s)} \right| \right. \\
& \left. - (r_s)^{d-1} \le \frac{f_1'(r_s) f_2(r_s)}{f_1(r_s)} - f_2'(r_s) \ri 
- (d-1) f_1(r_b) \, (r_b)^{d-2} \log \left| \frac{f_1(r_b)}{f_2(r_b)} \right| \right. \\
& \left. 
+ (r_b)^{d-1} \le \frac{f_2'(r_b) f_1(r_b)}{f_2(r_b)} - f_1'(r_b) \ri
- \frac{(r^{\rm st}_1)^{d-2}}{2} \le \frac{f_2(r_s)}{f_1(r_s)} + 1 \ri
\left[ 2(d-1) f_1(r^{\rm st}_1) + r^{\rm st}_1 f'_1(r^{\rm st}_1) \right] \right. \\
& \left. + \frac{(r^{\rm st}_2)^{d-2}}{2} \le \frac{f_1(r_b)}{f_2(r_b)} + 1 \ri
\left[ 2(d-1) f_2(r^{\rm st}_2) + r^{\rm st}_2 f'_2(r^{\rm st}_2) \right] 
\right\rbrace \, ,
\end{aligned}
\eeq
obtained by differentiating eq.~\eqref{eq:bdy_term_special_config} with respect to a symmetric boundary time.
This result cannot be achieved from eq.~\eqref{eq:rate_bdy_intermediate} in a simple way because the change in shape of the WDW patch, depicted in fig.~\ref{fig:alternative4_WDWpatch}, implies that there are additional GHY terms compared to the standard case.

\subsubsection{Hyperfast growth}

A common feature to several complexity conjectures in dS space is the hyperfast growth, which corresponds to the complexity (and its rate) becoming divergent at a finite boundary time \cite{Jorstad:2022mls}.
While it is difficult in general to find an analytic expression for complexity, it is easier to work with its rate.
We therefore study the limit $t \rightarrow t_{c2}^-$ (approached from below) of the complexity rate to study whether there is a divergence.
According to the hierarchy reported in table \ref{tab:special_critical_times}, it is clear that this limit always happens when the WDW patch does \textit{not} assume the special configuration in fig.~\ref{fig:alternative4_WDWpatch}.\footnote{At the cost of being pedantic, we stress that the reasoning in this subsection will apply to both the standard configurations of the WDW patch in fig.~\ref{fig:critical_times}, and to the special configurations depicted in figs.~\ref{fig:alternative1_WDWpatch} and \ref{fig:alternative2_WDWpatch}. In other words, the critical times $t_{c,s}$ and $t_{c,b}$ in table~\ref{tab:special_critical_times} do not affect the present discussion.}
In such case, the bulk action is given by eq.~\eqref{eq:rate_Ibulk}, and the boundary term is \eqref{eq:rate_bdy_intermediate}. 
To check whether the complexity rate is divergent when the WDW patch approaches $\mathcal{I}^+$, we perform a series expansion around $r_{m2} = \infty$.
This gives
\beq
\lim_{t \rightarrow t_{c2}^-} \frac{d\mathcal{C}_A}{dt} = 
    \frac{\Omega_{d-1}}{32 \pi^2 G_N L^2} (d-1) (r_{m2})^d  \le  \frac{f_1(r_b)}{f_2(r_b)} +1 \ri   \log \left[ \frac{L^2}{\ell_{\rm ct}^2 (d-1)^2} \right]  + \mathrm{finite}
\, .
\label{eq:hyperfast_CA}
\eeq
The leading divergence contributing to the bulk action gets cancelled by part of the joint contribution coming from the top vertex of the WDW patch.
However, there is a residual contribution from the counterterm on null boundaries which makes the complexity rate divergent, precisely when $r_{m2}$ approaches the cutoff surface $r=r_{\rm max}$, and the regulator is removed ($\delta \rightarrow 0$) to reach future timelike infinity $\mathcal{I}^+$.
This shows that the hyperfast growth happens at the same critical time $t_{c2}$ as for the CV2.0 conjecture \cite{Baiguera:2023tpt}. 

One can check that the following identity holds
\beq
\frac{f_1(r_b)}{f_2(r_b)} +1  \geq 0 \, .
\eeq
This is trivial if $r_b \geq r_{c2}$, and can be confirmed numerically otherwise.
The complexity rate (and therefore complexity itself) is positively divergent if the counterterm length scale satisfies
\beq
\ell_{\rm ct} < \frac{L}{d-1} \, .
\label{eq:ct_positiveCA}
\eeq
This requirement is the same that fixes the positivity of complexity in empty dS space \cite{Jorstad:2022mls}.

A similar analysis can be carried out to show that CA has a hyperfast time derivative when approaching the first critical time from above, \ie $t \rightarrow t_{c1}^+$.
In such case, the condition \eqref{eq:ct_positiveCA} implies that the rate is negatively divergent, opposite to the previous limit.

\subsection{Explicit examples}
\label{ssec:examples_CA}

\begin{figure}[ht]
    \centering
  \subfigure[]{ \label{fig:integratedCASdS3} \includegraphics[scale=0.6]{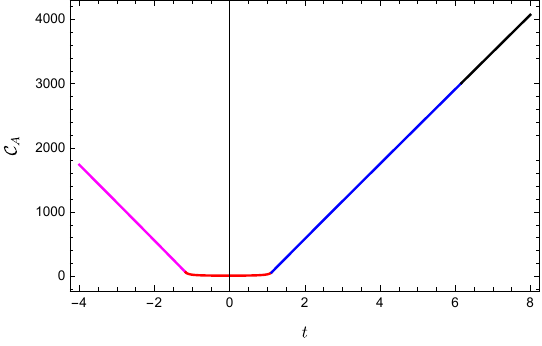}}
 \subfigure[]{\label{subfig:focus_CA_SdS3_plateau}  \includegraphics[scale=0.575]{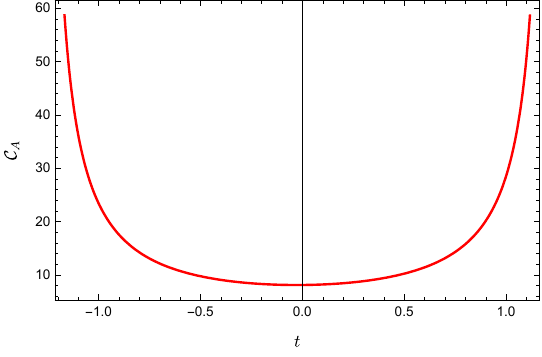}}
    \caption{(a) Complexity computed according to the CA proposal as a function of time in $d=2$. We fix $L=1, \rho=0.5, t_w=2, \delta=0.05, G_N\mathcal{E}_1=0.02, \ell_{\rm ct}=1/3$  and $\varepsilon=0.1$, according to the definition \eqref{eq:generic_epsilon_geometries}.
    (b) Focus on the plateau regime during the interval $t \in [t_{c1}, t_{c2}].$   }
    \label{fig:integratedCASdS3case1}
\end{figure}

\begin{figure}[ht]
    \centering
  \subfigure[]{ \label{fig:integratedCASdS3_new_st_long_pl} \includegraphics[scale=0.6]{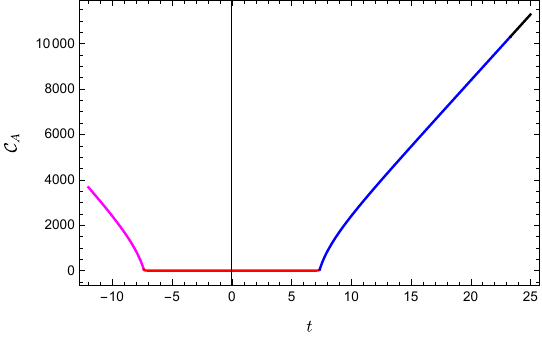}}
 \subfigure[]{\label{subfig:focus_CA_SdS3_plateau_new_st_long_pl}  \includegraphics[scale=0.57]{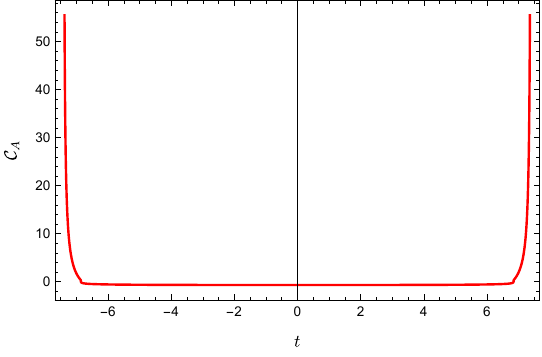}}
    \caption{(a) Complexity computed according to the CA proposal as a function of time in $d=2$. We fix $L=1, \rho=0.99, t_w=6, \delta=0.05, G_N\mathcal{E}_1=0.02,  \ell_{\rm ct}=1/3$ and the quantity $\varepsilon=0.01$ defined in eq.~\eqref{eq:generic_epsilon_geometries}. 
    The parameters are chosen such that the plateau is in the regime where its duration is growing linearly as a function of $t_w$ (see discussion in subsection \ref{ssec:CA_switchback_plateau}). (b) Focus on the intermediate interval $t \in [t_{c1}, t_{c2}].$}
    \label{fig:integrated_CA_SdS3_new_st_long_pl}
\end{figure}

\begin{figure}[ht]
    \centering
  \subfigure[]{\label{subfig:general_integratedCA_case4} \includegraphics[scale=0.6]{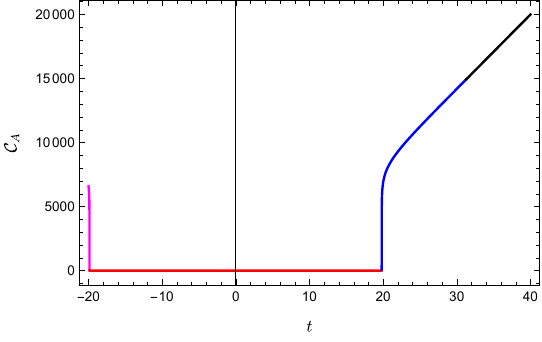}}
 \subfigure[]{\label{subfig:focus_CA_dS_plateau_case4}  \includegraphics[scale=0.57]{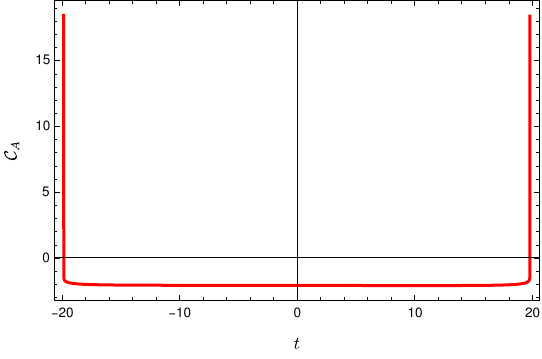}}
    \caption{(a) Complexity computed according to the CA proposal as a function of time in $d=2$. We fix $L=1, \rho=0.99, t_w=10, \delta=0.05, G_N\mathcal{E}_1=0.02,  \ell_{\rm ct}=1/3$  and the quantity $\varepsilon=0.1$ defined in eq.~\eqref{eq:generic_epsilon_geometries}. 
    (b) Focus on the time interval $t \in [t_{c1}, t_{c2}]$, when complexity can become negative. }
     \label{fig:integratedCA_case3}
\end{figure}

We provide numerical plots for the time dependence of CA, whose rate was computed in subsection \ref{ssec:general_CA} (at intermediate times) and in appendix \ref{app:ssec:other_regimes} (for later times).
We specialize to the three-dimensional setting ($d=2$) for practical convenience, but we checked that the qualitative behaviour of complexity is the same in other dimensions.
In the following, we consider various choices of the parameters describing the geometry and the shockwave insertion.
Let us summarize the main features of CA case by case:
\begin{itemize}
    \item In fig.~\ref{fig:integratedCASdS3case1} only the standard configuration of the WDW patch occurs.
    The plot is similar to empty dS space, with a plateau region at intermediate times and a linear behaviour for early and late times, as a consequence of the regularization with a cutoff surface close to timelike infinity \cite{Jorstad:2022mls}.
    \item In fig.~\ref{fig:integrated_CA_SdS3_new_st_long_pl} there is a regime when the special configuration in fig.~\ref{fig:alternative4_WDWpatch} of the WDW patch appears. This makes the plateau region longer, and there is a kink when the joints of the WDW patch cross the stretched horizons. 
    \item Fig.~\ref{fig:integratedCA_case3} shows a situation similar to the previous bullet, but here we further increase the insertion time of the shockwave $t_w$ to clearly show that complexity can become negative in the plateau region. 
    We will discuss in detail this peculiar phenomenon below eq.~\eqref{eq:CA_formation_general} and in the conclusions (subsection~\ref{ssec:conclusions}).
\end{itemize}

In general, the rate of the action (\textit{not} reported explicitly in the plots) is discontinuous at the critical times $t_{c1}, t_{c2}$.
Technically, this is a consequence of the fact that the rate depends on the counterterm length scale at intermediate times $t \in [t_{c1}, t_{c2}]$, but the corresponding terms in the action get replaced by GHY terms independent of $\ell_{\rm ct}$ for early and later times (see appendix~\ref{app:ssec:other_regimes} for the computation of the boundary complexity at early and late times).
One can fine-tune the counterterm scale such that the rate is continuous, and use this criterion as a way to remove the ambiguity in the definition of CA.\footnote{The same phenomenon happens in empty dS space without shockwaves \cite{Jorstad:2022mls}.}
In the remainder of the paper, we will keep $\ell_{\rm ct}$ generic and show that the switchback effect occurs independently of its specific value.

\subsection{Cosmological switchback effect (action)}
\label{ssec:CA_switchback_plateau}

We employ the results obtained for CA conjecture to show two manifestations of the switchback effect in asymptotically dS geometries.

\subsubsection{Plateau of complexity}

\begin{figure}[ht]
\centering
\subfigure[]{ \includegraphics[scale=0.64]{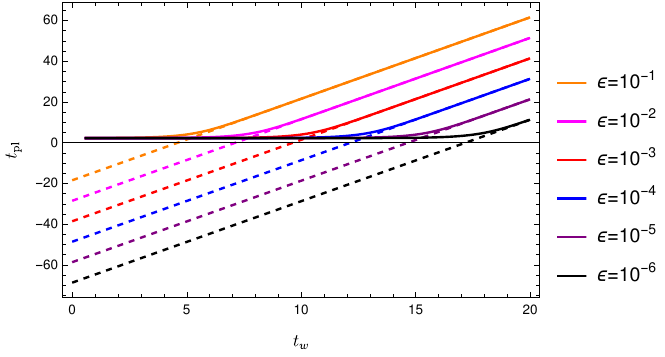}}
\subfigure[]{\label{subfig:plateau_SdS_rho099} \includegraphics[scale=0.64]{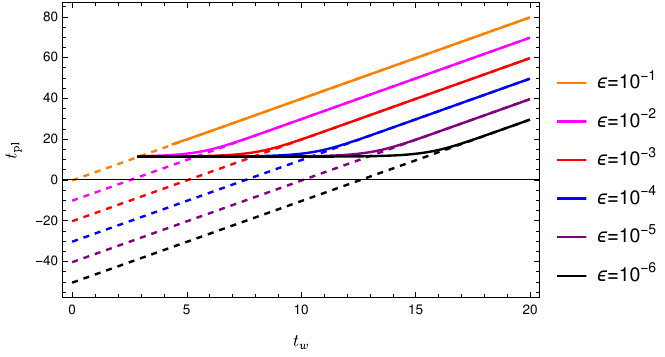}}
\caption{Duration of the plateau regime \eqref{eq:def_plateau_time} as a function of the insertion time of the shockwave, for various choices of $\varepsilon$ defined in eq.~\eqref{eq:generic_epsilon_geometries} in $d=2$. (a) We fix $L=1,\delta =0.05, G_N\mathcal{E}_1 = 0.02$ and $\rho=0.5$ . The dotted lines corresponds to the curves $t_{\rm pl} = 4 (t_w - t_*)$ with scrambling time in eq.~\eqref{eq:duration_plateau_SdS3}.
(b) Same plot, but using the parameter $\rho=0.99$ instead. }
\label{fig:plateau_rho099_SdS3}
\end{figure}

We have shown in eq.~\eqref{eq:hyperfast_CA} (and text below) that the rate of growth of CA is divergent at the critical times $t_{c1}, t_{c2}$ in the limit when the regulator $\delta$ is removed.
Since holographic complexity is not divergent during the interval $t \in [t_{c1}, t_{c2}]$, and it is much smaller compared to the other phases of the evolution, we refer to the behaviour of CA in the intermediate time regime as a \textit{plateau} of complexity.
We measure the duration of the plateau using the quantity
\beq
t_{\rm pl} \equiv t_{c2} - t_{c1} \, .
\label{eq:def_plateau_time}
\eeq
Since the critical time $t_{c1} (t_{c2})$ is only characterized by the geometric feature that the bottom (top) joint of the WDW patch reaches timelike infinity $\mathcal{I}^- (\mathcal{I}^+)$, we conclude that the duration of the plateau in the CA case is the same as for CV2.0.
We refer to sections~6.1.1, 6.2.1 and 6.3 of \cite{Baiguera:2023tpt} for a detailed analysis of the latter, but we summarize here the main results. 
Using eqs.~\eqref{eq:tc1} and \eqref{eq:tc2}, we get the formal expression
\beq
t_{\rm pl} = -4 t_w - 4 \le r^*_2(r_{\rm max}) + r^*_1 (r_{\rm max})  \ri + 4 \le r^*_2(r_b) + r^*_1 (r_s) \ri+ 2 \le r^*_1(r_1^{\rm st}) - r^*_2 (r_2^{\rm st}) \ri \, ,
\label{eq:tpl_SdS3}
\eeq
where $r_s$ is computed using eq.~\eqref{eq:rstc1_identity1}, while $r_b$ is taken from eq.~\eqref{eq:rbtc2_identity1}.
Numerical plots for the duration of the plateau can be obtained in any dimensions, but for simplicity we show two examples for $d=2$ in fig.~\ref{fig:plateau_rho099_SdS3}.

Notice that numerical ranges when the critical times satisfy $t_{c1} < t_{c0}$ are excluded in these plots because they involve a computation of CA with time-dependent stretched horizon (see discussion in subsection \ref{ssec:stretched_horizon}), which is technically challenging.
We highlight that the duration of the plateau starts from a constant value independent of $\varepsilon$, and then always increases when the shockwave is inserted earlier in the past, until a linear regime is asymptotically approached.
This linear regime at $t_w \gg L$ describes a \textit{scrambling time} $t_*$ given by\footnote{The scrambling time is the interval it takes for a perturbation to spread into the system.} 
\beq
t_{\rm pl} = 4 (t_w - t_*) \, .
\label{eq:linear_approx_plateau}
\eeq
One can analytically compute the scrambling time in dimensions $d=2,3$, see eqs.~(6.7) and (6.18) in \cite{Baiguera:2023tpt}.
For the purposes of this work, we report the three-dimensional result
\beq
 t^{\rm SdS_{3}}_* = \le\frac{L}{4a_1}+\frac{3L}{4a_2} \ri\log \le \frac{1-\rho}{1+\rho} \ri
+ \frac{L}{2} \le \frac{1}{a_1} + \frac{1}{a_2} \ri \log \le \frac{a_1+a_2}{a_2-a_1} \ri \, ,
\label{eq:duration_plateau_SdS3}
\eeq
where $a$ was defined in eq.~\eqref{eq:cosmological_horizon_SdS3}.
It is relevant to notice that this expression is valid for any choice of the parameters $(\rho, \varepsilon)$.

In general dimensions, the analytic expression for the scrambling time can only be obtained in the following double scaling limit
\beq
\varepsilon \rightarrow 0 \, , \qquad
\rho \rightarrow 1 \, , \qquad
\frac{1-\rho}{\varepsilon} \quad \mathrm{fixed} \, .
\label{eq:double_scaling_limit}
\eeq
The result reads
\beq
t^{\rm SdS_{d+1}}_* =  \frac{1}{2 \pi T_{c1}} 
\log \le   \frac{1-\rho}{\beta r_{\rm cr} \varepsilon} \le r_{c1}- r_{h1} \ri  \ri
+ \mathcal{O} (1-\rho, \varepsilon) \, ,
\label{rocker4}
\eeq
where the critical radius and the constant $\beta$ are defined by\footnote{Comparing with eq.~\eqref{eq:rc2_rc1_intro} in the introduction, here we defined $\alpha= \beta r_{\rm cr}$.} 
\beq
r_{\rm cr} \equiv L \sqrt{\frac{d-2}{d}}
 \, , \qquad
r_{c2} = r_{c1} + \beta r_{\rm cr} \, \varepsilon + \mathcal{O}(\varepsilon^2) \, .
\label{rocker3}
\eeq
The existence of a scrambling time that shortens the duration of the plateau is a manifestation of the switchback effect in the presence of shockwaves inserted at finite boundary time in a black hole geometry \cite{Stanford:2014jda,Chapman:2018dem,Chapman:2018lsv}.
The main features of this result are discussed in the introduction, below eq.~\eqref{eq:scrambling_Vaidya}.

\subsubsection{Complexity of formation}
\label{ssec:CA_formation}

Another geometric observable that acts as a diagnostic of the switchback effect is the complexity of formation, defined as the difference of complexity in the presence of a shockwave, compared to that of empty dS space (without a shock), \ie
\beq
\Delta \mathcal{C} = \mathcal{C} \left({\rm SdS}_{\mathrm{shockwave}}\right)\big|_{t_L=t_R=0} - 
\mathcal{C} ({\rm dS})\big|_{t_L=t_R=0} \, .
\label{eq:general_formation}
\eeq
Since the critical times of the WDW patch numerically satisfy the relation $t_{c1} < 0 < t_{c2}$, we need to evaluate the total complexity in the intermediate time regime, setting $t_L=t_R=0$.
In order to study the time delay corresponding to the switchback effect, we focus on the regime when the shockwave is inserted in the far past, \ie very close to the past cosmological horizon.
As it was observed for the computation of CV2.0 conjecture in reference \cite{Baiguera:2023tpt}, the limit $t_w \rightarrow \infty$ in the intermediate regime of evolution leads to the special configuration of the WDW patch depicted in fig.~\ref{fig:alternative4_WDWpatch}.

In dimensions $d \geq 2$, this implies that the bulk contribution to the action reads
\beq
\begin{aligned}
 I_{\mathcal{B}} (0) & =  \frac{d \Omega_{d-1}}{8 \pi G_N L^2} \left[   
\int_{r^{\rm st}_2}^{r_b} dr \, r^{d-1} \, \le t_w + 2 r^*_2(r^{\rm st}_2)  - 2 r^*_2 (r_b) \ri \right. \\
& \left. + \int_{r_b}^{r_s} dr \, r^{d-1} \, \le  t_w + 2 r^*_2(r^{\rm st}_2)  - 2 r^*_2 (r) + 2 r^*_1(r) - 2 r^*_1 (r_s) \ri  \right. \\
& \left. + \int_{r^{\rm st}_1}^{r_b} dr \, r^{d-1} \, \le  t_w+ r^*_1(r^{\rm st}_1)  + r^*_2(r^{\rm st}_2)  - 2 r^*_1 (r_s) \ri 
\right] \, ,
\end{aligned}
\label{eq:total_CV20_alt5_form}
\eeq
while the boundary term \eqref{eq:bdy_term_special_config} becomes
\beq
\begin{aligned}
 I_{\rm bdy} (0) &  = \frac{\Omega_{d-1}}{8 \pi G_N} \left\lbrace  (r_s)^{d-1} \log \left| \frac{f_2(r_s)}{f_1(r_s)} \right|
+ (r_b)^{d-1} \log \left| \frac{f_1(r_b)}{f_2(r_b)} \right|
\right. \\
& \left.  + \frac{(r^{\rm st}_2)^{d-2}}{2} \le t_w - 2 r^*_2(r_b) + 2 r^*_2(r^{\rm st}_2) \ri  \left[ 2(d-1) f_2(r^{\rm st}_2) + r^{\rm st}_2 f'_2(r^{\rm st}_2) \right]   \right. \\
& \left.  +  \frac{(r^{\rm st}_1)^{d-2}}{2}  \le t_w - 2 r^*_1(r_s) +  r^*_1(r^{\rm st}_1) +  r^*_2(r^{\rm st}_2) \ri \left[ 2(d-1) f_1(r^{\rm st}_1) + r^{\rm st}_1 f'_1(r^{\rm st}_1) \right]
\right\rbrace \, .
\end{aligned}
\label{eq:boundary_term_compl_formation}
\eeq
In the following, we show that in the regime $t_w \gg L$, there is a linear approximation with a delay characterizing the switchback effect.
Technical details of the approximation are collected in appendix~\ref{app:ssec:linear_CA_formation}.
In three dimensions, the computation can be analytically performed without any assumption on $(\rho, \varepsilon)$, giving
\begin{subequations}
    \beq
\left[\mathcal{C}_{A} \right]_{d=2} (0) \approx 
\frac{1}{8 \pi G_N}   \le  1 - 5 \rho^2 \ri \le a_1^2 + a_2^2 \ri \le t_w - t_* \ri \, ,
\label{eq:linear_approx_CA0}
\eeq
\beq
t_*   =  \frac{L}{a_1^2 + a_2^2 } \left[ 
 \frac{1}{2} \le a_1+ 2a_2 + \frac{a_1^2}{a_2} \ri  \log \le \frac{1-\rho}{1+\rho} \ri + \le a_1+a_2 \ri \log \le \frac{a_1+a_2}{a_2-a_1} \ri
\right] \, .
\label{eq:scrambling_time_CAf_SdS3}
\eeq
\end{subequations}
In higher dimensions, analytic results for large insertion times of the shockwave are only available in the case of small black holes $r_{h}/r_c \ll 1$ and after applying the double scaling limit \eqref{eq:double_scaling_limit}.
We find
\begin{subequations}
    \beq
\mathcal{C}_{A} (0) \approx 
\frac{\Omega_{d-1}}{16 \pi^2 G_N}   \left[ 5d (1-\rho^2) -8  \right]   \le t_w - t_* \ri \, ,
\label{eq:approx_CA_formation_gend}
\eeq
\beq
t_*  \approx   r_{c1} \log \le   \frac{ r_{c1}}{\beta r_{\rm cr}} \frac{1-\rho}{\varepsilon} \ri  = 
 \frac{1}{2 \pi T_{c1}} \log \le   \frac{ r_{c1}}{\beta r_{\rm cr}} \frac{1-\rho}{\varepsilon} \ri \, ,
 \label{eq:approx_scrambling_CAform}
\eeq
\label{eq:CA_formation_general}
\end{subequations}
where $\beta$ and $r_{\rm cr}$ were defined in eq.~\eqref{rocker3}.
First of all, one can check that the three-dimensional case \eqref{eq:scrambling_time_CAf_SdS3} reduces to eq.~\eqref{eq:approx_scrambling_CAform} if we perform the double scaling limit \eqref{eq:double_scaling_limit} in $d=2$.
Second, the scrambling time \eqref{eq:approx_scrambling_CAform} coincides with eq.~\eqref{rocker4} in the case of light black holes with $r_h/r_c \ll 1$.
Therefore, this result is yet another manifestation of the switchback effect in cosmological spacetimes.
In this case, it provides a delay in the increasing of complexity when the shockwave is inserted earlier from the right stretched horizon.

\begin{figure}[ht]
    \centering
    \subfigure[]{\label{subfig:CA0_case1}  \includegraphics[scale=0.82]{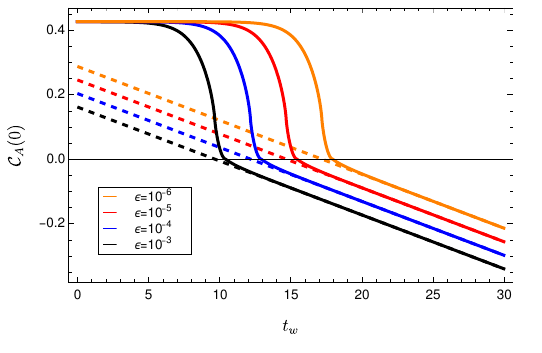}} 
     \subfigure[]{ \includegraphics[scale=0.82]{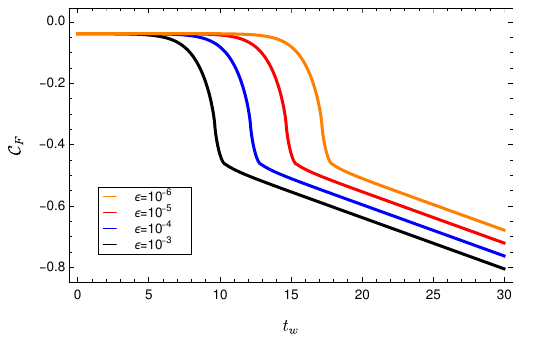}} 
    \caption{(a) Complexity at $t=0$ as a function of $ t_w,$ for various choices of $\varepsilon$ at fixed $d=2, L=1,\rho=0.5, G_N\mathcal{E}_1=0.02, \ell_{\rm ct}=1/3. $ The dashed lines represent the linear approximation in eqs.~\eqref{eq:linear_approx_CA0} and \eqref{eq:scrambling_time_CAf_SdS3}. (b) Same plot for the complexity of formation. }
    \label{fig:plot_CV20_at0_SdS3}
\end{figure}

\begin{figure}[ht]
    \centering
    \subfigure[]{ \label{subfig:CA0_case2} \includegraphics[scale=0.82]{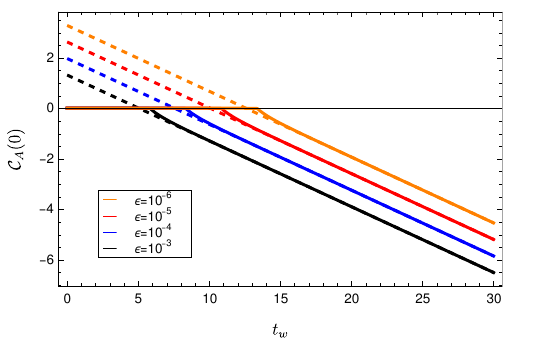}} 
     \subfigure[]{\includegraphics[scale=0.82]{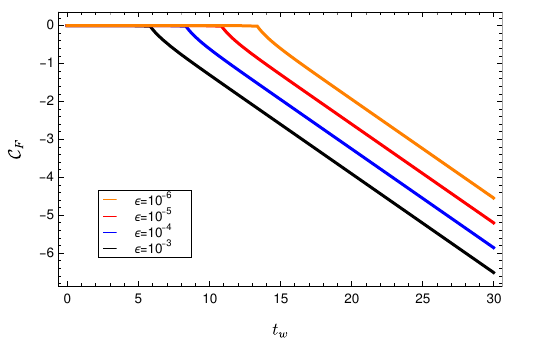}} 
    \caption{(a) Complexity at $t=0$ as a function of $ t_w,$ for various choices of $\varepsilon$ at fixed $d=2, L=1,\rho=0.99, G_N\mathcal{E}_1=0.02, \ell_{\rm ct}=1/3. $ The dashed lines represent the linear approximation in eqs.~\eqref{eq:linear_approx_CA0} and \eqref{eq:scrambling_time_CAf_SdS3}. (b) Same plot for the complexity of formation.  }
    \label{fig:plot_CV20_at0_SdS3_rho099}
\end{figure}

We report the numerical plots of the complexity of formation for two cases in fig.~\ref{fig:plot_CV20_at0_SdS3} and \ref{fig:plot_CV20_at0_SdS3_rho099}.
The action conjecture presents a switchback effect with the same scrambling time obtained for the CV2.0 conjecture \cite{Baiguera:2023tpt}.
The only change is in the overall prefactor of the complexity of formation, as a consequence of the inclusion of the boundary terms.
Interestingly, this quantity is always negative for big enough $\rho$, contrarily to the positivity that we found in the CV2.0 case.
If we interpret the complexity of formation as measuring the difficulty to build a state in a dual quantum picture, this would imply that the insertion of a shockwave makes this task easier to perform. 

We point out that a negative complexity of formation is not a novelty in the context of holography: in the CV case it happens in geometries violating the assumptions of the theorems proposed in \cite{Engelhardt:2021mju}, while in the CA case it happens for AdS black holes \cite{Chapman:2016hwi}.
The main difference in this setting is that CA itself (before any subtraction with other backgrounds) is negative for a stretched horizon satisfying $\rho^2 > \frac{5d-8}{5d}$, see eq.~\eqref{eq:approx_CA_formation_gend}.
This phenomenon does not happen in empty dS space \cite{Jorstad:2022mls}, instead it is a consequence of inserting the shockwave far enough in the past ($t_w \gg L$), such that the special configuration in fig.~\ref{fig:alternative4_WDWpatch} occurs.
This is clear by observing that CA in figs.~\ref{fig:plot_CV20_at0_SdS3} and \ref{fig:plot_CV20_at0_SdS3_rho099} becomes negative after a kink, which signals the transition to the special configuration of the WDW patch where the joints lie behind the stretched horizon.\footnote{Indeed, notice that at the critical times $t_{c, \rm st}$ (defined in subsection~\ref{ssec:special_configurations_WDW}) when the top and bottom joints of the WDW patch lie precisely on the stretched horizon, CA vanishes identically. }
This outcome might be a manifestation of the fact that shockwaves in asymptotically dS space bring the static patches into causal contact, making possible the communication between two otherwise spacelike-separated stretched horizons \cite{Gao:2000ga}.

\section{Complexity=volume}
\label{sec:CV_conjecture}

We compute the CV conjecture in the perturbed SdS background~\eqref{eq:dS_metric_shock_wave}, using static patch holography. General properties of the maximal codimension-one slice anchored at the stretched horizons are introduced in subsection \ref{ssec:general_analysis_volume}, and are then employed to determine the time evolution of the volume in subsection \ref{ssec:time_evo_CV}.
We analytically show the existence of a hyperfast growth. 
Subsection \ref{ssec:examples_CV} collects numerical plots confirming this trend, and exhibiting a regime where complexity is approximately constant. 
Finally, we show the existence of a switchback effect in subsection \ref{ssec:switchback_CV}, by analyzing the duration of the plateau regime.

\subsection{General analysis}
\label{ssec:general_analysis_volume}

In the shockwave geometry \eqref{eq:dS_metric_shock_wave} with blackening factor \eqref{eq:blackening_factor_shock_SdS}, we decompose the computation of holographic complexity according to the CV conjecture \eqref{eq:intro_CV} as
\beq
\mathcal{C}_V = \mathcal{C}_{V1} + \mathcal{C}_{V2} \, ,  \qquad 
\mathcal{C}_{Vi} \equiv \frac{\mathcal{V}_i }{G_N L}  \qquad
(i=1,2) 
\label{eq:CV_formula_gen}
\eeq
Despite the discontinuity when traversing the shockwave, each spacetime region (labelled as $i=1,2$) is a SdS background with its usual symmetries, that we can exploit to simplify the computation.

Any codimension-one surface with maximal volume in a geometry of the form \eqref{eq:asympt_dS} extremizes the following functional
\beq
\mathcal{V}_i = \Omega_{d-1} \int ds \, r^{d-1} \sqrt{-f_i\dot{v}^2+2\dot{v}\dot{r}}= \Omega_{d-1} \int ds \, r^{d-1} \sqrt{-f_i\dot{u}^2-2\dot{u}\dot{r}} \, ,
\label{eq:general_volume_functional}
\eeq
where we introduced a radial paramater $\sigma$ along the surface, running from the left to the right side of the Penrose diagram, and we defined the derivative $\cdot \equiv d/d\sigma$. 
Due to the reparametrization invariance of the problem, we select the convenient gauge choice
\beq
\sqrt{-f_i \dot{v}^2+2\dot{v}\dot{r}} = \sqrt{-f_i \dot{u}^2-2\dot{u}\dot{r}}  = r^{d-1} \, .
\label{eq:parametrization_volume}
\eeq
In each black hole region the variable $u$ (or $v$) is cyclic, therefore there exists a conserved momentum given by
\beq
P_i =\frac{r^{d-1} \le \dot{r}-f_i\dot{v} \ri}{\sqrt{-f_i \dot{v}^2+2\dot{v}\dot{r}} }= \frac{r^{d-1} \le -\dot{r}-f_i\dot{u} \ri}{\sqrt{-f_i \dot{u}^2-2\dot{u}\dot{r}}} \, ,
\label{eq:conserved_momenta_gen}
\eeq
which, after using the parametrization choice \eqref{eq:parametrization_volume}, simplifies to
\beq
P_i =\dot{r}-f_i\dot{v}= -\dot{r}-f_i\dot{u} \, .
\label{eq:conserved_momenta}
\eeq
We use the identities \eqref{eq:parametrization_volume} and \eqref{eq:conserved_momenta} to solve for the derivatives
\begin{subequations}
\beq
\dot{r}_{\pm} [P_i, r] = \pm \sqrt{f_i(r) r^{2(d-1)} + P_i^2} \, , 
\label{eq:rdot_volume}
\eeq
\beq
\dot{u}_{\pm}[P_i, r] = \frac{-\dot{r} -P_i}{f_i(r)} = \frac{1}{f_i(r)} \le -P_i \mp \sqrt{f_i(r) r^{2(d-1)} +P_i^2} \ri \, ,   
\label{eq:udot_volume}
\eeq
\beq
\dot{v}_{\pm}[P_i, r] = \frac{\dot{r} -P_i}{f_i(r)} = \frac{1}{f_i(r)} \le -P_i \pm \sqrt{f_i(r) r^{2(d-1)} +P_i^2} \ri \, ,
\label{eq:vdot_volume}
\eeq
\end{subequations}
where the symbols $\pm$ always refer to an increasing (decreasing) value of the radial coordinate $r$ when moving from the left to the right side of the extremal surface.
The maximization problem is determined by two boundary conditions: the coordinate times $t_L$ and $t_R$ along the stretched horizons where the surface is attached. 
However, in the symmetric case \eqref{eq:symmetric_times}, it will be convenient to trade the choice of $t_L=t_R$ with fixing $P_1$ instead.

By integrating in the vicinity of the shockwave the second-order differential equations induced from the functionals \eqref{eq:general_volume_functional} with blackening factor $F(u,r)$ in eq.~\eqref{eq:blackening_factor_shock_SdS},
we find that $\dot{u}$ is continuous, while $\dot{r}$ jumps as follows:
\beq
\dot{r}_2 (r_{\rm sh}) =  \dot{r}_1 (r_{\rm sh}) - \frac{\dot{u}(r_{\rm sh})}{2} \left[ f_2(r_{\rm sh}) - f_1(r_{\rm sh}) \right]  \, ,
\eeq
where $r_{\rm sh}$ denotes the intersection of the extremal surface with the shockwave.
By exploiting the definition \eqref{eq:conserved_momenta}, this implies a jump of the conserved momentum when crossing the shockwave, \ie
\beq
P_2 = P_1 - \frac{\dot{u}(r_{\rm sh})}{2} \left[ f_2(r_{\rm sh}) - f_1(r_{\rm sh}) \right] \, .
\label{eq:relation_P2_P1}
\eeq
To determine the shape of the maximal surface, we need to consider various cases according to the existence or not of a \textit{turning point} (\ie a point where $\dot{r}=0$), and depending on the surface passing through the future or past exterior regions to the cosmological horizon.

Using eq.~\eqref{eq:rdot_volume}, if a turning point $r_{t,i}$ in the $i$-th region exists, it is defined by
\beq
P_i^2 + f_i(r_{t,i}) r_{t,i}^{2(d-1)} = 0 \, .
\label{eq:momentum_turning_point}
\eeq
Next, we compute the time dependence of the maximal volume.

\subsection{Time evolution of the volume}
\label{ssec:time_evo_CV}

The following observations on the shape that a maximal surface can take in the geometry \eqref{eq:dS_metric_shock_wave} hold:
\begin{enumerate}
    \item If a turning point exists, it can only be located in the exterior of the cosmological horizon, \ie $r_{t,i} \geq r_{c,i}$.
    This can be checked by analyzing the real and positive roots of eq.~\eqref{eq:momentum_turning_point} with blackening factor \eqref{eq:blackening_factor_shock_SdS}. 
    \item When $P_1>0$, the maximal surface passes through the future exterior of the cosmological horizon, while when $P_1<0$, it goes through the past exterior. When $P_1=0$, the slice crosses the bifurcation surface.
    This result is a direct consequence of expressing the conserved momentum \eqref{eq:conserved_momenta} as $P_i = -f_i(r) \dot{t}$, together with the observation that $f_i(r)<0$ outside the cosmological horizon.
    \item A maximal surface passing through the past exterior of the cosmological horizon necessarily admits a turning point in region 1, due to the boundary condition that the slice is attached to the stretched horizons on the two sides of the geometry. 
\end{enumerate}

\begin{figure}[ht]
    \centering
     \subfigure[Early times]{\label{subfig:caseE_resCV} \includegraphics[scale=0.35]{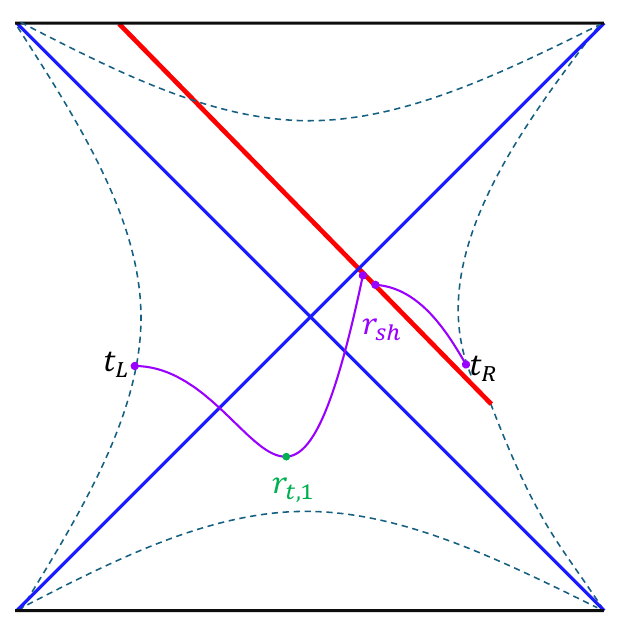} }
    \subfigure[Intermediate times]{\label{fig:caseD_resCV} \includegraphics[scale=0.35]{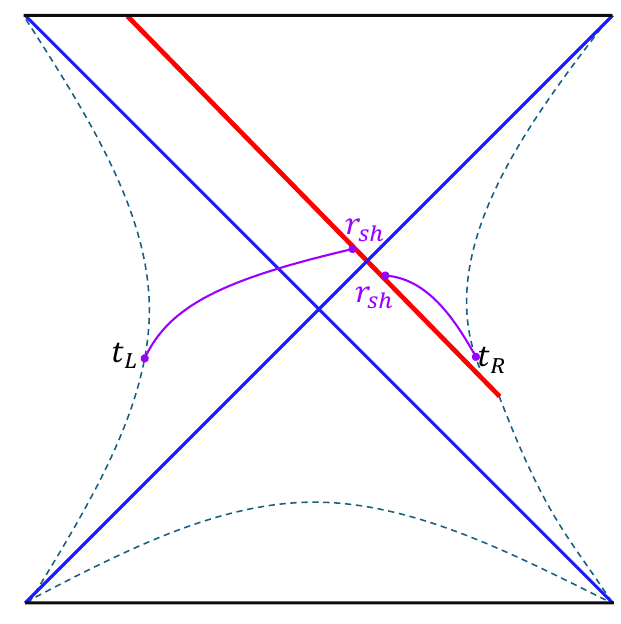} } 
     \subfigure[Late times]{\label{fig:caseC_resCV} \includegraphics[scale=0.35]{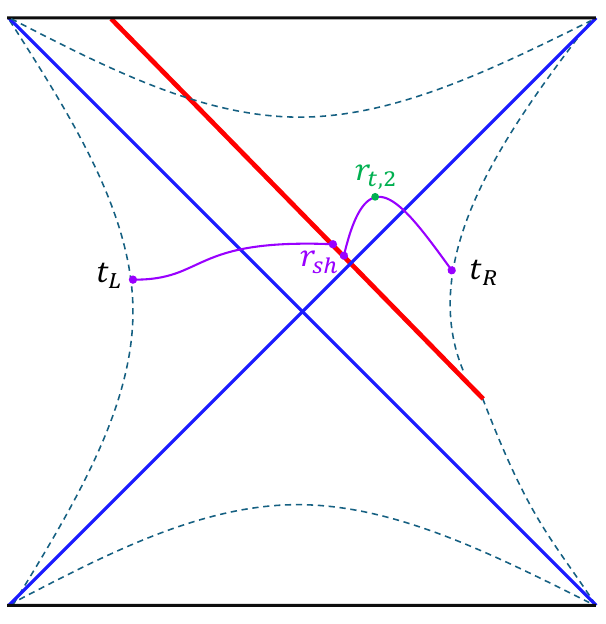} } 
    \caption{Relevant shapes for the time evolution of the extremal surface, corresponding to cases A-C in appendix~\ref{app:details_CV}. The configuration (b) only exists for certain choices of the parameters of the geometry, in particular when $t_w \gg L$.
    }
    \label{fig:relevant_extremal_surfaces_Penrose}
\end{figure}

Several configurations for the maximal surface consistent with the previous observations are possible, as classified in appendix \ref{app:details_CV}.
However, a numerical analysis shows that only few configurations occur during the time evolution. We depict them in fig.~\ref{fig:relevant_extremal_surfaces_Penrose}.

The configurations (a) and (c) always describe early ($t \gtrsim -2 t_w$) and late ($t \gg L$) times of the evolution, while case (b) may happen or not depending on the parameters $(\rho, \varepsilon, t_w)$ characterizing the geometry.
When the shock is inserted in the far past ($t_w \gg L$), the shape (b) always occurs at intermediate times ($t \simeq 0$).

We refer the reader to appendix~\ref{app:details_CV} for the details on the evaluation of the maximal surfaces.
Here we only report the relevant definitions and results.
For convenience, we introduce the quantities
\beq
\tau_i [P_i,r] \equiv  \frac{1}{f_i(r)} - 
\frac{P_i}{f_i(r) \sqrt{f_i(r) r^{2(d-1)} +P_i^2}} \, ,
\qquad 
R_i[P_i,r] \equiv \frac{r^{2(d-1)}}{\sqrt{f_i(r) r^{2(d-1)} +P_i^2}} \, .
\label{eq:deftau_CV}
\eeq
In this way, the boundary times can be expressed as
\begin{subequations}
\beq
t_R + t_w =
\begin{cases}
     - \int_{r_{\rm sh}}^{r_{t,2}} dr \, \tau_2 [-P_2,r] 
+ \int_{r^{\rm st}_2}^{r_{t,2}} dr \, \tau_2 [P_2, r] & 
\text{cases (a), (c)} \\
\int_{r^{\rm st}_2}^{r_{\rm sh}} dr \, \tau_2 [P_2,r]  & \text{case (b)}
\end{cases}
\label{eq:times_tL_CV}
\eeq
\beq
t_L - t_w =
\begin{cases}
r^*_1(r^{\rm st}_1) + r^*_2(r^{\rm st}_2) -  2 r^*_1(r_{t,1}) 
+ \int_{r^{\rm st}_1}^{r_{t,1}} dr \, \tau_1[P_1,r] +
 \int_{r_{\rm sh}}^{r_{t,1}} dr \, \tau_1 [P_1,r] & 
\text{cases (a), (b)} \\
r^*_1(r^{\rm st}_1) + r^*_2(r^{\rm st}_2) -  2 r^*_1(r_{s}) 
+ \int_{r^{\rm st}_1}^{r_{\rm sh}} dr \, \tau_1 [P_1, r]  & \text{case (c)} 
\end{cases}
\label{eq:times_tR_CV}
\eeq
\label{eq:times_CV}
\end{subequations}
Holographic complexity, computed according to the CV conjecture as in eq.~\eqref{eq:CV_formula_gen}, is given by
\begin{subequations}
    \beq
\mathcal{C}_{V1} \equiv \frac{\Omega_{d-1}}{G_N L}
\begin{cases}
    \int_{r^{\rm st}_1}^{r_{t,1}} dr\, R_1 [P_1,r]  +\int_{r_{\rm sh}}^{{r_{t,1}}} dr \, R_1[P_1,r] & \text{case (c)}  \\
    \int_{r^{\rm st}_1}^{r_{s}} dr \, R_1 [P_1,r]   & \text{cases (a), (b)}
\end{cases}
\label{eq:CV1_formula}
\eeq
\beq
\mathcal{C}_{V2} \equiv \frac{\Omega_{d-1}}{G_N L} 
\begin{cases}
 \int_{r^{\rm st}_2}^{r_{t,2}} dr \,R_2[P_2,r]  + \int_{r_{\rm sh}}^{r_{t,2}} dr \, R_2[P_2,r]    & \text{case (a)} \\
   \int_{r^{\rm st}_2}^{r_{\rm sh}} dr \, R_2[P_2,r]     & \text{cases (b), (c)}
\end{cases}
\label{eq:CV2_formula}
\eeq
\label{eq:CVtot_formula_gen}
\end{subequations}
Finally, one can show that the rate of evolution of complexity for all the configurations reads
\beq
\frac{d\mathcal{C}_V}{dt} = \frac{\Omega_{d-1}}{G_N L} \le P_1 \frac{dt_L}{dt} + P_2 \frac{dt_R}{dt}  \ri =
\frac{\Omega_{d-1}}{2 G_N L} \le P_1  + P_2 \ri\, ,
\label{eq:CV_rate}
\eeq
the latter equality corresponding to the symmetric case \eqref{eq:symmetric_times}.

\subsubsection{Hyperfast growth}
\label{ssec:hyperfast_CV}

We analytically show that the CV conjecture admits a hyperfast growth in the future (past) when a turning point reaches future (past) timelike infinity $\mathcal{I}^+ (\mathcal{I}^-)$.
To evaluate the volume precisely, we introduce a regulator $r=r_{\rm max}$ as in eq.~\eqref{eq:definition_rmax_shocks} to define the maximal radial coordinate that the turning point can reach, and we will send $r_{\rm max} \rightarrow \infty$ at the end of the computation.
By virtue of eq.~\eqref{eq:momentum_turning_point}, there is an associated critical momentum $\bar{P}_i$ given by
\beq
\bar{P}_i^2 = \frac{(r_{\rm max})^{2d}}{L^2} + \mathcal{O} \le r_{\rm max}^{2(d-1)} \ri \, , 
\label{eq:maximal_Pi}
\eeq
which clearly diverges ($|\bar{P}_i| \rightarrow \infty$) when the regulator is removed.
In this limit, it is clear from the definition \eqref{eq:deftau_CV} that any integral involving $R_i[\bar{P}_i,r]$ vanishes, unless it is evaluated close to timelike infinity.
Therefore, the only non-vanishing contributions to the volume \eqref{eq:CVtot_formula_gen} come from the following terms
\beq
 \int^{r_{t,i}} dr \, R_i [\bar{P}_i,r]  \approx 
\int^{r_{\rm  max}} dr \, \frac{L r^{2(d-1)}}{\sqrt{(r_{\rm max})^{2d} - r^{2d}}}  =  \frac{L (r_{\rm max})^{d-1}}{2d} \frac{\sqrt{\pi} \Gamma \le \frac{2d-1}{2d} \ri}{ \Gamma \le \frac{2d+1}{2d} \ri} \, .
\label{eq:divergent_integral_CV}
\eeq
The result \eqref{eq:divergent_integral_CV} manifestly shows that the CV 
conjecture \eqref{eq:CV_formula_gen} diverges when $r_{\rm max} \rightarrow \infty$.
Since in this regime $|\bar{P}_i| \rightarrow \infty$, we immediately conclude that the rate \eqref{eq:CV_rate} is divergent, too.\footnote{One can show by using eq.~\eqref{eq:relation_P2_P1} that whenever $P_1 \rightarrow \pm \infty$, the same is true for $P_2$.}

In conclusion, we have analytically argued that the CV proposal and its rate are divergent when the turning point approaches timelike infinity, similar to what happens in empty dS space \cite{Jorstad:2022mls}.
A numerical analysis revealed us that the time evolution of CV is described by maximal
surfaces with the shape depicted in fig.~\ref{fig:relevant_extremal_surfaces_Penrose}, where the early times are governed by the configuration
(a), while the late times by case (c). Both cases admit one turning point, therefore CV reaches a regime where it is divergent.

In relation to the CA proposal analyzed in section~\ref{sec:CA}, we point out that the limit $|P_i| \rightarrow \infty$ leads to a configuration where the extremal surface becomes null, approaching the same shape given by the null boundaries of the WDW patch.
For this reason, the configuration (a) in fig.~\ref{fig:relevant_extremal_surfaces_Penrose} approaches the form of the bottom boundaries of the WDW patch depicted in fig.~\ref{fig:WDW_patches}, with $r_{\rm sh} \rightarrow r_s$.
Similarly, in case (c) the extremal surface approaches the top boundaries of the WDW patch, with $r_{\rm sh} \rightarrow r_b$.

To complete the proof that the volume admits hyperfast growth, we need to show that the divergences occur at finite critical times.
We will prove this statement in subsection \ref{ssec:switchback_CV}.
Before that, we will provide further evidence for the hyperfast growth via a numerical analysis of CV in subsection \ref{ssec:examples_CV}.

\subsection{Explicit examples}
\label{ssec:examples_CV}

\begin{figure}[ht]
    \centering
    \subfigure[]{ \label{subfig:CV_3d_tw4} \includegraphics[scale=0.78]{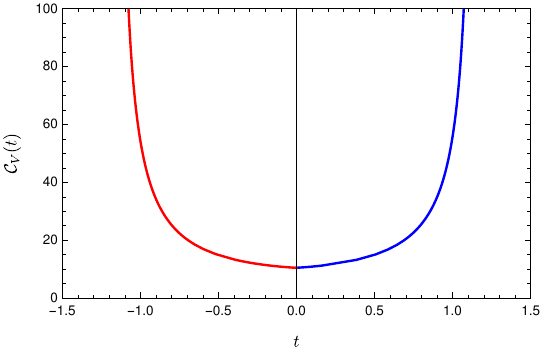} }
    \subfigure[]{ \label{subfig:CVrate_3d_tw4} \includegraphics[scale=0.78]{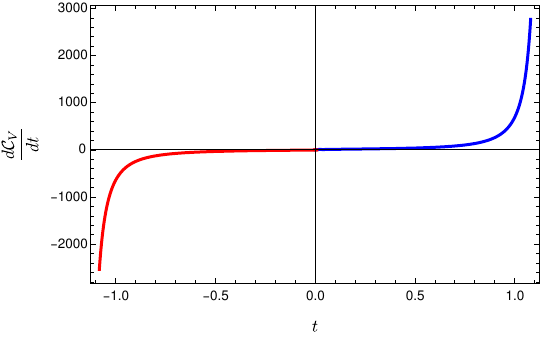} }
    \caption{Time dependence of CV conjecture for symmetric times \eqref{eq:symmetric_times} in the SdS$_3$ background with shockwaves. We set $L=1, 8\pi G_N \mathcal{E}_1=0.01, \varepsilon=0.01, t_w=4, \rho=0.5$. 
    The red part refers to the maximal surface assuming the configuration (a), while the blue part refers to case (c) in fig.~\ref{fig:relevant_extremal_surfaces_Penrose}. }
\end{figure}

We numerically investigate the CV conjecture \eqref{eq:CV_formula_gen} in the case of the metric \eqref{eq:dS_metric_shock_wave}.
Since the numerical analysis turns out to be challenging, we will implement the following list of simplifying assumptions:
\begin{itemize}
    \item We focus on the case $d=2$, where an analytic solution to the integrals in eqs.~\eqref{eq:times_CV} can be obtained in terms of the incomplete elliptic function of the third kind $\Pi$.
    The qualitative features of CV that we are going to discuss persist independently of the number of dimensions.
    \item We restrict to times $t \geq -2t_w$, when the right stretched horizon is time-independent.
    Notice that the time-dependent regime of the stretched horizon is less interesting because it occurs before the shockwave insertion.
    \item We will only plot the CV conjecture during the interval $t \in [t_{c1}, t_{c2}]$ when the turning points do not reach timelike infinity $\mathcal{I}^{\pm}$. 
    This regime is relevant to study the switchback effect.\footnote{The precise definition of $t_{c1}, t_{c2}$ in the CV case will be given in subsection \ref{ssec:switchback_CV}.}
    To explore times outside such regime, one should modify the CV prescription as in section~5 of \cite{Jorstad:2022mls}. 
    \item We will \textit{not} provide a numerical analysis of the case $t_{c1} \leq -2 t_w$, since it is difficult to achieve it by fine-tuning the geometric parameters, and it does not lead to additional physical insights.
\end{itemize}
We report a numerical plot of the time dependence of complexity in fig.~\ref{subfig:CV_3d_tw4} and of its time derivative in fig.~\ref{subfig:CVrate_3d_tw4} for a certain choice of the parameters.
The volume becomes positively divergent in correspondence of two critical times $t_{c1}$ and $t_{c2}$, as anticipated above.
Accordingly, the rate is negatively divergent in the past and positively divergent in the future, while it is very small in the middle part of the time evolution.
The maximal surface only assumes two configurations: (a) in the past and (c) in the future. The former case corresponds to the existence of a turning point in the region before the shockwave insertion, while the latter to a turning point after the shockwave.
These results are similar to the cases of empty dS without shocks \cite{Jorstad:2022mls} and to the regime without special configurations of the WDW patch in the CV2.0 and CA computation for SdS with shockwaves (see \cite{Baiguera:2023tpt} and fig.~\ref{fig:integratedCASdS3case1}).

\begin{figure}[ht]
    \centering
    \subfigure[]{ \label{subfig:CV_3d_tw8} \includegraphics[scale=0.78]{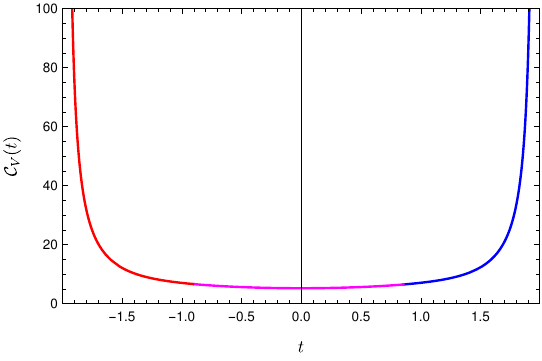} }
    \subfigure[]{ \label{subfig:CVrate_3d_tw8} \includegraphics[scale=0.78]{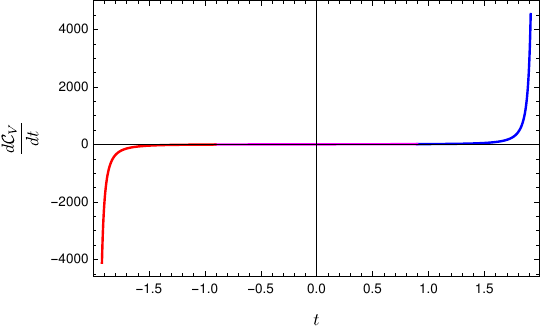} }
    \caption{Time dependence of CV conjecture for symmetric times \eqref{eq:symmetric_times} in the SdS$_3$ background with shockwaves. We set $L=1, 8\pi G_N \mathcal{E}_1=0.01, \varepsilon=0.01, t_w=8, \rho=0.5$.
    The red part refers to the maximal surface assuming the configuration (a), the magenta refers to case (b), and the blue part to case (c) in fig.~\ref{fig:relevant_extremal_surfaces_Penrose}. }
\end{figure}

We depict the time dependence of the volume and its rate for another choice of the parameters with bigger insertion time $t_w$ of the shockwave in fig.~\ref{subfig:CV_3d_tw8} and fig.~\ref{subfig:CVrate_3d_tw8}.
In this case, at intermediate times we find the existence of an additional regime governed by the shape (b) for the maximal surface, whose intersection with the shockwave lies in the interval $r_{\rm sh} \in [r_{c1}, r_{c2}]$.
This regime makes the plateau region longer, and occurs as a consequence of the Penrose diagram of asymptotically dS space getting taller in the presence of a positive pulse of null energy.
Therefore, this result parallels the similar phenomenon that happens for CV2.0 and CA when the special configurations of the WDW patch appear (see \cite{Baiguera:2023tpt} and fig.~\ref{fig:integrated_CA_SdS3_new_st_long_pl}).\footnote{Heuristically, the absence of turning points in configuration (b) for the CV case parallels the fact that the joints of the WDW patch move behind the stretched horizons in the configuration~\ref{fig:alternative4_WDWpatch} for the CA case.}

In summary, a numerical analysis revealed the following main features:
\begin{itemize}
    \item A hyperfast growth, \ie a divergent complexity and rate at finite times $t_{c1}, t_{c2}$.
    This result is analytically validated by the discussion below eq.~\eqref{eq:divergent_integral_CV}, where we have shown that configurations (a) and (c) lead to a hyperfast growth.
    \item A plateau regime where complexity is approximately constant. Its duration increases when the shockwave is inserted earlier in the past, \ie $t_w \gg L$.
\end{itemize}
These results match the analogous behaviour observed in the CV2.0 and CA case, see \cite{Baiguera:2023tpt} and subsection \ref{ssec:examples_CA}.

\subsection{Cosmological switchback effect (volume)}
\label{ssec:switchback_CV}

In this subsection we show that the hyperfast growth occurs at finite boundary time, and we analytically compute the duration of the plateau regime in general dimensions ($d \geq 2$) when $t_w \gg L$.
This investigation will allow us to motivate the existence of a switchback effect for the CV case.

We define the beginning of the plateau regime as the critical time $t_{c1}$ when the turning point in the configuration (a) of fig.~\ref{fig:relevant_extremal_surfaces_Penrose} reaches past timelike infinity $\mathcal{I}^-$, \ie $r_{t,1} = \infty$.
As discussed in subsection \ref{ssec:hyperfast_CV}, this implies that the momenta achieve the critical values $\bar{P}_1, \bar{P}_2 \rightarrow -\infty$.
In this limit, we find
\begin{subequations}
\beq
r^*_1(\infty) = 0 \, , \qquad
\lim_{\bar{P}_1 \rightarrow -\infty} \int_{r_1}^{r_2} dr \, \tau_1 [-\bar{P}_1, r] = 0 \, , 
\eeq
\beq
\qquad
\lim_{\bar{P}_2 \rightarrow -\infty}  \int_{r_1}^{r_2} dr\, \tau_2 [\bar{P}_2, r] 
= \int_{r_1}^{r_2} dr\, \frac{2}{f_2(r)} 
= 2 r^*_2(r_{2}) - 2 r^*_2(r_1) \, ,
\eeq
\label{eq:approximations_tc1}
\end{subequations}
where the definitions of tortoise coordinate \eqref{eq:general_tortoise_coordinate} and of the integrand \eqref{eq:deftau_CV} have been used, and $(r_1, r_2)$ denote two generic radial coordinates.
By summing and subtracting eqs.~\eqref{eq:times_tL_CV}--\eqref{eq:times_tR_CV}, and plugging in the limits \eqref{eq:approximations_tc1}, we get
  \beq
\begin{cases}
 2 t_w - r^*_1(r^{\rm st}_1) + 3 r^*_2(r^{\rm st}_2) - 2 r^*_1(r_{\rm sh1}) 
- 2 r^*_2(r_{\rm sh1}) = 0    \\
 t_{c1} = 2 t_w -4 r^*_1(r_{\rm sh1}) - 2 r^*_1(r^{\rm st}_1) + 2 r^*_2(r^{\rm st}_2) 
\end{cases}
\label{eq:final_tc1_CV}
    \eeq
where we denoted with $r_{\rm sh1}$ the intersection of the maximal surface with the shockwave.

Next, we define the end of the plateau regime as the critical time $t_{c2}$ when the turning point in the configuration (c) of fig.~\ref{fig:relevant_extremal_surfaces_Penrose} reaches future timelike infinity $\mathcal{I}^+$, \ie $r_{t,2}= \infty$.
In this case, the momenta approach the critical values $\bar{P}_1, \bar{P}_2 \rightarrow \infty$, leading to
\begin{subequations}
\beq
r^*_2(\infty) = 0 \, , \qquad
\lim_{\bar{P}_1 \rightarrow \infty} \int_{r_1}^{r_2} dr \, \tau_1 [\bar{P}_1, r] = \lim_{\bar{P}_2 \rightarrow \infty}  \int_{r_1}^{r_2} dr \, \tau_2 [\bar{P}_2, r] = 0 \, , 
\eeq
\beq
\lim_{\bar{P}_2 \rightarrow \infty}  \int_{r_{1}}^{r_{2}} dr\, \tau_2 [-\bar{P}_2, r] = \int_{r_1}^{r_2} dr \, \frac{2}{f_2(r)}  = 2 r^*_2(r_{2}) - 2 r^*_2 (r_1) \, .
\eeq  
\label{eq:approximations_tc2}
\end{subequations}
Summing and subtracting eqs.~\eqref{eq:times_tL_CV}--\eqref{eq:times_tR_CV}, together with the identities \eqref{eq:approximations_tc2}, gives
 \beq
\begin{cases}
    2t_w + r^*_1(r^{\rm st}_1) + r^*_2(r^{\rm st}_2) - 2 r^*_1(r_{\rm sh2}) - 2 r^*_2(r_{\rm sh2}) = 0 \, ,    \\
     t_{c2} = - 2 t_w + 4r^*_2(r_{\rm sh2})  \, , 
\end{cases}
\label{eq:final_tc2_CV}
    \eeq
where $r_{\rm sh2}$ is the intersection of the maximal surface with the shockwave in this configuration.
If we define the duration of the plateau as in eq.~\eqref{eq:def_plateau_time} and we plug in the results \eqref{eq:final_tc1_CV} and \eqref{eq:final_tc2_CV}, we find
\beq
t_{\rm pl} = - 4 t_w +  4 r^*_1(r_{\rm sh1}) + 4 r^*_2(r_{\rm sh2})  +2 r^*_1(r^{\rm st}_1)- 2 r^*_2(r^{\rm st}_2) \, .
\eeq
Remarkably, this expression and the definitions of $t_{c1}, t_{c2}$ coincide with the definitions of the corresponding critical times of the WDW patch, see eqs.~\eqref{eq:tc1}, \eqref{eq:tc2} and reference \cite{Baiguera:2023tpt}, after we take the strict limit $r_{\rm max} \rightarrow \infty$.
The only difference between the definitions of critical times in the CV and CA cases is that in the former one $r_{\rm sh1}$ and $r_{\rm sh2}$ are the intersection of the maximal surface with the shockwave, while in the latter case $r_s$ and $r_b$ correspond to the intersection of the WDW patch with the shockwave.
However, as we discussed below eq.~\eqref{eq:divergent_integral_CV}, in the limit $|P_i| \rightarrow \infty$ these special positions coincide, \ie $r_{\rm sh1} \rightarrow r_s$ and $r_{\rm sh2} \rightarrow r_b$.
For this reason, by repeating the same analysis performed in sections~6.1.1, 6.2.1 and 6.3 of \cite{Baiguera:2023tpt}, we conclude that the duration of the plateau coincides with the CV2.0 and CA cases, leading to the results collected in eqs.~\eqref{eq:linear_plateau_intro} and \eqref{eq:scrambling_time_SdS_intro}.
This remarkable observation shows that a switchback effect also occurs for the CV case, with exactly the same scrambling time as the codimension-zero observables.

\section{Discussion}
\label{sec:discussion}

\subsection{Conclusions}
\label{ssec:conclusions}
We investigated the reaction of the CV and CA conjectures to the insertion of a shockwave at finite boundary time in the SdS background.
This analysis provides a follow-up of the computation performed in reference \cite{Baiguera:2023tpt} for CV2.0 proposal, and extends the study carried out in \cite{Anegawa:2023dad} for a shockwave inserted along the cosmological horizon.
Remarkably, we found that CV, CV2.0 and CA conjectures all present a plateau regime around $t_L=t_R=0$ where they are approximately constant.
The duration of the plateau increases when the shockwave is inserted earlier in the past, until it approaches a linear asymptotic growth \eqref{eq:linear_plateau_intro} with the universal delay \eqref{eq:scrambling_time_SdS_intro}.
The shift provided by the scrambling time $t_*$ is a manifestation of the switchback effect, which occurs for all the above-mentioned complexity conjectures in dS space.
A similar linear increase as a function of the insertion time $t_w$ of the shockwave is also displayed by the complexity of formation in the CV2.0 (see \cite{Baiguera:2023tpt}) and CA cases (see subsection~\ref{ssec:CA_switchback_plateau}).\footnote{We expect that the same phenomenon happens for the complexity of formation in the CV case. However, given the difficulty to perform a precise numerical analysis of this problem, we leave this topic for future investigations.}
In the limit when the shockwave is sent along the cosmological horizon, we find the same conclusion obtained in reference \cite{Anegawa:2023dad}, \ie the beginning of the hyperfast growth is always delayed.\footnote{More specifically, by choosing a special value of the energy along the shockwave as in eq.~(A.6) of \cite{Baiguera:2023tpt}, with the aim to send a light shockwave along the cosmological horizon, the scrambling time \eqref{eq:scrambling_time_SdS_intro} matches with the computations performed in reference \cite{Anegawa:2023dad}.}

As reported in the discussion of \cite{Baiguera:2023tpt}, it is also relevant to observe that the scrambling time \eqref{eq:scrambling_time_SdS_intro} further simplifies in the limit of small black holes $r_h \ll r_c$, leading to the result 
\beq
t^{\rm SdS_{d+1}}_* \approx  \frac{1}{2 \pi T_{c1}} 
\log \le  (1-\rho) \frac{(d-1)\, S_{c1}}{\Delta S_{c1}}   \ri \, ,
\label{eq:scrambling_S}
\eeq
where $S_{c1}$ is the area entropy of the first cosmological horizon, and $\Delta S_{c1} = S_{c2} - S_{c1}$ its variation after the shockwave insertion. 
When the strength $\varepsilon$ of the shockwave is chosen to change the energy by a few thermal quanta $\Delta S_{c1} \sim d$, then the scrambling time reduces to\footnote{The factor of $d$ inside $\Delta S_{c1} \sim d$ is conventional, see eqs.~(7.3)--(7.4) in reference \cite{Baiguera:2023tpt}.}
\beq
t^{\rm SdS_{d+1}}_* \approx  \frac{1}{2 \pi T_{c1}} 
\log \le\frac{d-1}{d}\,(1-\rho) \, S_{c1}   \ri\,.
\label{smarter8b}
\eeq
In particular, the scrambling time is of order $t_* \sim \beta_{c1} \log (1/G_N)$, with $\beta_{c1}$ being the inverse temperature.
This result describes a fast scrambling behaviour similar to black holes in AdS, and it consistent with the computation of two-point functions in a dS setting analyzed in \cite{Milekhin:2024vbb}, and with the out-of-order-correlators investigated in \cite{Geng:2020kxh}.

Our results share some analogies with AdS-Vaidya geometries \cite{Chapman:2018lsv}, but they have a different geometrical origin, as summarized in table~\ref{tab:geom_reasons}.
Looking at the table by columns, we read that the plateau regime is determined by the location of the turning point of the maximal surface in the volume case, and by the top (bottom) joint of the WDW patch in the action case.
Reading by rows, we see that while in AdS geometry the plateau arises because these special positions approach regions at finite radial coordinate (either the event horizon or the singularity), instead in the dS geometry the plateau ends when the same special positions reach timelike infinities $\mathcal{I}^{\pm}$. 

\begin{table}[t!]   
\begin{center}    
\begin{tabular}  {|c|c|c|} \hline  & \textbf{Complexity=volume}  & \textbf{Complexity=action} \\ \hline
\rule{0pt}{4.9ex} \textbf{AdS}   & \makecell{Turning points of the maximal surface \\ approach the event horizons}     & \makecell{Top and bottom joints of the WDW \\ patch lie behind the singularities } \\
\rule{0pt}{4.9ex} \textbf{dS} & \makecell{Turning points of the maximal surface  \\ reach timelike infinities $\mathcal{I}^{\pm}$}  & \makecell{Top or bottom joints of the WDW \\ patch reach timelike infinities $\mathcal{I}^{\pm}$ }   \\[0.2cm]
\hline
\end{tabular}   
\caption{Geometrical reason for the plateau regime of CV and CA conjectures for black holes perturbed by shockwaves in AdS and dS spacetimes.} 
\label{tab:geom_reasons}
\end{center}
\end{table}

Focusing on the linear increase \eqref{eq:linear_plateau_intro} for the duration of the plateau, we notice that a crucial role is played by the fact that a light ray crossing the shockwave brings the left and right static patches into causal contact, thus allowing communication between the two stretched horizons \cite{Gao:2000ga}.
In the case of the codimension-zero complexity proposals (CV2.0 studied in \cite{Baiguera:2023tpt} and CA in section~\ref{sec:CA}), the plateau regime at large insertion times $t_w \gg L$ is characterized by the existence of special configurations of the WDW patch depicted in fig.~\ref{fig:alternative_WDWpatch}.
In the CV setting (section \ref{sec:CV_conjecture}), the plateau regime for a shockwave inserted in the far past is governed by the special shape (b) of the maximal surface in fig.~\ref{fig:extremal_surfaces_Penrose}.
In both cases, the above-mentioned geometrical configurations only arise because in a dS geometry the cosmological horizon grows after a shockwave insertion ($r_{c1} \leq r_{c2}$), contrarily to what happens in AdS space.
For this reason, we may ultimately associate the switchback effect with the causal properties of dS space under matter perturbations with positive energy.

Finally, we discuss another important novelty carried by shockwaves in asymptotically dS space: the CA observable is negative when the special configuration of the WDW patch in fig.~\ref{fig:alternative4_WDWpatch} occurs, see figs.~\ref{fig:integratedCA_case3}, \ref{subfig:CA0_case1} and \ref{subfig:CA0_case2}.
First of all, we notice that this phenomenon does \textit{not} happen in asymptotically AdS spacetime, since in that case there is a time-independent UV divergence coming from the fact that the WDW patch always reaches the timelike boundaries.
Since the prefactor of the leading divergence is controlled by the counterterm length scale, it is always possible to choose an appropriate value of $\ell_{\rm ct}$ such that complexity is positive.
In asymptotically dS space, a similar divergence arises when the WDW patch reaches timelike infinities $\mathcal{I}^{\pm}$, with its prefactor controlled again by $\ell_{\rm ct}$, as observed in eq.~\eqref{eq:hyperfast_CA}.
Since this divergence is absent in the intermediate time regime $t \in [t_{c1}, t_{c2}]$, it cannot be used to impose a positive CA.
Nonetheless, one can check that CA remains positive even in the intermediate time regime, if the unperturbed empty dS or SdS solutions are considered \cite{Jorstad:2022mls,Aguilar-Gutierrez:2024rka}.
Negative CA only arises when we first take a limit where the shockwave is inserted in the far past ($t_w \gg L$), and then we consider a stretched horizon close enough to the cosmological one ($\rho \rightarrow 1$).\footnote{While in general dimension this result is only obtained in the double scaling limit \eqref{eq:double_scaling_intro}, in three dimensions we do not need to take any assumption on the energy $\varepsilon$ carried by the shockwave.}
We may attribute this peculiar behaviour to the causal properties of dS space under perturbations, since these limits lead to the special configuration in fig.~\ref{fig:alternative4_WDWpatch} of the WDW patch.

If we aim to interpret the action as a holographic dual of complexity, an overall negative sign is an unwanted feature because the number of gates generating a target state (or operator) is positive.
There are several way outs to this problem.
One possibility is that quantum corrections might become important, since the regime in which CA becomes negative corresponds to the case where the geometric region delimited by the WDW patch is becoming small.
Another possibility is that holographic complexity in dS space is deeply different than in AdS.
For instance, it may happen that the desired reference state to which complexity is compared is unusual (\ie not a product state).
We leave the interesting study of this puzzle for future investigations.

\subsection{Future developments}
\label{ssec:future_developments}

The analysis presented in this work allowed us to further show that the switchback effect is a universal feature of holographic complexity in asymptotically dS space.
However, there are several research questions that remain open, which we would like to address in the future:
\begin{enumerate}
    \item \textbf{Circuit interpretation.}
    General features of the time evolution of complexity in fast-scrambling systems are encoded by simple circuit models that reproduce the essential properties observed in AdS geometries \cite{Susskind:2014jwa,Brown:2016wib,Chapman:2018lsv,Susskind:2018pmk,Chapman:2021jbh}.
    In references \cite{Susskind:2021esx,Lin:2022nss}, a circuit model reproducing certain features related to the hyperfast growth of dS space was proposed. We aim to reproduce the switchback effect and the scaling behaviour of its scrambling time in this setting.
    \item \textbf{Dual quantum theory.}
    It is believed that the dual quantum description of dS space is a quantum mechanical theory with a finite-dimensional Hilbert space in a maximally mixed state. 
    Recently, two proposals to understand the dual description of three-dimensional dS space have been investigated: either in terms of $T\bar{T}$ deformations \cite{Lewkowycz:2019xse,Shyam:2021ciy,Coleman:2021nor,Batra:2024kjl}, or by means of two copies of doubled-scaled SYK model with an energy condition \cite{Narovlansky:2023lfz,Rahman:2023pgt,Verlinde:2024znh,Verlinde:2024zrh,Rahman:2024vyg}.
    It would be interesting to exploit these techniques to get a better understanding of the dual state, possibly reproducing the switchback effect. 
    For instance, it is not clear whether the microscopic realization of a shockwave in dS space should be understood as a small perturbation to the original system.
    Some proposals for complexity observables in the quantum mechanical setting have been advanced in \cite{Aguilar-Gutierrez:2024nau}.
    \item \textbf{Universality of the cosmological switchback effect.}
    While the program developed in this paper and in \cite{Baiguera:2023tpt} exhausted the study of ``traditional" holographic complexity conjectures (CV, CV2.0 and CA) in a SdS background perturbed by shockwaves at finite boundary time, it would be interesting to consider the case of more general CAny observables.
    The analysis performed in \cite{Aguilar-Gutierrez:2023pnn} showed that when the shockwave is inserted along the cosmological horizon, the switchback effect arises independently of the late-time behaviour of the complexity observables in dS space.
   We would like to check whether the formula  \eqref{eq:scrambling_time_SdS_intro} for the scrambling time is also valid for all the other CAny observables, and if it holds when the shockwave is inserted at finite boundary time.
    This points towards the possibility to prove a universal theorem in general relativity (along the lines of \cite{Engelhardt:2021mju}) that encodes the geometric properties leading to the switchback effect.
    \item \textbf{Holographic complexity with conformal boundary conditions.}
    It was recently observed that fixing the induced metric on a timelike surface does not lead to a well-posed initial boundary value problem in general relativity \cite{An:2020nfw,An:2021fcq}.
    In particular, the same problem arises when imposing Dirichlet boundary conditions on the stretched horizon in dS space \cite{Anninos:2022ujl,Anninos:2023epi,Anninos:2024wpy}.
    On the other hand, it was conjectured (and verified in some cases, see, \eg \cite{Witten:2018lgb,An:2021fcq,Anninos:2022ujl,Anninos:2023epi,Anninos:2024wpy}) that conformal boundary conditions lead instead to a well-posed problem.
    It would be interesting to properly define holographic complexity observables consistent with these novel requirements. 
     \item \textbf{Centaur geometries and shockwaves.}
    Centaur geometries provide a setting where dS space is embedded inside AdS spacetime, which disposes of the standard timelike boundary where the dual observables are naturally defined.
    Shockwaves have been studied in these geometries in order to compute out-of-order correlators \cite{Anninos:2018svg} (for the original applications to the AdS case, see \cite{Shenker:2013pqa,Shenker:2013yza,Maldacena:2015waa}).
    We aim to analyze holographic complexity in centaur geometries perturbed by a shockwave in order to understand whether the switchback effect is a solid property of dS space, in a different context than static patch holography.
    \item \textbf{Other generalizations of the holographic setting.}
    An immediate extension of the holographic analysis that we performed would be to compute the complexity conjectures at earlier times than the insertion of the shock ($t_R<-t_w$). 
    This regime could be relevant because the shock may still influence the geometric observables through the fact that the stretched horizon is time-dependent. 
    Another natural generalization of our studies would be to consider the two-dimensional case. Since gravity is not dynamical, shockwave solutions are different and do not involve a jump in the mass of the black hole.
    \item \textbf{Causality properties of dS space.}
    Since the switchback effect arose as a consequence of geometric properties of dS space under the insertion of a null pulse of energy, we plan to better understand the causality properties of dS space.
    For instance, a realization of Gao-Wald theorems in terms of the Shapiro time delay of light rays crossing a shockwave in SdS space was considered in \cite{Bittermann:2022hhy}. We would like to elaborate on the relation between this phenomenon and the behaviour of holographic complexity in the same setting.
\end{enumerate}

\section*{Acknowledgements}

We are happy to thank Shira Chapman and Rob Myers for initial collaboration on the project, and for valuable discussions.
We gratefully acknowledge D.~A.~Galante for interesting comments on the draft.
The work of SB and RB is supported by the Israel Science Foundation (grant No.~1417/21), the German Research Foundation through a German-Israeli Project Cooperation (DIP) grant ``Holography and the Swampland'' and by  Carole and Marcus Weinstein through the BGU Presidential Faculty Recruitment Fund. 
SB is grateful to the Azrieli foundation for the award of an Azrieli fellowship.
RB is grateful to the Kreitman School of
Advanced Graduate Studies for the award of Negev doctoral studies scholarship.

\appendix

\section{Details for complexity=action}
\label{app:details_CA}

This appendix collects technical material associated with the computation of CA conjecture in the geometry \eqref{eq:dS_metric_shock_wave}.
We calculate the boundary terms during the intermediate time regime in subsection~\ref{app:ssec:bdy_CA_intermediate}.
We consider the other regimes of the evolution of complexity in subsection~\ref{app:ssec:other_regimes}.

\subsection{Computation of the boundary terms in the intermediate regime}
\label{app:ssec:bdy_CA_intermediate}

Let us consider the boundary terms \eqref{eq:def_Ibdy} of the gravitational action at a fixed instant $t \in [t_{c1}, t_{c2}]$.
We assume that the WDW patch takes the configuration depicted in fig.~\ref{subfig:intermediate_regime_dS}.
The following computation also applies to any other shape reported in fig.~\ref{fig:alternative_WDWpatch}, except for the special case \ref{fig:alternative4_WDWpatch} in which the top and bottom joints move behind the stretched horizons.
The latter scenario will be studied separately below.

\subsubsection{Standard configuration of the WDW patch}
\label{app:ssec:standard_WDW}

\paragraph{Codimension-one boundaries.}
In this configuration, there are no GHY terms.
We choose to describe the congruence of null geodesics generating the null boundaries of the WDW patch and the shockwave with an affine parameter so that the acceleration vanishes.\footnote{See eqs.~\eqref{eq:list_null_normals} and \eqref{eq:list_normal_shockwave} below for the specific parametrization.}
As anticipated in the list of bullets below eq.~\eqref{eq:CA_conjecture}, this implies that $I_{\mathcal{N}}=0$.

\paragraph{Joint terms.}
There are several joints, all obtained by intersections of codimension-one null boundaries.
The outward-directed normal one-forms to the boundaries of the WDW patch are given as follows\footnote{In order to pick the correct orientation for the normal one-forms, we remind that exactly one among the null coordinates $(u,v)$ flips sign whenever a horizon is crossed.}
\beq
\begin{aligned}
&  \mathrm{TR}: \qquad
k_{\mu}^{\rm TR} dx^{\mu} = \alpha du\Big|_{u = t_R - r^*_2(r^{\rm st}_2)} \\
&  \mathrm{TL}: \qquad
k_{\mu}^{\rm TL} dx^{\mu} = \begin{cases}
- \alpha' \le  du + \frac{2}{f_1(r)} dr \ri\Big|_{v = -t_L + r^*_1(r^{\rm st}_1)}  & \mathrm{if} \, \, r \leq r_b \\
- \tilde{\alpha}' \le  du + \frac{2}{f_2(r)} dr \ri\Big|_{v = -t_w - r^*_2(r^{\rm st}_2) + 2r^*_2(r_b) }  & \mathrm{if} \, \, r >  r_b 
\end{cases}  \\
&  \mathrm{BR}: \qquad
k_{\mu}^{\rm BR} dx^{\mu} = \begin{cases}
-\tilde{\beta}' \le  du + \frac{2}{f_1(r)} dr \ri\Big|_{v = -t_w - r^*_2(r^{\rm st}_2) + 2 r^*_1(r_s) }  & \mathrm{if} \, \, r > r_s \\
- \beta' \le  du + \frac{2}{f_2(r)} dr \ri\Big|_{v = t_R + r^*_2(r^{\rm st}_2)}  & \mathrm{if} \, \, r \leq  r_s 
\end{cases}  \\
&  \mathrm{BL}: \qquad
k_{\mu}^{\rm BL} dx^{\mu} =  \beta du\Big|_{u = -t_L - r^*_1(r^{\rm st}_1)} \\
\end{aligned}
\label{eq:list_null_normals}
\eeq
where R denotes right, L left, T top, B bottom.
In the previous expressions, $\alpha,\alpha',\tilde{\alpha}',\beta,\beta',\tilde{\beta}'$ are positive constants which parametrize the ambiguity in normalizing the null normals.
In order for the previous null normals to be affinely parametrized when crossing the shockwave, we further require the conditions\footnote{The conditions \eqref{eq:identities_affine_parametrization} concretely come from requiring that the null-null joints evaluated at the special positions $r_s$ (respectively $r_b$) of the WDW patch vanish. 
See \cite{Chapman:2018dem} for the derivation of this condition in the asymptotically AdS case.}
\beq
\frac{\beta'}{\tilde{\beta}'} = \frac{f_2(r_s)}{f_1(r_s)} \, , \qquad
\frac{\tilde{\alpha}'}{\alpha'} = \frac{f_2(r_b)}{f_1(r_b)} \, .
\label{eq:identities_affine_parametrization}
\eeq
The shockwave itself provides a null surface, whose corresponding normal one-forms are
\beq
k_{\mu}^{\rm s} dx^{\mu} = \begin{cases}
 \gamma du\Big|_{u = -t_w - r^*_2(r^{\rm st}_2) }  & \mathrm{if} \, \, u > -t_w-r^*_2(r^{\rm st}_2) \\
 - \gamma  du \Big|_{u = -t_w - r^*_2(r^{\rm st}_2)}  & \mathrm{if} \, \, u \leq  -t_w - r^*_2(r^{\rm st}_2)
\end{cases} 
\label{eq:list_normal_shockwave}
\eeq
where $\gamma>0$, and the parametrization is chosen to be affine.
However, it is important to stress that the shockwave does not contribute to the codimension-one boundaries for the evaluation of CA, because the bulk region inside the WDW patch is continuous.
The jump in the Penrose diagram (for instance, in fig.~\ref{fig:WDW_patches}) is just an artifact of the way in which we depict the geometry, \ie to draw the horizons continuously.

Following the recipe \eqref{eq:integrand_a_actionjoints} and the sign prescription in \cite{Lehner:2016vdi}, we compute all the non-vanishing joint contributions:
\begin{subequations}
    \beq
I_{\mathcal{J}}^{r_{m1}} = \frac{\Omega_{d-1}}{8 \pi G_N} (r_{m1})^{d-1} \log \left| \frac{\beta \beta' f_1(r_s)}{f_2(r_s) f_1(r_{m1})}  \right| \, , \qquad
I_{\mathcal{J}}^{r_{m2}} = \frac{\Omega_{d-1}}{8 \pi G_N} (r_{m2})^{d-1} \log \left| \frac{\alpha \alpha' f_2(r_b)}{f_1(r_b) f_2(r_{m2})}  \right| \, ,
\eeq
\beq
I_{\mathcal{J}}^{r^{\rm st}_1} = - \frac{\Omega_{d-1}}{8 \pi G_N} (r^{\rm st}_{1})^{d-1} \log \left| \frac{\alpha' \beta}{f_1(r^{\rm st}_{1})}  \right| \, ,
\qquad
I_{\mathcal{J}}^{r^{\rm st}_2} = - \frac{\Omega_{d-1}}{8 \pi G_N} (r^{\rm st}_{2})^{d-1} \log \left| \frac{\alpha \beta'}{f_2(r^{\rm st}_{2})}  \right| \, ,
\eeq
\end{subequations}
where the superscript denotes the radial coordinate of the joint under consideration. 

\paragraph{Counterterm on null boundaries.}
To compute the counterterm \eqref{eq:counterterm_action} on null boundaries, we evaluate the expansion parameter $\Theta$ for all the congruences of null geodesics generating the boundaries of the WDW patch.
The results read
\beq
\begin{aligned}
&  \mathrm{TR}: \qquad
\Theta^{\rm TR}  = - \frac{\alpha (d-1)}{r} \\
&  \mathrm{TL}: \qquad
\Theta^{\rm TL}   = \begin{cases}
- \frac{\alpha' (d-1)}{r} & \mathrm{if} \, \, r \leq r_b \\
- \frac{\tilde{\alpha}' (d-1)}{r} & \mathrm{if} \, \, r >  r_b 
\end{cases}  \\
&  \mathrm{BR}: \qquad
\Theta^{\rm BR}   = \begin{cases}
-\frac{\tilde{\beta}' (d-1)}{r}  & \mathrm{if} \, \, r > r_s \\
- \frac{\beta' (d-1)}{r}  & \mathrm{if} \, \, r \leq  r_s 
\end{cases}  \\
&  \mathrm{BL}: \qquad
\Theta^{\rm BL}  = - \frac{\beta (d-1)}{r} \\
\end{aligned}
\label{eq:list_expansions}
\eeq
By using these expressions, we correspondingly get the contributions
\begin{subequations}
    \beq
I_{\rm ct}^{\rm TR} 
 = - \frac{\Omega_{d-1}}{8 \pi G_N} \left\lbrace (r_{m2})^{d-1} \left[ \log \left| \frac{\alpha \ell_{\rm ct} (d-1)}{r_{m2}} \right| + \frac{1}{d-1} \right] 
- (r^{\rm st}_{2})^{d-1} \left[ \log \left| \frac{\alpha \ell_{\rm ct} (d-1)}{r^{\rm st}_{2}} \right| + \frac{1}{d-1} \right] \right\rbrace \, ,
\eeq
\beq
I_{\rm ct}^{\rm BL}  = - \frac{\Omega_{d-1}}{8 \pi G_N} \left\lbrace (r_{m1})^{d-1} \left[ \log \left| \frac{\beta \ell_{\rm ct} (d-1)}{r_{m1}} \right| + \frac{1}{d-1} \right] 
-  (r^{\rm st}_{1})^{d-1} \left[ \log \left| \frac{\beta \ell_{\rm ct} (d-1)}{r^{\rm st}_{1}} \right| + \frac{1}{d-1} \right] \right\rbrace \, ,
\eeq
\beq
\begin{aligned}
I_{\rm ct}^{\rm BR}& = - \frac{\Omega_{d-1}}{8 \pi G_N} \left\lbrace (r_{m1})^{d-1} \left[ \log \left| \frac{\tilde{\beta}' \ell_{\rm ct} (d-1)}{r_{m1}} \right| + \frac{1}{d-1} \right] \right. \\
& \left . -  (r^{\rm st}_{2})^{d-1} \left[ \log \left| \frac{\beta' \ell_{\rm ct} (d-1)}{r^{\rm st}_{2}} \right| + \frac{1}{d-1} \right]
+ (r_s)^{d-1} \log \left| \frac{\beta'}{\tilde{\beta}'} \right|
\right\rbrace \, ,
\end{aligned}
\eeq
\beq
\begin{aligned}
I_{\rm ct}^{\rm TL} & = - \frac{\Omega_{d-1}}{8 \pi G_N} \left\lbrace (r_{m2})^{d-1} \left[ \log \left| \frac{\tilde{\alpha}' \ell_{\rm ct} (d-1)}{r_{m2}} \right| + \frac{1}{d-1} \right] \right. \\
& \left . -  (r^{\rm st}_{1})^{d-1} \left[ \log \left| \frac{\alpha' \ell_{\rm ct} (d-1)}{r^{\rm st}_{1}} \right| + \frac{1}{d-1} \right]
+ (r_b)^{d-1} \log \left| \frac{\alpha'}{\tilde{\alpha}'} \right|
\right\rbrace \, .
\end{aligned}
\eeq
\end{subequations}

\paragraph{Total boundary term.}
Combining all the boundary terms as in eq.~\eqref{eq:CA_conjecture}, we get
\beq
\begin{aligned}
& I_{\rm bdy} = \frac{\Omega_{d-1}}{8 \pi G_N} \left\lbrace  
(r_{m1})^{d-1} \left[ \log \left| \frac{(r_{m1})^2}{f_1(r_{m1}) \ell_{\rm ct}^2 (d-1)^2} \right| - \frac{2}{d-1} \right] \right. \\
& \left.  + (r_{m2})^{d-1} \left[ \log \left| \frac{(r_{m2})^2}{f_2(r_{m2}) \ell_{\rm ct}^2 (d-1)^2} \right| - \frac{2}{d-1} \right]
- (r^{\rm st}_{1})^{d-1} \left[ \log \left| \frac{(r^{\rm st}_{1})^2}{f_1(r^{\rm st}_{1}) \ell_{\rm ct}^2 (d-1)^2} \right| - \frac{2}{d-1} \right] \right. \\
& \left.  - (r^{\rm st}_{2})^{d-1} \left[ \log \left| \frac{(r^{\rm st}_{2})^2}{f_2(r^{\rm st}_{2}) \ell_{\rm ct}^2 (d-1)^2} \right| - \frac{2}{d-1} \right]
- (r_s)^{d-1} \log \left| \frac{f_2(r_s)}{f_1(r_s)} \right| 
- (r_b)^{d-1} \log \left| \frac{f_1(r_b)}{f_2(r_b)} \right| 
\right\rbrace \, .
\end{aligned}
\label{eq:total_boundary_intermediate_regime}
\eeq
As expected, the normalization of null normals all cancel in this computation, and the only arbitrary scale entering the result is the counterterm parameter $\ell_{\rm ct}.$
In order to derive the previous result, we used the identities \eqref{eq:identities_affine_parametrization}.

\subsubsection{Special configuration of the WDW patch}

The configuration in fig.~\ref{fig:alternative4_WDWpatch} requires a separate analysis.

\paragraph{GHY term.}
First of all, there are now timelike boundaries corresponding to the stretched horizons.
The spacelike one-form normal to them reads
\beq
n_{\mu} dx^{\mu} = \frac{dr}{\sqrt{f_i(r^{\rm st}_i)}} \, , \qquad
(i=1,2) \, .
\eeq
Consequently, we obtain the GHY contributions
\begin{subequations}
    \beq
I_{\rm GHY}^{R} = \frac{\Omega_{d-1}}{8 \pi G_N} \le t_R + t_w - 2 r^*_2(r_b) + 2 r^*_2(r^{\rm st}_2) \ri \frac{(r^{\rm st}_2)^{d-2}}{2} \left[ 2(d-1) f_2(r^{\rm st}_2) + r^{\rm st}_2 f'_2(r^{\rm st}_2) \right] \, ,
\eeq
\beq
I_{\rm GHY}^{L} = \frac{\Omega_{d-1}}{8 \pi G_N} \le t_w - t_L - 2 r^*_1(r_s) +  r^*_1(r^{\rm st}_1) +  r^*_2(r^{\rm st}_2) \ri \frac{(r^{\rm st}_1)^{d-2}}{2} \left[ 2(d-1) f_1(r^{\rm st}_1) + r^{\rm st}_1 f'_1(r^{\rm st}_1) \right] \, ,
\eeq
\end{subequations}
where $L(R)$ denote the left (right) stretched horizons.

\paragraph{Joint terms.}
The null boundaries of the WDW patch are now composed by four parts, corresponding to the following normal one-forms:
\beq
\begin{aligned}
&  \mathrm{TR}: \qquad
k_{\mu}^{\rm TR} dx^{\mu} = \beta' \le du + \frac{2}{f_2(r)} dr \ri\Big|_{v= t_R + r^*_2(r^{\rm st}_2)} \\
&  \mathrm{TL}: \qquad
k_{\mu}^{\rm TL} dx^{\mu} = 
- \tilde{\beta}' \le  du + \frac{2}{f_1(r)} dr \ri\Big|_{v = -t_w - r^*_2(r^{\rm st}_2) + 2 r^*_1(r_s)} 
 \\
&  \mathrm{BL}: \qquad
k_{\mu}^{\rm BL} dx^{\mu} = 
 \alpha' \le  du + \frac{2}{f_1(r)} dr \ri\Big|_{v = -t_L + r^*_1(r^{\rm st}_1) } \\
&  \mathrm{BR}: \qquad
k_{\mu}^{\rm BR} dx^{\mu} = - \tilde{\alpha}' \le du + \frac{2}{f_2(r)} dr \ri\Big|_{v = -t_w - r^*_2(r^{\rm st}_2) + 2 r^*_2(r_b)} \\
\end{aligned}
\label{eq:list_null_normals_joints_behind_stretched}
\eeq
Other than this difference, the calculation of the joint terms then proceeds in a similar way to appendix \ref{app:ssec:standard_WDW}, giving
\beq
I_{\mathcal{J}} = \frac{\Omega_{d-1}}{8 \pi G_N} \left[ (r^{\rm st}_2)^{d-1} \log \left| \frac{\beta'}{\tilde{\alpha}'} \right| + (r^{\rm st}_1)^{d-1} \log \left| \frac{\alpha'}{\tilde{\beta}'} \right|  \right] \, .
\eeq

\paragraph{Counterterm on null boundaries.}
In this case the expansion for the congruence of null geodesics is given by
\beq
\begin{aligned}
&  \mathrm{TR}: \qquad
\Theta^{\rm TR}  =  \frac{\beta' (d-1)}{r} \\
&  \mathrm{TL}: \qquad
\Theta^{\rm TL}  = - \frac{\tilde{\beta}' (d-1)}{r}  \\
&  \mathrm{BL}: \qquad
\Theta^{\rm BL}  =  \frac{\alpha' (d-1)}{r}  \\
&  \mathrm{BR}: \qquad
\Theta^{\rm BR}  = - \frac{\tilde{\alpha}' (d-1)}{r} \\
\end{aligned}
\eeq
The sum of all the counterterms reads
\beq
I_{\rm ct}   = \frac{\Omega_{d-1}}{8 \pi G_N} \left\lbrace  (r^{\rm st}_1)^{d-1} \log \left| \frac{\tilde{\beta}'}{\alpha'} \right|
+ (r^{\rm st}_2)^{d-1} \log \left| \frac{\tilde{\alpha}'}{\beta'} \right|
+ (r_s)^{d-1} \log \left| \frac{\beta'}{\tilde{\beta}'} \right|
+ (r_b)^{d-1} \log \left| \frac{\alpha'}{\tilde{\alpha}'} \right|
\right\rbrace \, .
\eeq

\paragraph{Total boundary term.}
Summing all the boundary terms as defined in eq.~\eqref{eq:def_Ibdy}, we obtain
\beq
\begin{aligned}
I_{\rm bdy} &  = \frac{\Omega_{d-1}}{8 \pi G_N} \left\lbrace  (r_s)^{d-1} \log \left| \frac{f_2(r_s)}{f_1(r_s)} \right|
+ (r_b)^{d-1} \log \left| \frac{f_1(r_b)}{f_2(r_b)} \right|
\right. \\
& \left.  + \frac{(r^{\rm st}_2)^{d-2}}{2} \le t_R + t_w - 2 r^*_2(r_b) + 2 r^*_2(r^{\rm st}_2) \ri  \left[ 2(d-1) f_2(r^{\rm st}_2) + r^{\rm st}_2 f'_2(r^{\rm st}_2) \right]   \right. \\
& \left.  +  \frac{(r^{\rm st}_1)^{d-2}}{2}  \le t_w - t_L - 2 r^*_1(r_s) +  r^*_1(r^{\rm st}_1) +  r^*_2(r^{\rm st}_2) \ri \left[ 2(d-1) f_1(r^{\rm st}_1) + r^{\rm st}_1 f'_1(r^{\rm st}_1) \right]
\right\rbrace \, .
\end{aligned}
\label{eq:bdy_term_special_config}
\eeq

\subsection{Complexity=action in other regimes}
\label{app:ssec:other_regimes}

The computation of the boundary terms \eqref{eq:def_Ibdy} during other regimes of the time evolution can be done in a similar way by using the strategy outlined in appendix \ref{app:ssec:bdy_CA_intermediate}.
Due to the analysis of critical times in subsection \ref{ssec:special_configurations_WDW}, we also conclude that the special configuration depicted in fig.~\ref{fig:alternative4_WDWpatch} never occur in these cases.
We directly report the results:
\begin{itemize}
    \item Regime $t_{c0} \leq t <  t_{c1}$:
    \begin{adjustwidth}{-\leftmargin}{0cm}
    \small
    \beq
\begin{aligned}
& I_{\rm bdy} = \frac{\Omega_{d-1}}{8 \pi G_N} \left\lbrace 
(r_{m2})^{d-1} \left[ \log \left| \frac{(r_{m2})^2}{f_2(r_{m2}) \ell_{\rm ct}^2 (d-1)^2} \right| - \frac{2}{d-1} \right] \right. \\
& \left. + (r_{\rm max})^{d-1} \left[ \log \left| \frac{(r_{\rm max})^2}{f_1(r_{\rm max}) \ell_{\rm ct}^2 (d-1)^2} \right| - \frac{2}{d-1} \right]
-  (r^{\rm st}_{1})^{d-1} \left[ \log \left| \frac{(r^{\rm st}_{1})^2}{f_1(r^{\rm st}_{1}) \ell_{\rm ct}^2 (d-1)^2} \right| - \frac{2}{d-1} \right] \right. \\
& \left. -  (r^{\rm st}_{2})^{d-1} \left[ \log \left| \frac{(r^{\rm st}_{2})^2}{f_2(r^{\rm st}_{2}) \ell_{\rm ct}^2 (d-1)^2} \right| - \frac{2}{d-1} \right] 
- (r_s)^{d-1} \log \left| \frac{f_2(r_s)}{f_1(r_s)} \right|
- (r_b)^{d-1} \log \left| \frac{f_1(r_b)}{f_2(r_b)} \right|
\right. \\
& \left.  - \frac{(r_{\rm max})^{d-2}}{2} 
\left[ 2(d-1) f_1(r_{\rm max}) + r_{\rm max} \, f_1'(r_{\rm max}) \right] 
\le t_w - t_L - r^*_1(r^{\rm st}_1) + r^*_2(r^{\rm st}_2) + 2 r^*_1(r_{\rm max}) - 2r^*_1(r_s) \ri 
\right\rbrace \, .
\end{aligned}
\label{eq:CA_bdy_tc0_tc1}
\eeq
\end{adjustwidth}
\normalsize
\item Regime $t_{c2} \leq t < t_{c3}$:
 \begin{adjustwidth}{-\leftmargin}{0cm}
    \small
\beq
\begin{aligned}
& I_{\rm bdy}  = \frac{\Omega_{d-1}}{8 \pi G_N} \left\lbrace 
(r_{m1})^{d-1} \left[ \log \left| \frac{(r_{m1})^2}{f_1(r_{m1}) \ell_{\rm ct}^2 (d-1)^2} \right| - \frac{2}{d-1} \right] \right. \\
& \left. + (r_{\rm max})^{d-1} \left[ \log \left| \frac{(r_{\rm max})^2}{f_1(r_{\rm max}) \ell_{\rm ct}^2 (d-1)^2} \right| - \frac{2}{d-1} \right]
-  (r^{\rm st}_{1})^{d-1} \left[ \log \left| \frac{(r^{\rm st}_{1})^2}{f_1(r^{\rm st}_{1}) \ell_{\rm ct}^2 (d-1)^2} \right| - \frac{2}{d-1} \right] \right. \\
& \left. -  (r^{\rm st}_{2})^{d-1} \left[ \log \left| \frac{(r^{\rm st}_{2})^2}{f_2(r^{\rm st}_{2}) \ell_{\rm ct}^2 (d-1)^2} \right| - \frac{2}{d-1} \right] 
- (r_s)^{d-1} \log \left| \frac{f_2(r_s)}{f_1(r_s)} \right|
- (r_b)^{d-1} \log \left| \frac{f_1(r_b)}{f_2(r_b)} \right|
\right. \\
& \left.  - \frac{(r_{\rm max})^2}{2} \left[ 2(d-1) f_2(r_{\rm max}) + r_{\rm max} f'_2(r_{\rm max})  \right] \le t_R + t_w  - 2 r^*_2(r_b) + 2 r^*_2(r_{\rm max}) \ri
\right\rbrace \, .
\end{aligned}
\label{eq:CA_bdy_tc2_tc3}
\eeq
\end{adjustwidth}
\normalsize
\item Regime $ t \geq t_{c3}$: 
 \begin{adjustwidth}{-\leftmargin}{0cm}
    \small
    \beq
\begin{aligned}
&  I_{\rm bdy}   = \frac{\Omega_{d-1}}{8 \pi^2 G_N} \left\lbrace 
(r_{m1})^{d-1} \left[ \log \left| \frac{(r_{m1})^2}{f_1(r_{m1}) \ell_{\rm ct}^2 (d-1)^2} \right| - \frac{2}{d-1} \right] - (r_s)^{d-1} \log \left| \frac{f_2(r_s)}{f_1(r_s)} \right| \right. \\
& \left. + (r_{\rm max})^{d-1} \left[ \log \left| \frac{r_{\rm max}}{\sqrt{-f_1(r_{\rm max}}) \ell_{\rm ct} (d-1)} \right|  
 - \frac{1}{d-1} \right]
+ (r_{\rm max})^{d-1} \left[ \log \left| \frac{r_{\rm max}}{\sqrt{-f_2(r_{\rm max})} \ell_{\rm ct} (d-1)} \right|  
 - \frac{1}{d-1} \right] \right. \\
& \left. -  (r^{\rm st}_{1})^{d-1} \left[ \log \left| \frac{(r^{\rm st}_{1})^2}{f_1(r^{\rm st}_{1}) \ell_{\rm ct}^2 (d-1)^2} \right| - \frac{2}{d-1} \right]   -  (r^{\rm st}_{2})^{d-1} \left[ \log \left| \frac{(r^{\rm st}_{2})^2}{f_2(r^{\rm st}_{2}) \ell_{\rm ct}^2 (d-1)^2} \right| - \frac{2}{d-1} \right]  
\right. \\
& \left.  - \frac{(r_{\rm max})^{d-2}}{2}  \le t_L - t_w - r^*_1(r^{\rm st}_1) - r^*_2(r^{\rm st}_2)  + 2 r^*_1(r_{\rm max})  \ri  
\left[ 2(d-1) f_1(r_{\rm max}) + r_{\rm max} \, f_1'(r_{\rm max}) \right] \right. \\
& \left. - \frac{(r_{\rm max})^{d-2}}{2}   \le t_R + t_w  \ri  
\left[ 2(d-1) f_2(r_{\rm max}) + r_{\rm max} \, f_2'(r_{\rm max}) \right] 
\right\rbrace \, .
\end{aligned}
\label{eq:CA_bdy_aftertc3}
\eeq
    \end{adjustwidth}
\normalsize
\end{itemize}

In the previous computations, it is relevant to point out that the total CA observable \eqref{eq:CA_conjecture} is always continuous.
On the contrary, as discussed in subsection~\ref{ssec:examples_CA}, the corresponding rate is in general discontinuous at the critical times $t_{c1}, t_{c2}$ due to the existence of additional GHY terms outside the intermediate time regime.

It is straightforward to determine the rate of evolution of complexity by applying the time derivatives \eqref{eq:derivatives_tR} and \eqref{eq:derivatives_tL} to the CA conjecture \eqref{eq:CA_conjecture}.
Since this is not explicitly required for the analysis performed in this work, we do not report its expression here.
For our purposes, it is only relevant to focus on the late time behaviour, which corresponds to the case when the geometric data become
\beq
r_{m1} \rightarrow r_{c1} \, , \quad
r_s \rightarrow r_{c2} \quad  \Rightarrow \quad
f_2(r_s) \rightarrow 0 \, .
\eeq
In this limit, one can show that the total rate reads
\beq
\frac{d \mathcal{C}_{A}}{dt} (t \gg t_{c3})  \approx \frac{\Omega_{d-1}}{8 \pi^2 G_N L^2} (d+1)
 \le  r_{\rm max} \ri^d   
 \approx 
\frac{\Omega_{d-1}}{8 \pi^2 G_N L^2} (d+1) 
\le  \frac{r_{c1}}{\delta} \ri^d  \, ,
\label{eq:late_time_rate_dS}
\eeq
where we approximated the results by assuming $\delta \ll 1$ and in the last step we used the definition \eqref{eq:definition_rmax_shocks}.
In summary, we found that the late time behaviour coincides with the case of empty dS without a shockwave for late times \cite{Jorstad:2022mls}.
In other words, this regime is dominated by the asymptotic structure close to timelike infinity, independently of the presence of a black hole in the spacetime.

\subsection{Linear approximation for the complexity of formation}
\label{app:ssec:linear_CA_formation}

In this subsection, we provide additional steps to determine the linear approximation of the complexity of formation in eqs.~\eqref{eq:linear_approx_CA0} (in three dimensions) and \eqref{eq:approx_CA_formation_gend} (in general dimensions, under the double-scaling limit \eqref{eq:double_scaling_limit}).
The expansion is valid when the shockwave is inserted in the far past ($t_w \gg L$), in which case the special positions and the joints of the WDW patch satisfy 
\beq
\lim_{t_w \rightarrow \infty} r_s = r_{c2} \, , \quad
\lim_{t_w \rightarrow \infty} r_b = r_{c1} \, , \quad 
\lim_{t_w \rightarrow \infty} r_{m1} = r^{\rm st}_1 \, , \quad
\lim_{t_w \rightarrow \infty} r_{m2} = r^{\rm st}_{2} \, .
\label{eq:simplifications_twinf_limit_CF_SdS3}
\eeq

\subsubsection{Three dimensions}

In three dimensions, the full CA computation can be carried out analytically.
In the case of the bulk term, the result can simply be inherited, by means of eq.~\eqref{eq:bulk_CV20_relation}, from the CV2.0 calculation performed in reference \cite{Baiguera:2023tpt}:
\begin{subequations}
    \beq
I_{\mathcal{B}} (0) \approx \frac{1}{8 G_N}   \le  1 - \rho^2 \ri \le a_1^2 + a_2^2 \ri \le t_w - t_* \ri \, ,
 \label{eq:Cf_largetw_SdS3}
\eeq
\beq
\begin{aligned}
t_*   =  \frac{L}{a_1^2 + a_2^2 } \left[ 
 \frac{1}{2} \le a_1+ 2a_2 + \frac{a_1^2}{a_2} \ri  \log \le \frac{1-\rho}{1+\rho} \ri + \le a_1+a_2 \ri \log \le \frac{a_1+a_2}{a_2-a_1} \ri
\right] \, .
\end{aligned}
\label{eq:scrambling_time_Cf_SdS3}
\eeq
\end{subequations}
Next, the approximation of the boundary term \eqref{eq:boundary_term_compl_formation} reads
\beq
 I_{\rm bdy}  (0) \approx 
- \frac{2 \rho^2}{2 G_N} a_1 L   \left[ 
\frac{a_1}{L} t_w +
\log \le \frac{1+\rho}{1-\rho} \frac{2G_N \mathcal{E}_1}{a_1^2} \, \varepsilon \ri   \right] \, ,
\label{eq:Ibdyform_3d}
\eeq
where the following expansion of the factor $a$ in eq.~\eqref{eq:cosmological_horizon_SdS3} has been used to simplify the result:
\beq
a_2 \approx
a_1 \le 1  + \varepsilon \, \frac{4 G_N \mathcal{E}_1}{a_1^2} \ri \, .
\eeq
Combining eqs.~\eqref{eq:Cf_largetw_SdS3} and \eqref{eq:Ibdyform_3d}, we finally obtain the formula \eqref{eq:linear_approx_CA0}.

\subsubsection{Higher dimensions}

In higher dimensions, it is not possible to find a closed expression for complexity.
However, many simplifications occur in the double-scaling limit \eqref{eq:double_scaling_limit}, and a linear expansion in $t_w$ can still be achieved.
The bulk term can be obtained as a simple consequence of the computation performed in \cite{Baiguera:2023tpt}, giving
\begin{subequations}
    \beq
I_{\mathcal{B}} (0) \approx \frac{d \Omega_{d-1}}{16 \pi G_N} (r_{c1})^{d-2} \le 1-\rho^2 \ri \le t_w - t_* \ri \, ,
\label{eq:IBform_gend}
\eeq
\beq
t_* \approx r_{c1} \log \le \frac{r_{c1}}{\beta r_{\rm cr}} \frac{1-\rho}{\varepsilon} \ri = 
\frac{1}{2 \pi T_{c1}}  \log \le \frac{r_{c1}}{\beta r_{\rm cr}} \frac{1-\rho}{\varepsilon} \ri  \, .
\label{eq:scrambling_formation}
\eeq
\end{subequations}
After manipulating the tortoise coordinate, we find
\beq
I_{\rm bdy} (0) \approx 
 \frac{\Omega_{d-1}}{4 \pi G_N} \left[ d(1-\rho^2) -2 \right] (r_{c1})^{d-2} \left\lbrace t_w -  r_{c1} \log \le   \frac{ r_{c1}}{\beta r_{\rm cr}} \frac{1-\rho}{\varepsilon} \ri \right\rbrace \, .
 \label{eq:Ibdyform_gend}
\eeq
Combining eqs.~\eqref{eq:IBform_gend} and \eqref{eq:Ibdyform_gend}, we obtain the result \eqref{eq:approx_CA_formation_gend}.

\section{Details for complexity=volume}
\label{app:details_CV}

\begin{figure}[ht]
    \centering
    \subfigure[]{\label{fig:case C} \includegraphics[scale=0.35]{Figures/caseE_diagram.pdf} }
     \subfigure[]{\label{fig:case D} \includegraphics[scale=0.35]{Figures/caseD_diagram.pdf} }
      \subfigure[]{\label{fig:case E} \includegraphics[scale=0.35]{Figures/caseC_diagram.pdf} }
  \subfigure[]{\label{fig:case A} \includegraphics[scale=0.35]{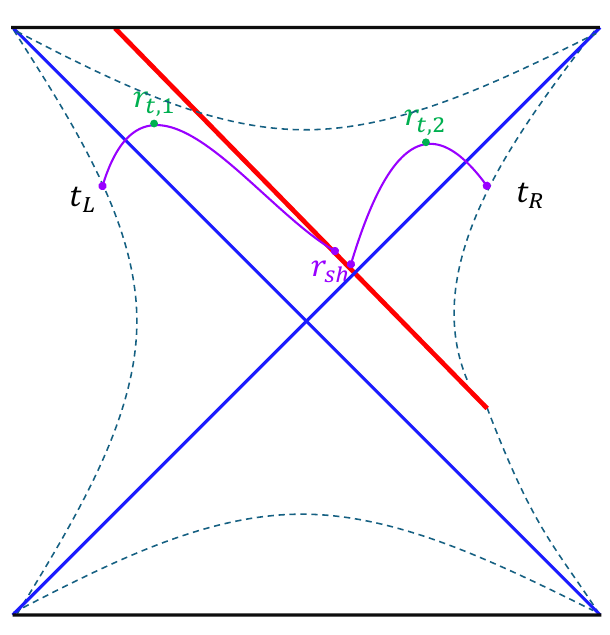} } 
   \subfigure[]{\label{fig:case B} \includegraphics[scale=0.35]{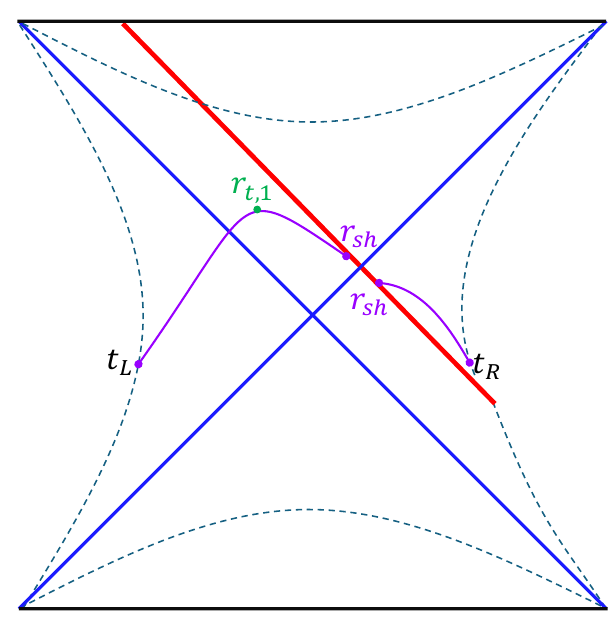} }
     \subfigure[]{\label{fig:case F} \includegraphics[scale=0.35]{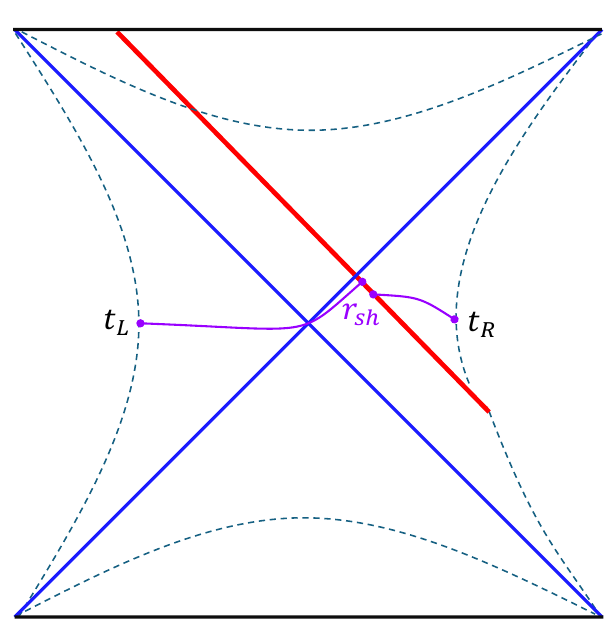} }
    \caption{Possible configurations for the extremal surface. A numerical analysis shows that cases (a)--(c) are the only configurations that occur during the time evolution, as reported in fig.~\ref{fig:relevant_extremal_surfaces_Penrose}.  }
    \label{fig:extremal_surfaces_Penrose}
\end{figure}

In this appendix, we provide additional technical details for the computation of CV. 
By applying the observations described in subsection~\ref{ssec:time_evo_CV}, we find that there exist six different possibilities for the shape that a maximal surface anchored at the stretched horizons can take.
They differ by the number of turning points, and whether the surface passes through the future or past exterior of the cosmological horizon, as depicted in fig.~\ref{fig:extremal_surfaces_Penrose} and summarized below:\footnote{For notational convenience, we denote with IP the inflationary patch of SdS space, \ie the region outside the cosmological horizon.}
\begin{itemize}
    \item \textbf{Case A.} The surface passes in the past IP$_1$ with $P_1<0$ and admits a turning point in IP$_1$, but not in BH$_2$.
      \item \textbf{Case B.} The surface passes in the future IP$_1$ with $P_1>0$ and does not admit turning points.
       \item \textbf{Case C.} The surface passes in the future IP$_1$ with $P_1>0$, admits a turning point in IP$_2$, but not in IP$_1$.
    \item \textbf{Case D.} The surface passes in the future IP$_1$ with $P_1>0$ and admits turning points both in IP$_1$ and IP$_2$.
    \item \textbf{Case E.} The surface passes in the future IP$_1$ with $P_1>0$, admits a turning point in IP$_1$, but not in IP$_2$.
       \item \textbf{Case F.} The surface has conserved momentum $P_1=0$ and passes through the bifurcation surface. There are no turning points.
\end{itemize}

\subsection{General strategy}
\label{ssec:extremal_surfaces_gen}

\begin{figure}[ht]
    \centering
    \includegraphics[scale=0.45]{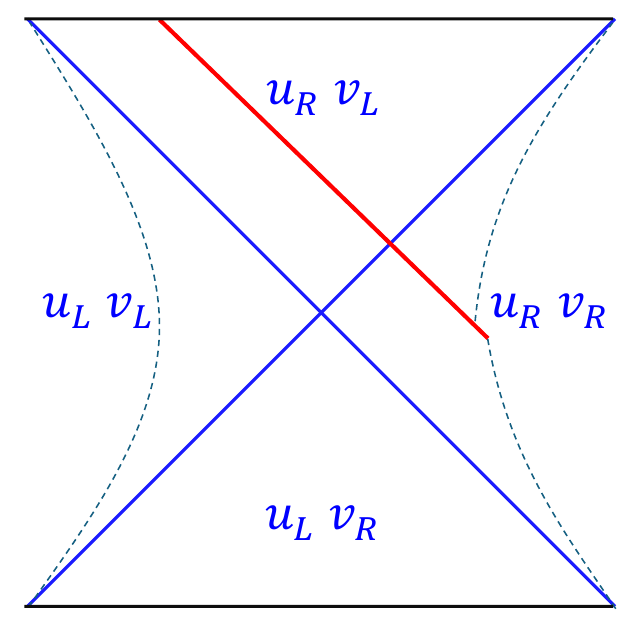} 
    \caption{Definition of the null coordinates in the various patches of the cosmological region of SdS space.}
    \label{fig:coordinate_system}
\end{figure}

In order to determine the shape of the extremal surfaces, we need to find a relation between the conserved momenta $P_i$ defined in eq.~\eqref{eq:conserved_momenta} and the boundary times $t_L, t_R$.
To this aim, we will integrate the extremal surface from the left to the right side of the Penrose diagram, and make use of the finite variations
\begin{subequations}
    \beq
\Delta u_{\pm} = \int \frac{\dot{u}_{\pm}}{\dot{r}_{\pm}} dr 
= - \int \tau[\mp P, r] dr \, ,
\label{eq:deltau_surface}
\eeq
 \beq
\Delta v_{\pm}  = \int \frac{\dot{v}_{\pm}}{\dot{r}_{\pm}} dr 
= \int \tau[\pm P, r] dr \, ,
\label{eq:deltav_surface}
\eeq
\end{subequations}
obtained by integrating eqs.~\eqref{eq:udot_volume}--\eqref{eq:vdot_volume} and plugging the definition \eqref{eq:deftau_CV} of $\tau$.
In the previous expressions, the $+$ sign is chosen when the radial coordinate increases towards the right side of the Penrose diagram, and the $-$ sign is chosen in the opposite case.
The null coordinates in all the quadrants are depicted in fig.~\ref{fig:coordinate_system}.

We outline the general method to compute the relation between boundary times and momenta:
\begin{enumerate}
    \item Starting from the left stretched horizon, we integrate the null coordinate $v$ (since it is always continuous across the past cosmological horizon $r_{c1}$) until we reach the shockwave.
    \item If there exists a turning point $r_{t1}$ in IP$_1$, we split the evaluation of \eqref{eq:deltav_surface} in two parts, since the radial coordinate will increase until the turning point and decrease afterwards. 
    \item The only exception to the previous rules is case A, where we decide to evaluate $\Delta u$ using eq.~\eqref{eq:deltau_surface} from the left stretched horizon until the turning point $r_{t,1}$, and we evaluate $\Delta v$ using eq.~\eqref{eq:deltav_surface} until the shock. 
    \item On the right side of the shockwave, we always evaluate $\Delta u$ by means of eq.~\eqref{eq:deltau_surface}, since it is always continuous in such region.
    \item We sum all the previous integrations, and we relate the result to the boundary times by using eq.~\eqref{eq:general_null_coordinates}.
\end{enumerate}

By following the previous steps, we get for each case an identity which relates the boundary times to the conserved momenta.
By imposing $t_L=t_R$ and fixing a choice of $P_1$, we can then use eq.~\eqref{eq:relation_P2_P1} to compute $P_2$, and then solve numerically the set of equations to obtain $r_{\rm sh}$.
This ultimately determines the time dependence of the extremal surfaces.
The maximal volume is then computed by integrating the function $R_i[P_i,r]$ in eq.~\eqref{eq:deftau_CV} from the left to the right stretched horizon, as we will do explicitly below.

\subsection{Analysis of the cases for the extremal surfaces}
\label{ssec:extremal_surfaces_cases}

A numerical analysis reveals that the only shapes that occur during the evolution correspond to configurations A--C.
Therefore, for practical convenience, we will only report these cases in the analysis below, following the bullet points outlined in appendix~\ref{ssec:extremal_surfaces_gen}.

\paragraph{Case A.}
In this case, depicted in fig.~\ref{fig:case C}, the intersection with the shockwave must lie in the region $r_{\rm sh} \leq r_{c1}$.
As specified in the third bullet in appendix~\ref{ssec:extremal_surfaces_gen},
on the left side we integrate the $u$ coordinate until the turning point, and then we integrate the $v$ direction. This gives
\begin{subequations}
    \beq
u_{t,1} - u^{\rm st}_L = - \int_{r^{\rm st}_1}^{r_{t,1}} dr \, \tau_1 [-P_1,r] \, ,
\eeq
\beq
v_s - v_{t,1} = \int_{r_{t,1}}^{r_{\rm sh}} dr \, \tau_1 [-P_1,r] \, ,
\eeq
\end{subequations}
which sum to
\beq
t_L - t_w + r^*_1(r^{\rm st}_1) - r^*_2(r^{\rm st}_2) - 2 r^*_1(r_{t,1}) + 2 r^*_1(r_{\rm sh})  =  - \int_{r^{\rm st}_1}^{r_{t,1}} dr \, \tau_1 [-P_1,r] 
+ \int_{r_{t,1}}^{r_{\rm sh}} dr \, \tau_1 [-P_1,r]  \, .
\label{eq:sumtimesL_caseE}
\eeq
On the right side of the Penrose diagram, we integrate the coordinate $u_R$ (which is continuous across the future horizon) to get
\beq
u^{\rm st}_R - u_s = t_R + t_w = - \int_{r_{\rm sh}}^{r^{\rm st}_2} dr \, \tau_2 [P_2,r] \, ,
\label{eq:sumtimesR_caseD}
\eeq
where $u_{s}$ is the null coordinate evaluated at the shockwave, and $u^{\rm st}_R$ is the value of the null coordinate at the right stretched horizon.
The volume reads
\beq
\frac{1}{\Omega_{d-1}} \mathcal{V} = 
\int_{r^{\rm st}_1}^{r_{t,1}} R_1 [P_1,r] dr  + \int_{r_{\rm sh}}^{r_{t,1}} R_1[P_1,r] dr  + \int_{r^{\rm st}_2}^{r_{\rm sh}} R_2[P_2,r] dr  \, .
\eeq

\begin{figure}[ht]
    \centering
    \subfigure[Case B1]{\includegraphics[scale=0.35]{Figures/caseD_diagram.pdf} }
    \subfigure[Case B2]{\includegraphics[scale=0.35]{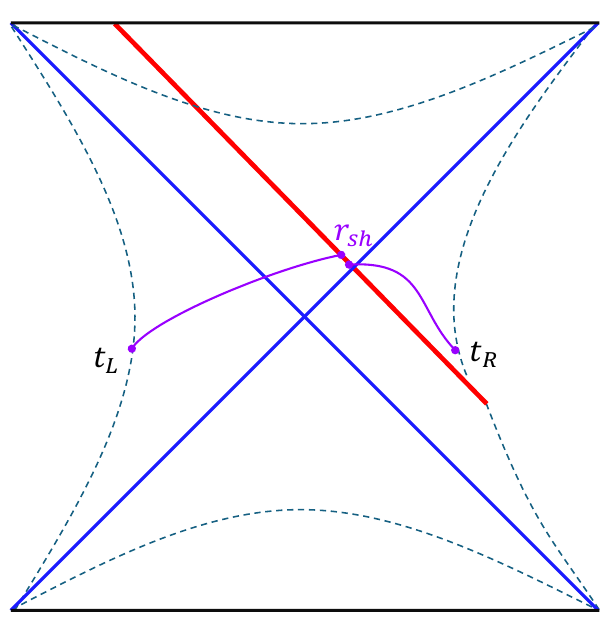} }
    \caption{Possible configurations of the maximal surface in case B.}
    \label{fig:casesD_maximal_surface}
\end{figure}

\paragraph{Case B.}
Without any turning point on the two sides, the intersection with the shockwave can either be located in the interval $r_{\rm sh} \in [r_{c1}, r_{c2}]$, or satisfy $r_{\rm sh} > r_{c2}$.
These possibilities are depicted in fig.~\ref{fig:casesD_maximal_surface}. %\rot{and the distinction comes from numerical reasons }.
Despite this distinction, the two cases can be formally treated together.

Starting from the left stretched horizon, in this case we do not have any turning point in the IP$_1$, therefore we obtain
\beq
v_s - v_L^{\rm st} = t_L - t_w -r^*_1(r^{\rm st}_1) -r^*_2(r^{\rm st}_2) + 2r^*_1(r_{\rm sh}) = \int_{r^{\rm st}_1}^{r_{\rm sh}} dr \, \tau_1 [P_1, r] \, ,
 \label{eq:sumtimesL_caseC}
\eeq
where $v^{\rm st}_L$ is the value of the null coordinate at the left stretched horizon and $v_{s}$ at the turning point. 
The right side is evaluated in the same way as case A, giving
\beq
t_R + t_w = - \int_{r_{\rm sh}}^{r^{\rm st}_2} dr \, \tau_2 [P_2,r] \, .
\label{eq:sumtimesR_caseE}
\eeq
The maximal volume is
\beq
\frac{1}{\Omega_{d-1}} \mathcal{V} = 
\int_{r^{\rm st}_1}^{r_{s}} dr \, R_1 [P_1,r]  + \int_{r^{\rm st}_2}^{r_{\rm sh}} dr \, R_2[P_2,r]   \, ,
\eeq

\paragraph{Case C.}
The Penrose diagram is depicted in fig.~\ref{fig:case E}.
In order for a turning point to exist after the shockwave, we need $r_{\rm sh} \geq r_{c2}$.
The evaluation of the left side coincides with case B, giving
\beq
 t_L - t_w -r^*_1(r^{\rm st}_1) -r^*_2(r^{\rm st}_2) + 2r^*_1(r_{\rm sh}) = \int_{r^{\rm st}_1}^{r_{\rm sh}} dr \, \tau_1 [P_1, r] \, ,
\label{eq:sumtimesL_caseD}
\eeq
On the right side of the Penrose diagram, we use the coordinate $u_R$, continuous across the cosmological horizon $r_{c2}$. Since by assumption there is a turning point in IP$_2$, we split the integration in two parts:
\begin{subequations}
\beq
u_{t,2} - u_s = - \int_{r_{\rm sh}}^{r_{t,2}} dr \, \tau_2 [-P_2,r] \, ,
\eeq    
\beq
u^{\rm st}_R - u_{t,2} = - \int_{r_{t,2}}^{r^{\rm st}_2} dr \, \tau_2 [P_2, r] \, .
\eeq
\end{subequations}
Summing the previous expressions in region 2 gives
\beq
t_R + t_w = - \int_{r_{\rm sh}}^{r_{t,2}} dr \, \tau_2 [-P_2,r] 
+ \int_{r^{\rm st}_2}^{r_{t,2}} dr \, \tau_2 [P_2, r] \, .
 \label{eq:sumtimesR_caseA}
\eeq
The volume of the maximal surface is easily obtained by integrating $R_i[P_i,r]$ in eq.~\eqref{eq:deftau_CV} along the three portions of the extremal surface:
\beq
\frac{1}{\Omega_{d-1}} \mathcal{V} = 
\int_{r^{\rm st}_1}^{r_{s}} dr \, R_1 [P_1,r]  + \int_{r_{\rm sh}}^{r_{t,2}}  dr \, R_2[P_2,r]  + \int_{r^{\rm st}_2}^{{r_{t,2}}} dr \, R_2[P_2,r]  \, .
\label{eq:caseA_vol}
\eeq

\subsection{Rate of growth of the volume}
\label{ssec:extremal_surfaces_rate}

We compute the rate of growth of the CV conjecture.
The calculation can be done for each case separately, but since it leads to a universal result, we will only present it for a representative shape, \ie in case C.
In view of the following computations, we list a set of useful definitions and identities:
\begin{subequations}
    \beq
    \tilde{R}[P,r] \equiv \frac{\sqrt{f(r) r^{2(d-1)} +P^2} - P }{f(r)} \, .
    \label{eq:def_R_Rtilde}
    \eeq
    \beq
R[P,r] = \tilde{R}[P,r] +P \, \tau[P,r] = \tilde{R}[-P,r] - P \, \tau[-P,r] \, ,
\label{eq:identity1_Rtau}
    \eeq
    \beq
\p_P \tilde{R}[P,r] = - \tau[P,r] \, , \qquad
\p_P \tilde{R}[-P,r] = \tau[-P,r] \, ,
\label{eq:identity2_Rtau}
    \eeq
    \beq
\tilde{R}[P,r_t] = - \frac{P}{f(r)} \, ,
\label{eq:identity3_Rtau}
    \eeq
    \beq
\tilde{R}[P,r] = \tilde{R}[-P,r] - \frac{2P}{f(r)} = - \dot{u}_+ [-P,r] = \dot{u}_- [P,r] =
\dot{v}_+[P,r] = - \dot{v}_-[-P,r] \, .
\label{eq:identity4_Rtau}
    \eeq
\end{subequations}

Starting from the volume~\eqref{eq:caseA_vol}, we use the identities \eqref{eq:sumtimesL_caseC}, \eqref{eq:sumtimesR_caseA} and \eqref{eq:identity1_Rtau} to obtain
\beq
\begin{aligned}
\frac{1}{\Omega_{d-1}} \mathcal{V} & = 
\int_{r^{\rm st}_1}^{r_{s}} \tilde{R}_1 [P_1,r] dr   + \int_{r_{\rm sh}}^{r_{t,2}} \tilde{R}_2[-P_2,r] dr  + \int_{r^{\rm st}_2}^{r_{t,2}} \tilde{R}_2[P_2,r] dr \\
& +P_1 \left[ t_L - t_w + 2 r^*_1(r_{s}) - r^*_1(r^{\rm st}_1) - r^*_2(r^{\rm st}_2) \right] + P_2 \le t_R + t_w \ri \, .
\end{aligned}
\label{eq:intermediate_volume_caseC}
\eeq
The continuity of $\dot{u}$ across the shockwave implies 
\beq
\dot{u}_+ [P_1,r] - \dot{u}_+ [P_2,r] = 0  \quad \Rightarrow
\quad
\tilde{R}_1 [P_1,r] + \frac{2 P_1}{f_1(r_{\rm sh})} = \tilde{R}_2 [-P_2,r] \, ,
\label{eq:identity5_caseC}
\eeq
where the identity \eqref{eq:identity4_Rtau} was used.

The time derivative of eq.~\eqref{eq:intermediate_volume_caseC} is then given by
\beq
\begin{aligned}
\frac{1}{\Omega_{d-1}} \frac{d\mathcal{V}}{dt} & = 
 \frac{dr_{t,2}}{dt} \le  \tilde{R}_2[-P_2,r_{t,2}] + \tilde{R}_2[P_2,r_{t,2}] \ri \\
& + \frac{dP_1}{dt} \left[ - \int_{r^{\rm st}_1}^{r_{s}} \tau_1 [P_1,r] dr  + \le  t_L - t_w + 2 r^*_1(r_{s}) - r^*_1(r^{\rm st}_1) - r^*_2(r^{\rm st}_2) \ri \right] \\
& + \frac{dP_2}{dt} \left[ \int_{r_{\rm sh}}^{r_{t,2}} \tau_2[-P_2,r] dr  - \int_{r^{\rm st}_2}^{r_{t,2}} \tau_2[P_2,r] dr  + \le t_R + t_w \ri \right] \\
& + \frac{dr_{\rm sh}}{dt} \le \tilde{R}_1 [P_1,r_{\rm sh}] - \tilde{R}_2[-P_2,r_{\rm sh}]  + \frac{2 P_1}{f_1(r_{\rm sh})} \ri
+ P_1 \frac{dt_L}{dt} + P_2 \frac{dt_R}{dt} \, ,
\end{aligned}
\eeq
where eq.~\eqref{eq:identity2_Rtau} was applied.
After using eqs.~\eqref{eq:identity3_Rtau}, \eqref{eq:identity5_caseC} and the identities determined in case C to relate the boundary times to the conserved momenta, we find that all the terms in parenthesis vanish.
This gives the simple result
\beq
\frac{d\mathcal{V}}{dt} = \Omega_{d-1} \le P_1 \frac{dt_L}{dt} + P_2 \frac{dt_R}{dt}  \ri \, ,
\eeq
which can be shown to hold for any shape of the extremal surface.

\bibliographystyle{JHEP}

\bibliography{bibliography}

\end{document}